\documentclass[%
 reprint,
superscriptaddress,
 amsmath,amssymb,
 aps,
pra
]{revtex4-1}
\usepackage{graphicx,amsmath}%
\usepackage{color,xcolor}
\usepackage[bookmarks=false]{hyperref}
\hypersetup{colorlinks=true,citecolor=cyan,linkcolor=blue,urlcolor=blue,pdfstartview=FitH,bookmarksopen=true}
\begin{document}

\title{Dipole-dipole interaction driven antiblockade of two Rydberg atoms}
 \author{Shi-Lei Su}
 \affiliation{School of Physics, Zhengzhou University, Zhengzhou 450001, China}
 \author{Weibin Li}
 \email{weibin.Li@nottingham.ac.uk}
 \affiliation{School of Physics and Astronomy, University of Nottingham, Nottingham NG7 2RD, United Kingdom}

\begin{abstract}
Resonant laser excitation of multiple Rydberg atoms are prohibited, leading to Rydberg blockade, when the long-range van der Waals interactions are stronger than the laser-atom coupling. Rydberg blockade can be violated, i.e. simultaneous excitation of more than one Rydberg atoms, by off-resonant laser excitation, causing an excitation antiblockade. Rydberg antiblockade gives rise to strongly correlated many-body dynamics and spin-orbit coupling, and also finds quantum computation applications. Instead of commonly used van der Waals interactions, we investigate antiblockade dynamics of two Rydberg atoms interacting via dipole-dipole exchange interactions. We study typical situations in current Rydberg atoms experiments, where different types of dipole-dipole interactions can be achieved by varying Rydberg state couplings. Effective Hamiltonian governing underlying antiblockade dynamics is derived. We illustrate that geometric gates can be realized with the Rydberg antiblockade which is robust against decay of Rydberg states.  Our study may stimulate new experimental and theoretical exploration of quantum optics and strongly interacting many-body dynamics with Rydberg antiblockade driven by dipole-dipole interactions.
\end{abstract}
\maketitle

\section{Introduction}\label{Sec1}
Highly excited Rydberg atoms with principal quantum number $n\gg1$ exhibit strong and long-range van der Waals (vdW) interactions due to their large polarizibility ($\sim n^{7}$) and strong interactions ($\sim n^{11}$)~\cite{Gallagher1994}. When excited from ground states with resonant laser lights, Rydberg blockade emerges in which excitation of two neighboring Rydberg atoms are prohibited due to energy shifts induced by vdW interactions. Rydberg blockade provides a mechanism in realizing quantum logic gates~\cite{djp2000,mmr2001,mtk2010,Comparat:10,li_entangling_2014,Saffman_2016}, which have been demonstrated experimentally~\cite{lex2010,xla2010,tac2010,kmt2015,ypx2017,Picken_2018,saa2018,levine2019parallel,graham2019rydberg,Omran570}. In contrast to Rydberg blockade, the interaction-induced excitation of two Rydberg atoms is referred to \textit{Rydberg antiblockade}~(RAB)~\cite{Ates2007}. Subsequently, the relevant experiment has also observed signatures of Rydberg antiblockade~\cite{Amthor2010}.  The strict condition for RAB was analyzed~\cite{Zuo2010,Lee2012}. RAB plays roles in the study of  motional effects~\cite{PhysRevLett.109.233003,Li2013}, dissipative dynamics~\cite{Carr2013,chen2018accelerated,*Li2019,*Yang2019,Lirui2020}, periodically driving~\cite{PhysRevLett.120.123204}, and quantum computation~\cite{Su2018,*Su2020,*Wu2020,*Zheng2020,Su201702}. RAB was also studied in detection of structural phase transitions~\cite{Gambetta20202}, Rydberg spin system~\cite{Mazza2020}, cold atom ensemble~\cite{taylor2019generation}, as well as in strongly interacting Rydberg atom experiment~\cite{Bai_2020}.
	
\begin{figure}[htb]\centering
	\includegraphics[width=\linewidth]{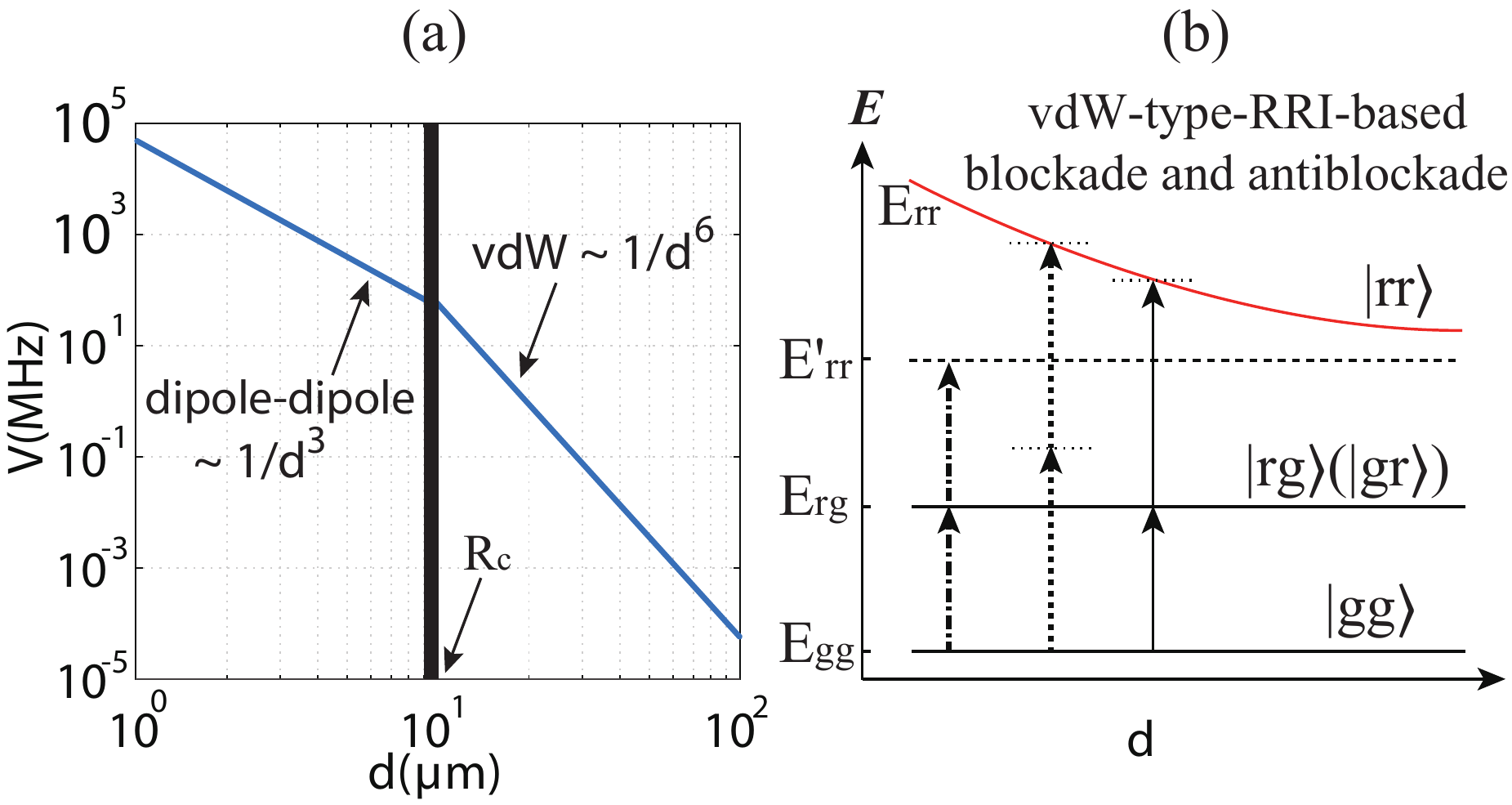}
	\caption{(a) Two-body interaction strength for Rb atoms excited to Rydberg state $|100s\rangle$ versus inter-atomic distance $d$. $R_{c}$ denotes the crossover distance between DD and vdW interactions~\cite{mtk2010}. (b) The dynamics of Rydberg blockade and antiblockade with vdW-type RRI. $E_{gg}$, $E_{rg}$ and $E_{rr}$ denote the energies of the two-atom state $|gg\rangle$, $|gr\rangle(|rg\rangle)$ and $|rr\rangle$, respectively. $E'_{rr}$ denotes the energy when both atoms are excited in Rydberg states but excluding two-body interactions. The resonant laser excitation~(dotted-dashed line) leads to the Rydberg blockade~\cite{djp2000,mmr2001,mtk2010}. The middle excitation process~(dotted line) is the conventional Rydberg antiblockade with simultaneous driving~\cite{Zuo2010,Lee2012,Li2013,Carr2013,chen2018accelerated,*Li2019,*Yang2019,Lirui2020,PhysRevLett.120.123204,Su2018,*Su2020,*Wu2020,*Zheng2020,Gambetta20202,Mazza2020,taylor2019generation}. The right one~(solid line) is the RAB with sequential driving~\cite{Su201702}.}\label{f001}
\end{figure}

The vdW and dipole-dipole~(DD) interactions exhibit different features. In Fig,~\ref{f001}(a), we show the strength and interaction range of DD and vdW interactions, respectively, focusing on one group of specific Rydberg states. The DD interaction is stronger at short distances, while the vdW interaction is stronger at long distances~\cite{Saffman_2016}. Most importantly, DD interactions typically involve two or more Rydberg states in the dynamics. In Fig.~\ref{f001}(b), we show regimes to realizing Rydberg blockade~\cite{djp2000,mmr2001,mtk2010}, conventional RAB with simultaneous driving~\cite{Zuo2010,Lee2012,Li2013,Carr2013,chen2018accelerated,*Li2019,*Yang2019,Lirui2020,PhysRevLett.120.123204,Su2018,*Su2020,*Wu2020,*Zheng2020,Gambetta20202,Mazza2020,taylor2019generation} as well as sequential-driving-based RAB~\cite{Su201702}, where the excitation conditions can be controlled by laser detuning.  When using DD interactions, the density-density as well as spin flip-flop interactions co-exist~\cite{Cidrim2020,Williamson2020}, leading to complicated many-body dynamics~\cite{orioli_relaxation_2018}. The resonant DD interactions are considered to construct two-~\cite{Beterov2016} and three-qubit~\cite{Beterov2018} quantum logic gates by using of experimentally observed F\"orster resonance~\cite{Tretyakov2017}. It has been shown that RAB can be used to limit the blockade error~\cite{Shi2017}, and to construct the multiple qubit Toffoli and Fan-out gates in a fast way~\cite{Klaus2020}. Recently, it has been shown that non-adiabatic dynamics around a conical intersection can be studied under the RAB condition with trapped Rydberg ions~\cite{PhysRevLett.126.233404}. 

Although there are different level schemes to achieve DD interactions, it is not clear how to achieve the RAB condition for the many types of DD interactions between Rydberg atoms. Moreover existing schemes typically require two or more Rydberg atoms in the Rydberg state simultaneously for a period of time~\cite{Beterov2016,Tretyakov2017,Beterov2018,Klaus2020} or to stay in a dark state~\cite{Pohl2009}. This could reduce coherence of the system due to, e.g. motional effects~\cite{PhysRevLett.109.233003,Li2013}. 

In this work  we study  RAB driven by different types of Rydberg DD interactions. We propose new schemes to realize RAB efficiently for three types of DD interactions that are typically encountered in various experiments. The first type is the F\"orster resonance, such as transitions given by $|d\rangle|d\rangle\leftrightarrow |p\rangle|f\rangle+|f\rangle|p\rangle$~\cite{Nipper2012,Ravets2014,Ravets2015,Bohlouli2007}, and $|p\rangle|p\rangle\leftrightarrow |s\rangle|s'\rangle+|s'\rangle|s\rangle$~\cite{Yakshina2016,Liu2020}. The second type is spin-exchange type RRIs via $|s\rangle|p\rangle\leftrightarrow |p\rangle|s\rangle$~\cite{Browaeys_2016, Ates_2008}, $|p\rangle|d\rangle\leftrightarrow |d\rangle|p\rangle$~\cite{Barredo2015} or $|s\rangle|p'\rangle\leftrightarrow |p\rangle|s'\rangle$~\cite{young2020asymmetric}. The third type is collective exchange interaction, i.e.  $|s\rangle|s'\rangle\leftrightarrow |p\rangle|p'\rangle$~\cite{Gallagher1998,Gorniaczyk2016,Beterov2016,Petrosyan2017,Liu2020,beterov2018resonant,Klaus2020}. Effective Hamiltonians of the different types of DD interactions are provided. When applying the proposed schemes in realizing quantum logic gates, the main feature is that only a one-step Rabi oscillation between the ground states and the multi-excited Rydberg states is required, without staying in the Rydberg states for a long period of time, avoiding disadvantages found in other schemes. We also discuss impacts of dissipation on the RAB and propose parameters to probe RAB.

The rest content of the manuscript is organized as follows: In Sec.~\ref{Sec3}, we show details on how to achieve RAB regime. The effective Hamiltonian is given, and the respective dynamics influenced by dissipation is studied with a quantum master equation. We show population evolution of different models. In Sec.~\ref{sec:comparison}, the main difference between the vdW and DD interactions are shown, which gives distinctive dynamics. In Sec.~\ref{Sec4}, we show the potential applications of the proposed RAB in building two-qubit quantum gates and creating steady state entanglement. The conclusion is given in Sec.~\ref{Sec5}.

\section{Antiblockade with different types of DD interactions}\label{Sec3}

\subsection{RAB with the F\"orster resonance}\label{s3.a}
\subsubsection{Level scheme and model}\label{s3.a.1}
\begin{figure}[http!]\centering
	\includegraphics[width=\linewidth]{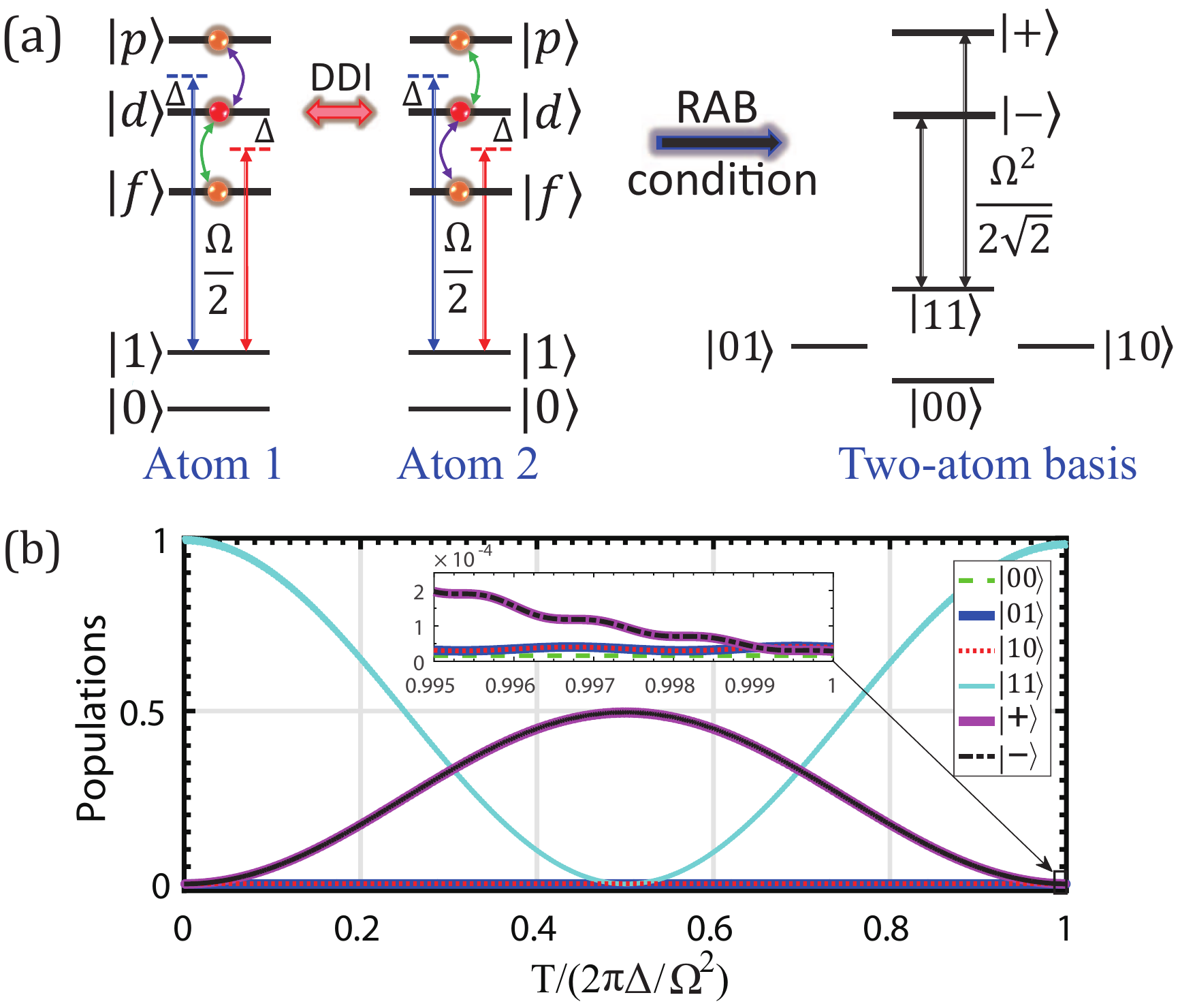}
	\caption{(a), Left panel shows two Rydberg atoms with resonant RRI. $|0\rangle$ and $|1\rangle$ are two ground states. $|p\rangle$, $|d\rangle$ and $|f\rangle$ are three Rydberg states with the F\"orster resonance interaction $\hat{H}_d=V_{d}(|dd\rangle\langle pf| + |dd\rangle\langle fp|+{\rm H.c.})$. Right panel gives the effective RAB process in the dressed state basis. (b), Populations of difference states for RAB scheme in Sec.~\ref{s3.a} during one evolution period $T=2\pi\Delta/\Omega^2$ with the consideration of practical atomic spontaneous emission $\gamma_p=1.89$~kHz, $\gamma_{d}=4.55$~kHz and $\gamma_{f}=7.69$~kHz. The inset shows that the dressed state decays to zero at the end of the laser pulse. Parameters are  $\Omega=2\pi\times5$~MHz and $\Delta$ is set to satisfy the antiblockade condition. The initial state is $|11\rangle$ and the inter-atomic distance is  3~$\mu$m.}\label{f002}
\end{figure}
To realize F\"oster resonance we consider the experimental configuration~\cite{Ravets2014} $|p\rangle\equiv|61P_{1/2}, m_J=1/2\rangle$, $|d\rangle\equiv|59D_{3/2},m_{J}=3/2\rangle$ and $|f\rangle\equiv|57F_{5/2},m_{J}=5/2\rangle$ of two $^{87}$Rb atoms, as shown in Fig.~\ref{f002}. By applying an electric fields $\epsilon=32$~mV cm$^{-1}$, these Rydberg states can be brought to exact resonance. One of the states in computational space is chosen as $|1\rangle\equiv|5S_{1/2},F=2,m_{F}=2\rangle$~\cite{Ravets2014} and the other state $|0\rangle$ in computational subspace is decoupled with the excitation process and may be chosen as $|0\rangle\equiv|5S_{1/2},F=1,m_{F}=0\rangle$. The excitation is accomplished by a two-photon process with two lasers with wavelengths 795 nm($\pi$ polarization) and 474 nm($\sigma_{+}$ polarization)~\cite{Ravets2014}. Bichromatic classical fields are imposed on these two atoms to off-resonantly drive the transition $|1\rangle\leftrightarrow|d\rangle$ with an identical Rabi frequency $\Omega$ but opposite detuning $\Delta$. 
With the rotating-wave approximation, the Hamiltonian for this system can be written as $\hat{H}=\hat{H}_{\Omega}+\hat{H}_{d}$ ($\hbar\equiv1$), where
\begin{eqnarray}\label{e1}
\hat H_{\Omega}&=&\frac{\Omega}{2}\left(e^{i\Delta t}+e^{-i\Delta t}\right)(|1\rangle_{1}\langle d|\otimes\mathcal{I}_{2}+\mathcal{I}_{1}\otimes|1\rangle_{2}\langle d|)
+{\rm H.c.}
\cr\cr&=&\frac{\Omega}{2}\left(e^{i\Delta t}+e^{-i\Delta t}\right)(|10\rangle\langle d0| + |11\rangle\langle d1| + |1p\rangle\langle dp| \cr\cr&& + |1d\rangle\langle dd| +|1f\rangle\langle df| + |01\rangle\langle0d| + |11\rangle\langle1d| \cr\cr&& +|p1\rangle\langle pd| +|d1\rangle\langle dd| + |f1\rangle\langle fd|)+{\rm H.c.}
\cr\cr \hat H_{d}&=&\sqrt{2}V_{d}|dd\rangle\langle r_{pf}|+{\rm H.c.}
\end{eqnarray}
where $\mathcal{I}_j$ denotes the identity matrix of atom \emph{j}, $V_{d}=C_3/r_{d}^3$ denotes the DD interaction strength. Here $C_3 =2.54$~GHz$\cdot\mu m^3$ ~\cite{Ravets2014,Singer_2005,*PhysRevA.84.041607,*SIBALIC2017319} and $r_d$ denotes the interatomic distance. $|mn\rangle$ denotes two atom state $|m\rangle_{1}\otimes|n\rangle_{2}$ and will be used throughout this manuscript. We have defined two atom state $|r_{pf}\rangle\equiv(|pf\rangle+|fp\rangle)/\sqrt{2}$. 

\subsubsection{Effective Hamiltonian}
To simplify the calculation, we first derive a Hamiltonian using the dressed state basis. It should be mentioned that the dressing here is different from Rydberg dressing of the ground state, which mainly generates long-range  interactions between ground state atoms~\cite{PhysRevLett.104.223002,Jau2016,Zeiher2016,PhysRevLett.105.160404,PhysRevLett.114.173002,PhysRevA.82.033412,PhysRevLett.114.243002,Balewski_2014,PhysRevLett.108.265301,PhysRevLett.115.093002,Tanasittikosol_2011,PhysRevA.89.011402,PhysRevA.87.052314,W_ster_2011,PhysRevA.91.012337,PhysRevLett.111.165302,li_probing_2012,PhysRevLett.116.243001,PhysRevLett.113.123003,mukherjee_phase-imprinting_2015,PhysRevA.95.013403,li_supersolidity_2018,PhysRevA.95.041801,zhou_quench_2020,mccormack_dynamical_2020,li_many-body_2020}.
One can diagonalize $\hat{H}_d$ as $\sqrt2V_{d}(|+\rangle\langle +|-|-\rangle\langle -|)$ with $|\pm\rangle\equiv(|dd\rangle\pm|r_{pf}\rangle)/\sqrt2$ being the dressed states. Then the  Hamiltonian can be written as
\begin{eqnarray}\label{e2}
{\hat H}_{\Omega}&=&\frac{\Omega}{2}\left(e^{i\Delta t}+e^{-i\Delta t}\right)\big[\sqrt{2}| 11\rangle\langle \Psi|+|\Psi\rangle(\langle +|+\langle -|)\big]+{\rm H.c.}
\cr\cr&&+\frac{\Omega}{2}\left(e^{i\Delta t}+e^{-i\Delta t}\right)(|01\rangle\langle 0d|+|10\rangle\langle d0|)+{\rm H.c.}\cr\cr
\hat{H}_{d}&=&\sqrt2V_{d}(|+\rangle\langle +|-|-\rangle\langle -|),
\end{eqnarray}
in which $|\Psi\rangle\equiv(|1d\rangle+|d1\rangle)/\sqrt2$.
From Eq.~(\ref{e2}), Hamiltonian $\hat{H}_{\Omega}$ itself describes resonant interactions when $\Delta=0$. However, when $V_d\gg\Omega$, after rotating the total Hamiltonian $\hat{H}$ with respect to $\hat{H}_d$, one can see that the two-excitation Rydberg states would be coupled off-resonantly with large detuning. Thus the Rydberg blockade is produced. In the following we would show how to achieve the RAB even when $V_d\gg\Omega$.

When the RRI strength is much stronger than Rabi frequency, the aim is to use the laser detuning to compensate the energy shift induced by the RRI~\cite{Su_2020epl}. And it is precisely from this point that one always rotates the whole Hamiltonian with respect to the RRI-related Hamiltonian, which is convenient to get the relation between laser detuning and RRI strength since $V_{d}$ is also moved to the exponential part (i.e. contributing to the phase)~\cite{Su2016,Su_2020epl}. With this at hand, one can employ the second-order perturbation theory to obtain the RAB condition. After rotating the whole Hamiltonian $\hat{H}_{\Omega}+\hat{H}_{d}$ with respect to $e^{i\hat{H}_{d}t}$, this yields~\cite{PhysRevLett.120.123204}
\begin{widetext}
\begin{eqnarray}\label{e3}
\hat {\mathcal{H}}&=&\big\{\frac{\Omega}{2}\big[\sqrt2\left(e^{i\Delta t}+e^{-i\Delta t}\right)|11\rangle\langle \Psi|+\left(e^{i(\Delta-\sqrt{2}V_{d}) t}+e^{-i(\Delta+\sqrt{2}V_{d}) t}\right)|\Psi\rangle\langle +|+\left(e^{i(\Delta+\sqrt{2}V_{d}) t}+e^{-i(\Delta-\sqrt{2}V_{d}) t}\right)|\Psi\rangle\langle -| \cr\cr&&+ \left(e^{i\Delta t}+e^{-i\Delta t}\right)(|01\rangle\langle 0d|+|10\rangle\langle d0|)\big]+{\rm H.c.}\big\}
\end{eqnarray}
\end{widetext}
If  the conditions $\{\Delta, \Delta\pm\sqrt{2}V_{d}\} \gg \Omega$, and $V_{d}=\sqrt2\Delta$ are satisfied, the effective form of Hamiltonian~(\ref{e3}) can be achieved through the second-order perturbation calculation~\cite{effective1,effective2,effective3,effective4} as ~[See the Appendix A for details]
\begin{eqnarray}\label{eq06}
\hat H_{\rm e}=&&\frac{\Omega^2}{2\Delta}|11\rangle\langle +|-|11\rangle\langle-|+{\rm H.c.}\cr\cr&&+\frac{\Omega^2}{3\Delta}(|+\rangle\langle+|-|-\rangle\langle-|).
\end{eqnarray} From Eq.~(\ref{eq06}), one can see that the collective ground state $|11\rangle$ is resonantly coupled with the two-excitation Rydberg state $|r_{pf}\rangle$ with effective Rabi frequency $\Omega_{\rm eff}\equiv\Omega^2/\Delta$, leading to the RAB. Here the Stark shift in Eq.~(\ref{eq06}) would no doubt influence the dynamics. One can remove the Stark shift by modifying the condition $V_{d}=\sqrt2\Delta$ to 
\begin{equation*}
V_{d}=\sqrt2\Delta-\Omega^2/(3\sqrt{2}\Delta),
\end{equation*}
the effective Hamiltonian~(\ref{eq06}) would be changed to
\begin{eqnarray}\label{e5}
\hat H_{\rm e}=\frac{\Omega^2}{2\sqrt{2}\Delta}|11\rangle(\langle+|-\langle-|)+{\rm H.c.}
\end{eqnarray}

Here we should mention that in Ref.~\cite{Ravets2014}, the ground state $|gg\rangle$ (Corresponding to $|11\rangle$ in our manuscript) is excited to Rydberg state $|dd\rangle$ firstly via $\pi$ pulse through the detuned laser. Then the electric field is tuned to make state $|dd\rangle$ resonant with $(|pf\rangle+|fp\rangle)/\sqrt{2}$. We consider the strong F\"orser resonant interactions from the beginning, and designed schemes to achieve the Rabi oscillation from collective ground state to two-excitation Rydberg state $(|pf\rangle+|fp\rangle)/\sqrt{2}$. Meanwhile, the states $|00\rangle$, $|01\rangle$ and $|10\rangle$ are decoupled with the two-excitation Rydberg states, which is convenient when applying this model for quantum information processing.

\subsubsection{Population dynamics at the RAB regime}
The effective Hamiltonian~(\ref{e5}) shows that perfect Rabi oscillation between the ground state and the doulbly excited Rydberg state can happen. In this section, we check the validity of the effective Hamiltonian~(\ref{e5}) by comparing dynamics obtained from the original Hamiltonian in the RAB regime. We furthermore take into account of spontaneous emission of Rydberg states. The dynamics of the system is governed by the master equation
\begin{eqnarray}\label{master1}
\dot{\hat{\rho}}=i[\hat{\rho},\hat {H}]+\frac{1}{2}\sum_{k}[ 2\hat{\mathcal{L}}_{k}\hat{\rho}\hat{\mathcal{L}}_{k}^{\dag}-\hat{\mathcal{L}}_{k}^{\dag}\hat{\mathcal{L}}_{k}\hat{\rho}-\hat{\rho}\hat{\mathcal{L}}_{k}^{\dag}\hat{\mathcal{L}}_{k}]
\end{eqnarray} 
where $\hat{\rho}$ denote the density matrix of system state, $\mathcal{\hat{L}}_{k}$ is the $k$-th Lindblad operator describing the dissipation process, and $\hat{H}=\hat{H}_{\Omega}+\hat{V}_d$ is the original Hamiltonian~(\ref{e1}).
 The lifetimes for $|p\rangle$, $|d\rangle$ and $|f\rangle$ are about 0.53 ms, 0.22 ms and 0.13 ms, respectively~\cite{lifetime1,lifetime2}. The Lindblad operators are given explicitly as
\begin{eqnarray}
\hat{\mathcal{L}}_{1}&=&\sqrt{\gamma_p/2}|0\rangle_{1}\langle p|,~~ \hat{\mathcal{L}}_{2}=\sqrt{\gamma_p/2}|1\rangle_{1}\langle p|,\cr\cr \hat{\mathcal{L}}_{3}&=&\sqrt{\gamma_d/2}|0\rangle_{1}\langle d|,~~ \hat{\mathcal{L}}_{4}=\sqrt{\gamma_d/2}|1\rangle_{1}\langle d|,\cr\cr \hat{\mathcal{L}}_{5}&=&\sqrt{\gamma_f/2}|0\rangle_{1}\langle f|,~~ \hat{\mathcal{L}}_{6}=\sqrt{\gamma_f/2}|1\rangle_{1}\langle f|,\cr\cr \hat{\mathcal{L}}_{7}&=&\sqrt{\gamma_p/2}|0\rangle_{2}\langle p|,~~ \hat{\mathcal{L}}_{8}=\sqrt{\gamma_p/2}|1\rangle_{2}\langle p|,\cr\cr \hat{\mathcal{L}}_{9}&=&\sqrt{\gamma_d/2}|0\rangle_{2}\langle d|,~~ \hat{\mathcal{L}}_{10}=\sqrt{\gamma_d/2}|1\rangle_{2}\langle d|,\cr\cr \hat{\mathcal{L}}_{11}&=&\sqrt{\gamma_f/2}|0\rangle_{2}\langle f|,~~ \hat{\mathcal{L}}_{12}=\sqrt{\gamma_f/2}|1\rangle_{2}\langle f|,
\end{eqnarray} where $\gamma_{j}$ denotes the atomic spontaneous emission rate.

Numerical results by solving the master equation are shown in Fig.~\ref{f002}(b). The evolution of the state under the given RAB condition is plotted, where the calculation takes into account of practical atomic spontaneous emission rates. It can be seen that the initial state can be fully converted to the dressed state, as described by the effective Hamiltonian. Here we should point out that, the original Hamiltonian rather than the effective Hamiltonian is used in evolving the master equation. This means that ideal RAB can be achieved with the DD interaction through the F\"orster resonance.

\subsection{RAB with spin-exchange interaction}\label{s3.b}
\subsubsection{Level scheme and model}
\begin{figure}[http!]\centering
	\includegraphics[width=\linewidth]{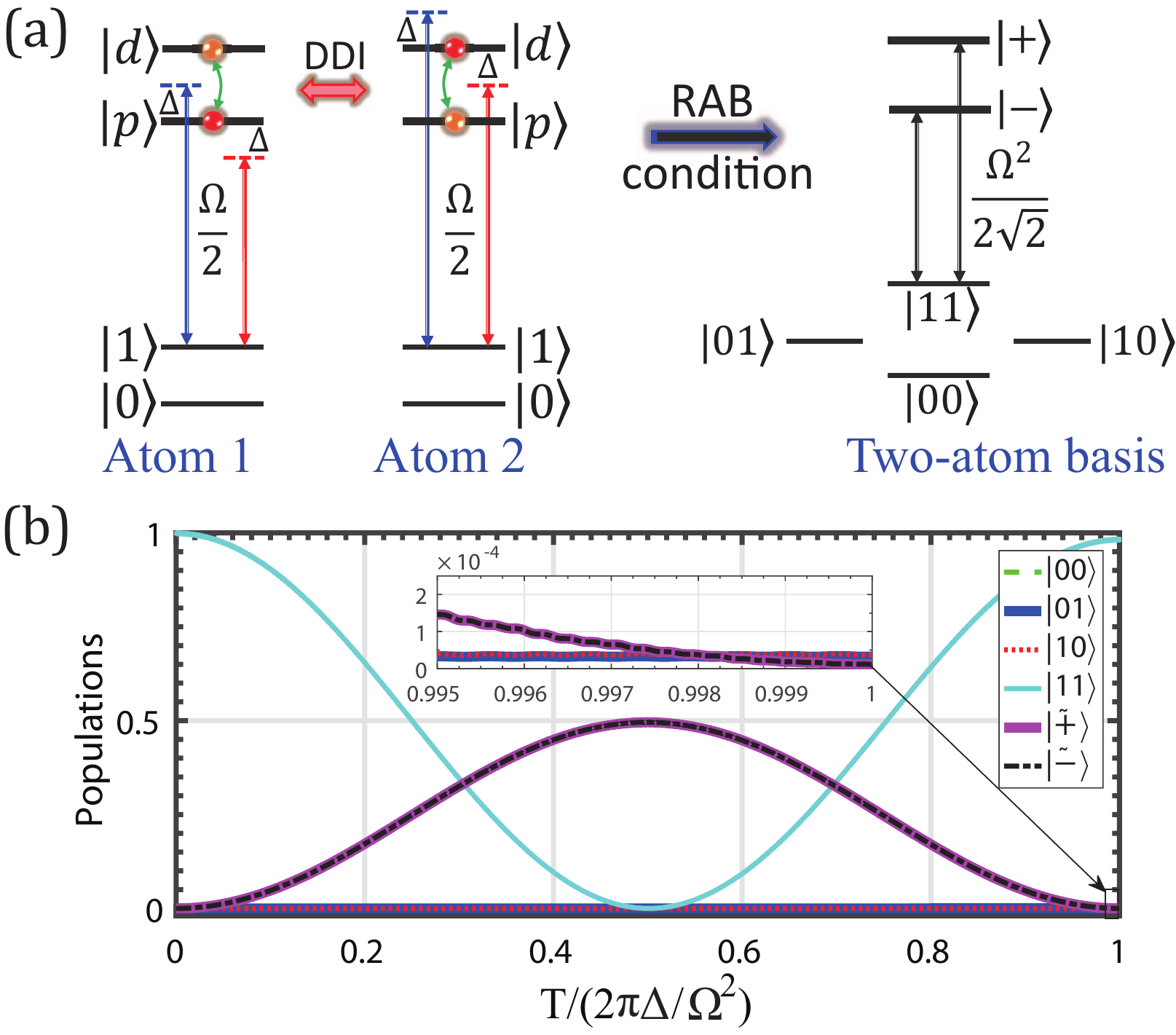}
	\caption{(a), Left panel shows two Rydberg atoms with DD interactions. $|0\rangle$ and $|1\rangle$ are two ground states. $|p\rangle$ and $|d\rangle$ are two Rydberg states with the spin-exchange interaction $\hat{H}_d=V_{d}(|pd\rangle\langle dp| + {\rm H. c.})$. Right panel: The effective RAB process in the dressed state  basis. (b), Populations of the states for RAB scheme in Sec.~\ref{s3.b} under one evolution period $T=2\pi\Delta/\Omega^2$ with the consideration of practical atomic spontaneous emission $\gamma_p=1.69$~kHz and $\gamma_{d}=4$~kHz. Parameters are chosen as $\Omega=2\pi\times5$~MHz and $\Delta$ is set to satisfy the antiblockade condition. The initial state is set as $|11\rangle$ and the inter-atomic distance is set as 3~$\mu$m.}\label{f003}
\end{figure}
 As shown in Fig.~\ref{f003}, we consider the experimental configuration as~\cite{Barredo2015} $|d\rangle\equiv|62D_{3/2},m_{J}=3/2\rangle$, $|p\rangle\equiv|63P_{1/2},m_{J}=1/2\rangle$. These two Rydberg states are resonant with each other. One of the ground states are chosen as $|1\rangle\equiv|5S_{1/2},F=2,m_{F}=2\rangle$~\cite{Barredo2015} and the rest computational state can be chosen as $|0\rangle\equiv|5S_{1/2},F=1,m_{F}=0\rangle$. The excitation process from $|1\rangle$ to state $|d\rangle$ is accomplished by a two-photon transition with wavelengths 795~nm ($\pi$ polarization) and 474~nm ($\sigma_{+}$ polarization), respectively. We also consider the single-photon excitation process from $|1\rangle$ to $|p\rangle$~\cite{Barredo2015}. For left~(right) Rydberg atom in the left panel, bichromatic classical fields are imposed to off-resonantly drive the transition $|1\rangle\leftrightarrow|p(d)\rangle$ through single(two)-photon process with an identical Rabi frequency $\Omega$ but opposite detuning $\Delta$. 
After the rotating-wave approximation, the Hamiltonian for this concrete system can be written as
\begin{eqnarray}\label{eq07}
\hat H_{\Omega}&=&\frac{\Omega}{2}\left(e^{i\Delta t}+e^{-i\Delta t}\right)(|1\rangle_{1}\langle p|\otimes\mathcal{I}_2+\mathcal{I}_1\otimes|1\rangle_{2}\langle d|)
+{\rm H.c.} 
\cr\cr &=& \frac{\Omega}{2}\left(e^{i\Delta t}+e^{-i\Delta t}\right)(|10\rangle\langle p0| + |11\rangle\langle p1| + |1p\rangle\langle pp| \cr\cr&&+ |1d\rangle\langle pd| + |01\rangle\langle 0d| + |11\rangle\langle 1d| + |p1\rangle\langle pd| \cr\cr&& + |d1\rangle\langle dd|) + {\rm H.c.}
\cr\cr \hat H_{d}&=&V_{d}|pd\rangle\langle dp|+{\rm H.c.},
\end{eqnarray}
in which $V_d = C_3/r_{d}^3$ with $C_3$ being 7.965 GHz$\cdot\mu m^3$ here~\cite{Barredo2015,Singer_2005,*PhysRevA.84.041607,*SIBALIC2017319} and $r_d$ that denotes the interatomic distance.
In the following we will show how to achieve the RAB with this Hamiltonian.

\subsubsection{Effective Hamiltonian}
We first define the dressed states $|\widetilde{\pm}\rangle\equiv(|pd\rangle\pm|dp\rangle)/\sqrt{2}$ by diagonalizing  the RRI Hamiltonian. Using the dressed states, one can rewrite Eq.~(\ref{eq07}) as 
\begin{eqnarray}\label{eq08}
\hat H_{\Omega}&=&\frac{\Omega}{\sqrt2}\left(e^{i\Delta t}+e^{-i\Delta t}\right)[|11\rangle\langle \Phi|+\frac{1}{\sqrt{2}}|\Phi\rangle(\langle \widetilde{+}|+\langle \widetilde{-}|)]\cr\cr&&+\frac{\Omega}{2}\left(e^{i\Delta t}+e^{-i\Delta t}\right)(|01\rangle\langle 0d|+|10\rangle\langle d0|)
+{\rm H.c.}
\cr\cr \hat{H}_{d}&=&V_{d}(|\widetilde{+}\rangle\langle \widetilde{+}|-|\widetilde{-}\rangle\langle \widetilde{-}|)
\end{eqnarray}
with $|\Phi\rangle\equiv(|1d\rangle+|p1\rangle)/\sqrt{2}$.
Follow the similar process used in Sec.~\ref{s3.a}, and considering $\Delta\gg\Omega$ and RAB condition $V_d=2\Delta-\Omega^2/(3\Delta)$, we obtain the respective effective Hamiltonian [See Appendix B for details]
\begin{equation}\label{eq09}
\hat{H}_e=\frac{\Omega^2}{2\sqrt{2}\Delta}|11\rangle(\langle \widetilde{+}|-\langle\widetilde{-}|)+{\rm H.c.},
\end{equation}
which means the Rabi oscillation between collective ground state $|11\rangle$ and two-excitation Rydberg state $|pf\rangle$ emerges and the RAB can be achieved with an effective $\pi$-pulse, i.e. $\Omega^2t/\Delta=\pi$.

Now we discuss the differences of excitation process between our scheme and that in Ref.~\cite{Barredo2015}. In Ref.~\cite{Barredo2015}, the atoms are excited step by step. Firstly, one of the Rydberg atoms is excited to $|d\rangle$ state through two-photon process. Then the state of the excited Rydberg atom is transferred from $|d\rangle$ to $|p\rangle$ through the microwave field coupling. Immediately, the rest Rydberg atom is excited to state $|d\rangle$ with $\Omega\simeq5.76V_d$ and along with this process the spin-exchange process also happens. The blockade effect in Ref.~\cite{Barredo2015} does not work because $V_d$ is less than $\Omega$. In our scheme in Sec.~\ref{s3.b}, by using the dressed state and appropriately choosing parameters, RAB can be accomplished in one step under condition $V_d\gg\Omega$.

\subsubsection{Population dynamics}
The full Hamiltonian of the model in Sec.~\ref{s3.b}  is shown in Eq.~(\ref{eq07}). The lifetimes for $|p\rangle$ and $|d\rangle$ are around 0.59 ms and 0.25 ms, respectively~\cite{lifetime1,lifetime2}. The resulting master equation is similar. Due to the change of Rydberg levels, the Lindblad operators are changed to
\begin{eqnarray}
\hat{\mathcal{L}}_{1}&=&\sqrt{\gamma_p/2}|0\rangle_{1}\langle p|,~~ \hat{\mathcal{L}}_{2}=\sqrt{\gamma_p/2}|1\rangle_{1}\langle p|,\cr\cr \hat{\mathcal{L}}_{3}&=&\sqrt{\gamma_d/2}|0\rangle_{1}\langle d|,~~ \hat{\mathcal{L}}_{4}=\sqrt{\gamma_d/2}|1\rangle_{1}\langle d|,\cr\cr \hat{\mathcal{L}}_{5}&=&\sqrt{\gamma_p/2}|0\rangle_{2}\langle p|,~~ \hat{\mathcal{L}}_{6}=\sqrt{\gamma_p/2}|1\rangle_{2}\langle p|,\cr\cr \hat{\mathcal{L}}_{7}&=&\sqrt{\gamma_d/2}|0\rangle_{2}\langle d|,~~ \hat{\mathcal{L}}_{8}=\sqrt{\gamma_d/2}|1\rangle_{2}\langle d|.
\end{eqnarray}
In Fig.~\ref{f003}(b), we plot the evolution of the state for the above RAB regime under the given RAB condition by taking into account of decay in Rydberg states. When numerically solving the master equation~(\ref{master1}), the original Hamiltonian~(\ref{eq07}) rather than the effective Hamiltonian is used. As shown in Fig.~\ref{f003}(b), we achieve the RAB such that only the initial and the dressed state participate in the dynamics.

\subsection{RAB with collective-exchange interaction}\label{s3.c}
\subsubsection{Level scheme and model}
\begin{figure}[htp!]\centering
	\includegraphics[width=\linewidth]{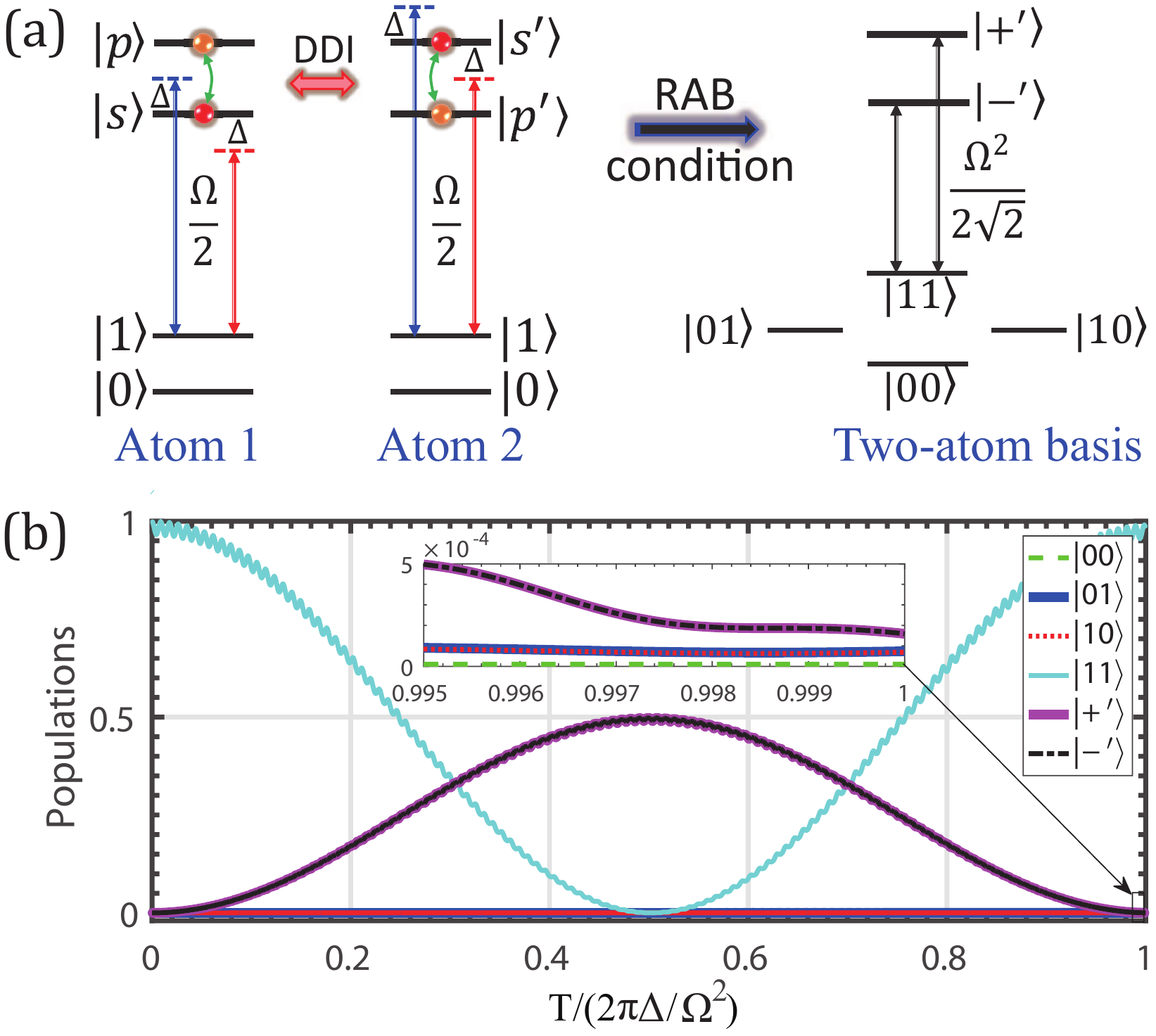}
	\caption{(a). Left panel shows level scheme of the two Rydberg atoms. $|0\rangle$ and $|1\rangle$ are two ground states. $|s\rangle$ and $|p\rangle$ are two Rydberg states for the left atom, and $|s'\rangle$ and $|p'\rangle$ are two Rydberg states for the right atom. These two Rydberg atoms are interacting each other through the spin-exchange interaction $\hat{H}_d=V_{d}(|ss'\rangle\langle pp'| + {\rm H. c.})$. Right panel: the effective RAB process in the dressed state basis. (b). Population dynamics for RAB scheme in Sec.~\ref{s3.c} under one evolution period $T=2\pi\Delta/\Omega^2$ with atomic spontaneous emission rate $\gamma_s=8.33$~kHz, $\gamma_{s'}=7.69$~kHz, $\gamma_p=4$~kHz, $\gamma_{p'}=3.7$~kHz. Parameters are $\Omega=2\pi\times5$~MHz and $\Delta$ is set to satisfy the antiblockade condition. The initial state is $|11\rangle$ and the inter-atomic distance is 2~$\mu$m.}\label{f004}
\end{figure}
As shown in Fig.~\ref{f004}, we consider two Rydberg atoms, and each has two ground states $|0\rangle$ and $|1\rangle$. The left~(right) atom has two Rydberg states $|s\rangle$ and $|p\rangle$ ($|s'\rangle$ and $|p'\rangle$). The experimental configuration is considered as~\cite{Gorniaczyk2016}, $|s\rangle\equiv|48S_{1/2},m_{J}=1/2\rangle$, $|p\rangle\equiv|48P_{1/2},m_{J}=1/2\rangle$, $|s'\rangle\equiv|50S_{1/2},m_{J}=1/2\rangle$, $|p'\rangle\equiv|49P_{1/2},m_{J}=1/2\rangle$. These states are resonant with each other when applying an electric fields $\epsilon=710$ mV cm$^{-1}$ and choosing $r_d=2\mu m$ with $C_3$ to be 0.6 GHz $\mu m^3$. Two ground states in computational subspace can be chosen as $|1\rangle\equiv|5S_{1/2},F=2,m_{F}=0\rangle$ and $|0\rangle\equiv|5S_{1/2},F=1,m_{F}=0\rangle$~\cite{Gorniaczyk2016}. The excitation  from $|1\rangle$ to $|s\rangle$ or $|s'\rangle$ can be implemented by a two-photon process~\cite{Gorniaczyk2016}.

We consider two bichromatic classical fields are imposed to off-resonantly drive the transition $|1\rangle\leftrightarrow|s(s')\rangle$ with an identical Rabi frequency $\Omega$ but opposite detuning $\Delta$. 
With the rotating-wave approximation, the Hamiltonian for this system can be written by $\hat{H}=\hat{H}_{\Omega} + \hat{H}_d$, where 
\begin{eqnarray}\label{eq10}
\hat H_{\Omega}&=&\frac{\Omega}{2}\left(e^{i\Delta t}+e^{-i\Delta t}\right)(|1\rangle_{1}\langle s|\otimes\mathcal{I}_{2}+\mathcal{I}_1\otimes|1\rangle_{2}\langle s'|)
+{\rm H.c.} \cr\cr&=&\frac{\Omega}{2}\left(e^{i\Delta t}+e^{-i\Delta t}\right)(|10\rangle\langle s0| + |11\rangle\langle s1| + |1s'\rangle\langle ss'| \cr\cr&&+|1p'\rangle\langle sp'|+ |01\rangle\langle 0s'| + |11\rangle\langle 1s'| + |s1\rangle\langle ss'| \cr\cr&& + |p1\rangle\langle ps'|) + {\rm H.c.}
\cr\cr \hat H_{d}&=&V_{d}|ss'\rangle\langle pp'|+{\rm H.c.}
\end{eqnarray}
As in previous sections, $\hat{H}_{\Omega}$ and $\hat{H}_{d}$ describe the laser-atom coupling and the dipole-dipole interaction, respectively.

\subsubsection{Effective Hamiltonian}
To derive the effective Hamiltonian, one can diagonalize the Hamiltonian $\hat{H}_d$ to get the dressed states. Using the respective dressed state, Hamiltonian~(\ref{eq10}) can be reformulated to be, 
\begin{eqnarray}\label{eq11}
\hat H_{\Omega}&=&\frac{\Omega}{\sqrt2}\left(e^{i\Delta t}+e^{-i\Delta t}\right)[|11\rangle\langle \Xi|+\frac{1}{\sqrt{2}}|\Xi\rangle(\langle +'|+\langle -'|)]\cr\cr&&+\frac{\Omega}{2}\left(e^{i\Delta t}+e^{-i\Delta t}\right)(|01\rangle\langle 0s'|+|10\rangle\langle s0|)
+{\rm H.c.}
\cr\cr \hat{H}_{d}&=&V_{d}(|+'\rangle\langle +'|-|-'\rangle\langle -'|)
\end{eqnarray}
with $|\Xi\rangle\equiv(|1s'\rangle+|s1\rangle)/\sqrt{2}$ and $|\pm'\rangle = (|ss'\rangle\pm|pp'\rangle)/\sqrt{2}$.
We find that the RAB condition $V_d=2\Delta-\Omega^2/(3\Delta)$ can be obtained when $\Delta\gg\Omega$. This leads to the effective Hamiltonian [See Appendix B for details]
\begin{equation}\label{e10}
\hat{H}_e=\frac{\Omega^2}{2\sqrt{2}\Delta}|11\rangle(\langle +'|-\langle-'|)+{\rm H.c.},
\end{equation}
which indicates the Rabi oscillation between collective ground state $|11\rangle$ and two-excitation Rydberg state $|pp'\rangle$. The ground state is completed transferred to the dressed state when $\Omega^2t/\Delta=\pi$ is fulfilled. In addition to the cases discussed here, the RAB with the resonant DD interaction discussed in Ref.~\cite{van2008, Paris_Mandoki_2016} can also be realized in the similar way.

In Ref.~\cite{Gorniaczyk2016}, optically trapped cloud of $2\times10^4$ $^{87}$Rb gate and source atoms are used for studying the enhancement of single-photon nonlinearity. At zero electric field, the interaction between the $|ss'\rangle$ pair which is of vdW type and much less than the DD interaction. Thus the collective ground state can be excited to $|ss'\rangle$ and the single-photon nonlinearity was observed to be enhanced by electrically tuning $|ss'\rangle$ and $|pp'\rangle$ pair states into resonant interactions~\cite{Gorniaczyk2016}. In this subsection, the resonant DD interaction is an initial consideration and on that basis we design the pulse to achieve the RAB in one step with the condition $V_d\ll\Omega$.

\subsubsection{Population dynamics}
For the model considered in Sec.~\ref{s3.c}, the full Hamiltonian is shown in Eq.~(\ref{eq10}). When including the lifetimes for $|s\rangle$, $|s'\rangle$, $|p\rangle$ and $|p'\rangle$, which are 0.12 ms, 0.13 ms, 0.25 ms and 0.27 ms, respectively~\cite{lifetime1,lifetime2}, the dynamics of the system can be obtained by solving the master equation with the following modified Lindblad operators
	\begin{eqnarray}
 \hat{\mathcal{L}}_{1}&=&\sqrt{\gamma_s/2}|0\rangle_{1}\langle s|,~~~ \hat{\mathcal{L}}_{2}=\sqrt{\gamma_s/2}|1\rangle_{1}\langle s|,\cr\cr \hat{\mathcal{L}}_{3}&=&\sqrt{\gamma_p/2}|0\rangle_{1}\langle p|,~~~ \hat{\mathcal{L}}_{4}=\sqrt{\gamma_p/2}|1\rangle_{1}\langle p|,\cr\cr \hat{\mathcal{L}}_{5}&=&\sqrt{\gamma_{s'}/2}|0\rangle_{2}\langle s'|,~~ \hat{\mathcal{L}}_{6}=\sqrt{\gamma_{s'}/2}|1\rangle_{2}\langle s'|,\cr\cr \hat{\mathcal{L}}_{7}&=&\sqrt{\gamma_{p'}/2}|0\rangle_{2}\langle p'|,~~ \hat{\mathcal{L}}_{8}=\sqrt{\gamma_{p'}/2}|1\rangle_{2}\langle p'|.
 \end{eqnarray}

In Fig.~\ref{f004}(b), we plot the evolution of the state for the RAB regime realized with the collective exchange interaction. The finite lifetime in the Rydberg state is taken into account in the simulation. The numerical simulation agrees with the prediction by the effective Hamiltonian nicely, indicating that an ideal RAB regime can be achieved with this type of DD interaction.
\section{Comparison with vdW interaction based RAB}\label{sec:comparison}

\subsection{Characteristic interatomic distance}\label{s3.1}
For a given Rydberg state, the DD interaction dominates at shorter distances compared to the vdW interaction. Roughly one can separate the two interactions with a characteristic distance $R_c=[4(C_{3})^2/\delta^2]^{1/6}$~\cite{Walker2008}, where $C_3$ is the dispersion coefficient, and $\delta$ is the detuning of the relevant Rydberg pair states participating the DD interaction~\cite{Walker2008,mtk2010}. The vdW interaction plays dominant roles when the interatomic distance \emph{r} is larger than $R_c$. As a result, one should consider alternative theories to analyze RAB and related dynamics~\cite{Zuo2010,Lee2012,Li2013,Carr2013,chen2018accelerated,*Li2019,*Yang2019,Lirui2020,PhysRevLett.120.123204,Su2018,*Su2020,*Wu2020,*Zheng2020,Gambetta20202,Mazza2020,taylor2019generation}. The present work focuses on the regime where interatomic distance \emph{r} is less than $R_c$.
As an example,  we show the characteristic interatomic distance for $|nS_{1/2},~m_{J}=1/2\rangle$ versus principle quantum number \emph{n} in Fig.~\ref{fsystematicerror}(a). By fitting the numerical data, it is found that the characteristic distance $R_c\propto n^{3.655}$, agreeing with the scaling analysis in Ref.~\cite{Walker2008}. It should be noted that here we suppose the channel $2|nS_{1/2}\rangle\rightarrow|nP_{1/2}\rangle+|(n-1)P_{1/2}\rangle$ is the dominate channel for simplicity and make numerical calculations. In practice, one might have to consider contributions from all the transition channels for evaluating the characteristic distance.

\subsection{Dependence on laser parameters}\label{s3.2}
\begin{figure}[htb!]\centering
	\includegraphics[width=\linewidth]{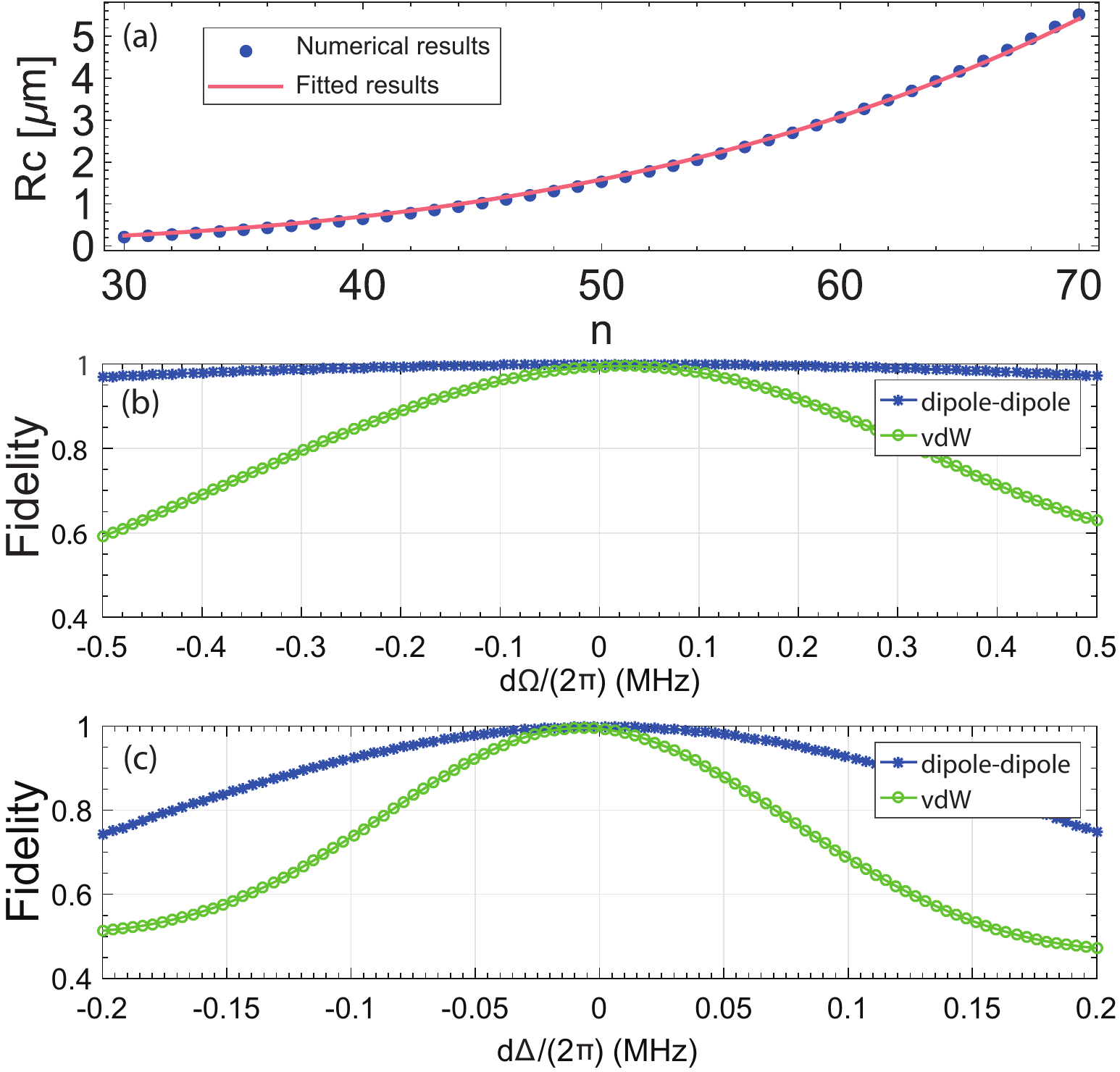}
	\caption{(a) Characteristic interatomic distance $R_c$ for the state $|nS_{1/2},~m_{J}=1/2\rangle$~\cite{Walker2008}. Fidelities of the controlled-Z gate versus  fluctuations of the laser parameters, i.e. deviations $d\Omega$ (b) and $d\Delta$ (c). For DD interaction, the energy level are same with that in Fig.~\ref{f002} with  $\Omega=2\pi\times$9.9~MHz, $\gamma_p=1.89$~kHz, $\gamma_{d}=4.55$~kHz and $\gamma_{f}=7.69$~kHz. The interatomic distance is 3~$\mu$m, and $\Delta$ is determined through the RAB condition given in Sec.~\ref{s3.a}. For vdW interaction, the energy levels are the same with that in Fig.~\ref{f002} without considering $|p\rangle$ and $|f\rangle$ states. The vdW interaction is given by the Hamiltonian $H_{\rm vdW} = C_6/r^6|d\rangle\langle d|\otimes|d\rangle\langle d|$ with $C_6$= 1700~GHz~$\cdot\mu m^6$~\cite{Singer_2005,*PhysRevA.84.041607,*SIBALIC2017319} with interatomic distance $6.6~\mu m$. Besides, $\Omega=2\pi\times2.2$~MHz, $\gamma_{d}=4.55$ kHz and the RAB condition in Ref.~\cite{Su2016} are considered. For panels (b) and (c), the gate time is determined by $T=2\pi\Delta/\Omega^2$.}\label{fsystematicerror}
\end{figure}

So far we have assumed that the laser parameters (Rabi frequency and detuning) are constant in deriving the Hamiltonian. In many experiments, fluctuations of the laser parameters can not be neglected. Importantly, the RAB based on DD and vdW interactions exhibits different responses to the parameter fluctuation. We illustrate the dependence of the two types of interactions by constructing a RAB-based controlled-Z gate. To qualify the fidelity of the gate, we consider the initial state as $|\psi(0)\rangle=(|00\rangle + |01\rangle + |10\rangle+ |11\rangle)/4$ and the ideal output state is $|\psi(t)\rangle=(|00\rangle + |01\rangle + |10\rangle-|11\rangle)/4$. The fidelity is then defined as $F = \langle\psi(t)|\rho(t)|\psi(t)\rangle$ throughout this manuscript. The fidelity of the controlled-Z gate versus the fluctuations of $\Omega(\Delta)$ is shown in Fig.~\ref{fsystematicerror}(b)[(c)]. The gate fidelity based on the DD interaction decreases slower than the vdW interaction when increasing the amplitudes of the fluctuations. This example shows that DD-interaction-based RAB has stronger robustness on the parameter fluctuations than vdW-based counterparts in construction of quantum logic gates.

When deriving the effective Hamiltonian via the second-order perturbation theory,  the laser detuning should be larger than Rabi frequency. Meanwhile, the RAB condition sets the relation between the RRI and laser detuning ($V\sim\Delta$). For given Rydberg states, the DD interaction is stronger than the vdW interaction [see Fig.~\ref{f001}a], where the range of the allowed Rabi frequency is larger when using the DD interaction than the vdW interaction.
\begin{figure}[htb!]\centering
	\includegraphics[width=\linewidth]{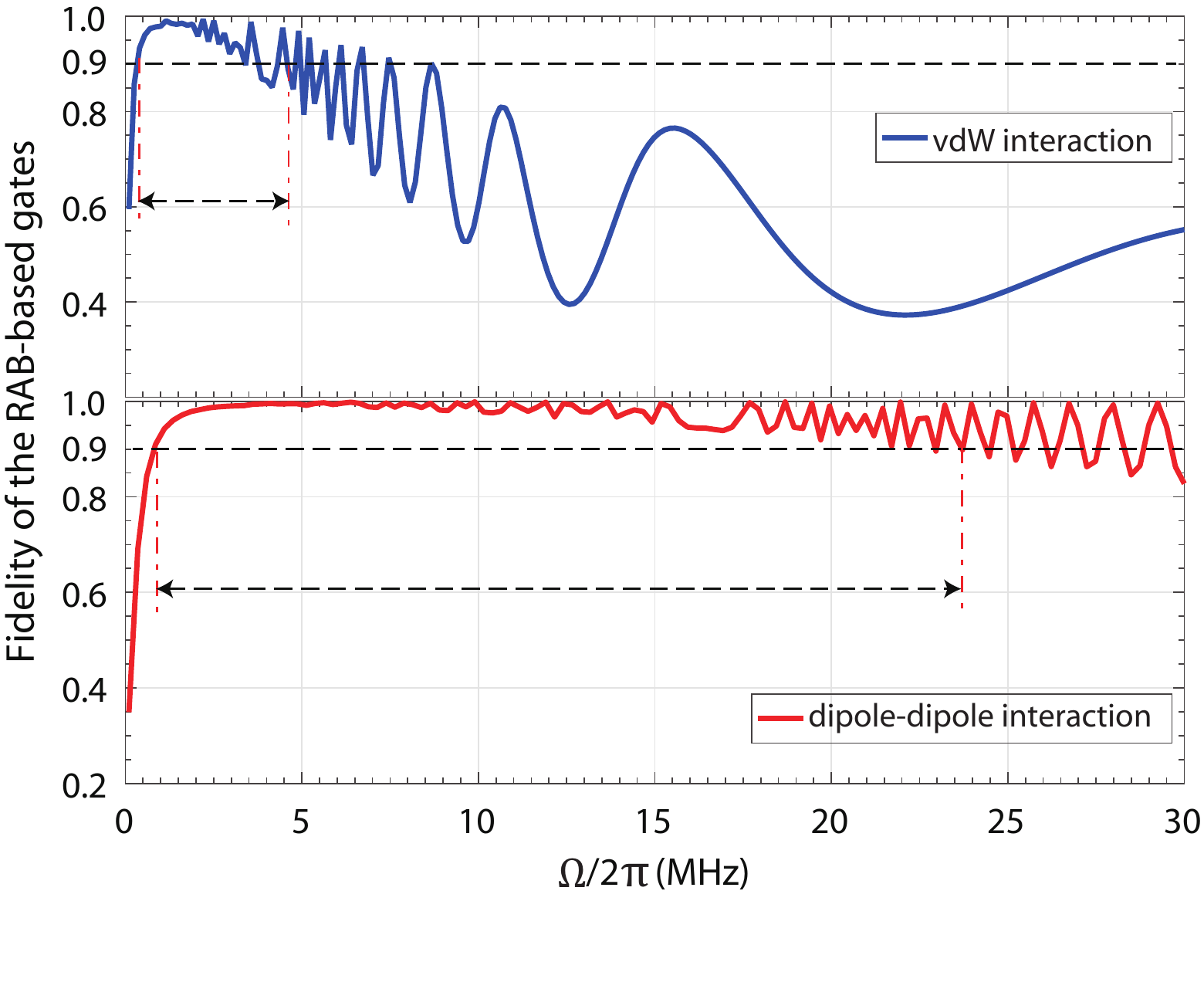}
	\caption{Fidelities of the controlled-Z gate versus Rabi frequency at the gate time $T=2\pi\Delta/\Omega^2$. For DD interactions, the energy level and parameters are chosen as that in Fig.~\ref{fsystematicerror}, except that the Rabi frequency are varied from $2\pi\times0.5$ to $2\pi\times30$~MHz. The detuning is chosen to satisfy the RAB condition given in Sec.~\ref{s3.a}.  For vdW interactions, the energy level is the same as that in Fig.~\ref{f002} without considering $|p\rangle$ and $|f\rangle$ and  the RAB condition in given in Ref.~\cite{Su2016}. In both cases, the arrow indicates the range of Rabi frequencies where the gate fidelity is larger than 0.9.}\label{fadd9}
\end{figure}
To illustrate this, we again examine the performance of the controlled-Z gate by using the DD and vdW interaction. As shown in Fig.~\ref{fadd9}(a), the gate fidelity drops apparently when increasing $\Omega$ in case of the vdW interaction. In contrast the fidelity decreases slowly with increasing $\Omega$ [Fig.~\ref{fadd9}(b)]. In fact, the fidelity in the latter case is greater than 0.9 for a large range of $\Omega$. This indicates that one can  achieve robust controlled Z-gate not only with flexible laser parameters, but can achieve high gate speed using the DD interaction.

\subsection{Dependence on fluctuations of the interatomic distance}\label{3d}
In this subsection, we discuss the influence of the deviation of atom-atom distance on the RAB without setting concrete energy level. The DD and vdW interactions are given by~\cite{mtk2010} 
\begin{equation}
V_d = \frac{C_3}{r_{d}^3}~~ {\rm and}~~ V_{\rm vdW}=\frac{C_6}{r_{\rm vdW}^6},
\end{equation}
where $C_3$ and $C_6$ denote the coefficients of the DD and vdW interactions, respectively. As the vdW and DD interaction have different length scales, we use $r_d$ and $r_{\rm vdW}$ to denote the interatomic distance. When there is a small deviation in the distance, one can find that the change of the interaction energy are,
\begin{equation}
dV_d = -\frac{3V_d}{r_d}dr_d ~~{\rm and}~~ dV_{\rm vdW} = -\frac{6V_{\rm vdW}}{r_{\rm vdW}}dr_{\rm vdW},
\end{equation}
where $dr_d$ and $dr_{\rm vdW}$ are the small deviation with respect to the interatomic distance. Recently it has been shown that this deviation can lead to interesting many-body phases~\cite{gambetta_engineering_2020,gambetta_long-range_2020,PhysRevLett.126.233404,zhang_submicrosecond_2020}.

\begin{figure}
	\includegraphics[width=\linewidth]{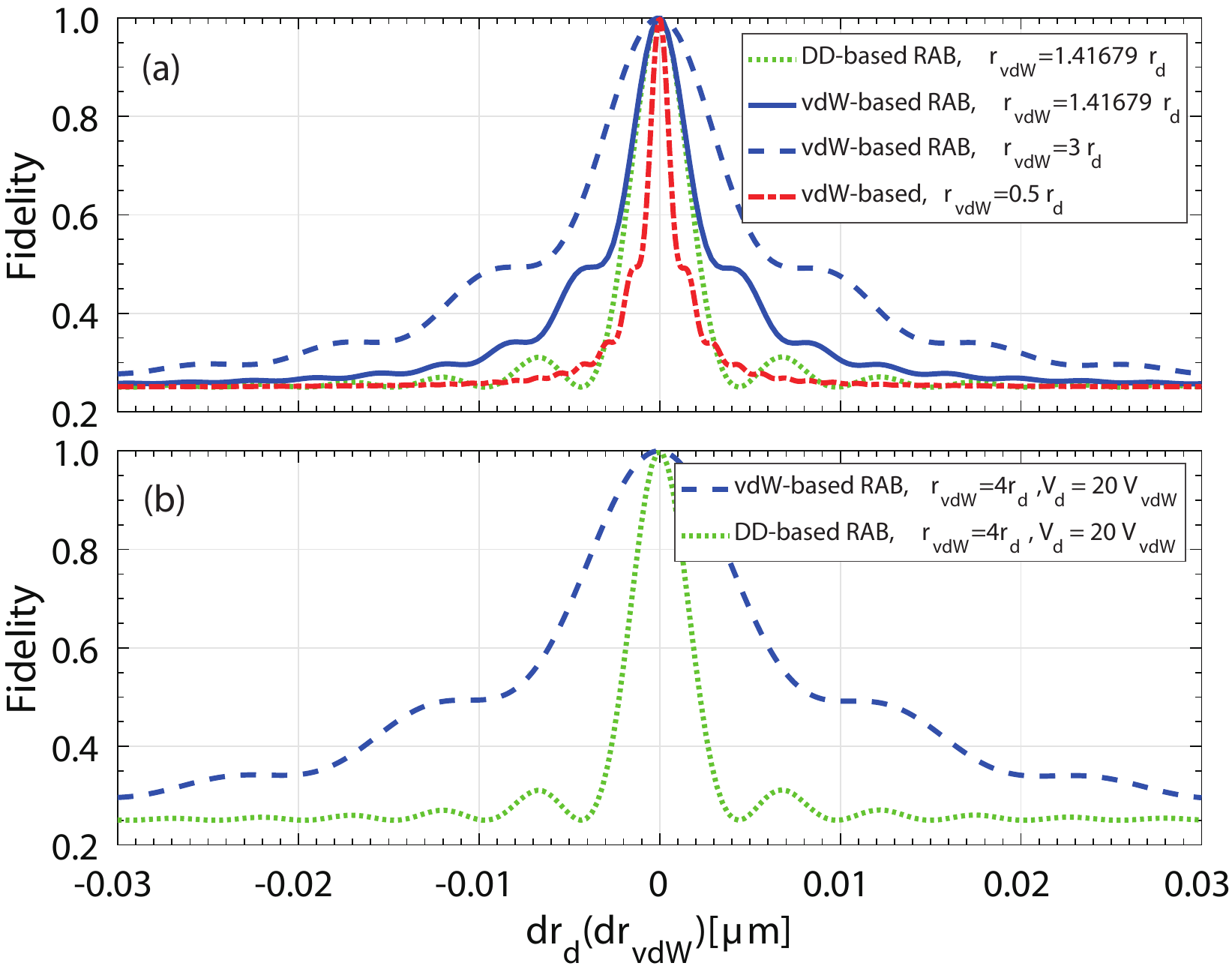}
	\caption{Fidelity of the RAB-based gates with respect to deviation of interatomic distances at the gate time $T=2\pi\Delta/\Omega^2$. Panel (a)[(b)] corresponds to case (i)[(ii)] in Sec.~\ref{3d}. Here, $\Omega=2\pi\times6.663$~MHz, $\Delta=10\Omega$, $r_d=3\mu m$. $V_{d}=2\pi\times94$~MHz. For panel (a) $V_{\rm vdw} = V_d$, while for panel~(b) $V_{\rm vdw}$ is shown in the legend of the figure.}\label{fdis}
\end{figure}

In order to achieve the RAB, the laser detuning has to satisfy the condition $V_d = \sqrt{2}\Delta-\Omega^2/(3\sqrt{2}\Delta)$ in this work, and $V_{\rm vdW} = 2\Delta-2\Omega^2/3\Delta$ in the vdW interaction discussed in Ref.~\cite{Su2016}. Thus, large interaction energy shift $dV_d$ and $dV_{\rm vdW}$ will invalidate the RAB condition. To determine the effect of $dr_d$ and $dr_{\rm vdW}$ quantitatively, we consider the following situation. When (i) $V_{d}=V_{\rm vdW}$ we consider identical deviations of the interatomic distance, i.e., $dr_d = dr_{\rm vdW}$. For the vdW interaction, the detuning $\Delta$ has to be adjusted by 0.50167$dV_{\rm vdW}$ to achieve the RAB. For the DD interaction, $\Delta$ need to be adjusted by 0.70818$dV_d$. Now if we force $0.50167dV_{\rm vdW}=0.70818dV_d$, one can derive the relation between the atomic distance, $r_{\rm vdW}~\simeq~1.41679r_{d}$. This means that $dV_{\rm vdW}$ is greater than $dV_d$ if $r_{\rm vdW}~\textless~1.41679r_{d}$, and vice versa. When one builds a controlled-Z gate, the DD (vdW) interaction based RAB leads to higher gate fidelity when $r_{\rm vdW}<~1.41679r_{d}$ ($r_{\rm vdW}\textgreater~1.41679r_{d}$), as depicted in Fig.~\ref{fdis}(a). In the second case (ii) we consider $V_d\gg V_{\rm vdW}$, and $r_d\sim r_{\rm vdW}$. In this case, one can see that $dV_d\gg dV_{\rm vdW}$ is achieved with the same deviation of interatomic distance ($dr_d=dr_{\rm vdW}$).  This means the vdW-based RAB is more robust than the DD interaction case. As depicted in Fig.~\ref{fdis}(b), the gate fidelity decreases relatively slow when using the vdW interaction.

\section{Applications of the DD interaction induced RAB}\label{Sec4}
Applications of Rydberg antiblockade have been discussed extensively recently~\cite{Ates2007,Amthor2010,Pohl2009,Qianjun2009,Zuo2010,Lee2012,Li2013,Carr2013,chen2018accelerated,*Li2019,*Yang2019,Lirui2020,PhysRevLett.120.123204,Su2018,*Su2020,*Wu2020,*Zheng2020,Su201702,Gambetta20202,Mazza2020,taylor2019generation,Bai_2020,Beterov2016,Beterov2018,Tretyakov2017,Shi2017,*Shi2019,Klaus2020}. Here we will illustrate the DD-interaction-based RAB can be applied in geometric quantum computation and dissipative dynamics. To be concrete, we will focus on the RAB scheme discussed in Sec.~\ref{s3.a}. It is possible to realize similar applications through other schemes.

\subsection{Two-qubit geometric quantum gate}
We first consider how to construct the controlled-arbitrary-phase geometric gate given by the following matrix
\begin{eqnarray}\label{e9}
\hat U_{\rm CP}=
\left(
\begin{array}{cccc}
1&0 &0 &0\\
0&1 &0 &0\\
0&0 &1 &0\\
0&0 &0 &e^{i\theta}\\
\end{array}
\right).
\end{eqnarray}
in the computational space \{$|00\rangle$, $|01\rangle$, $|10\rangle$, $|11\rangle$\}.
By modulating the Rabi frequencies of the initial Hamiltonian at the half evolution time ($T/2$) appropriately, one can achieve the effective Hamiltonian in the time interval $[T/2, T]$~\cite{effective1,effective2,effective3,effective4}
\begin{eqnarray}\label{eq15}
\hat H_{\rm e}=-\frac{e^{i\theta}\Omega^2}{2\sqrt{2}\Delta}|11\rangle(\langle +|-\langle-|)+{\rm H.c.},
\end{eqnarray}
where the phase is controlled by the laser.
\begin{figure}
	\includegraphics[width=\linewidth]{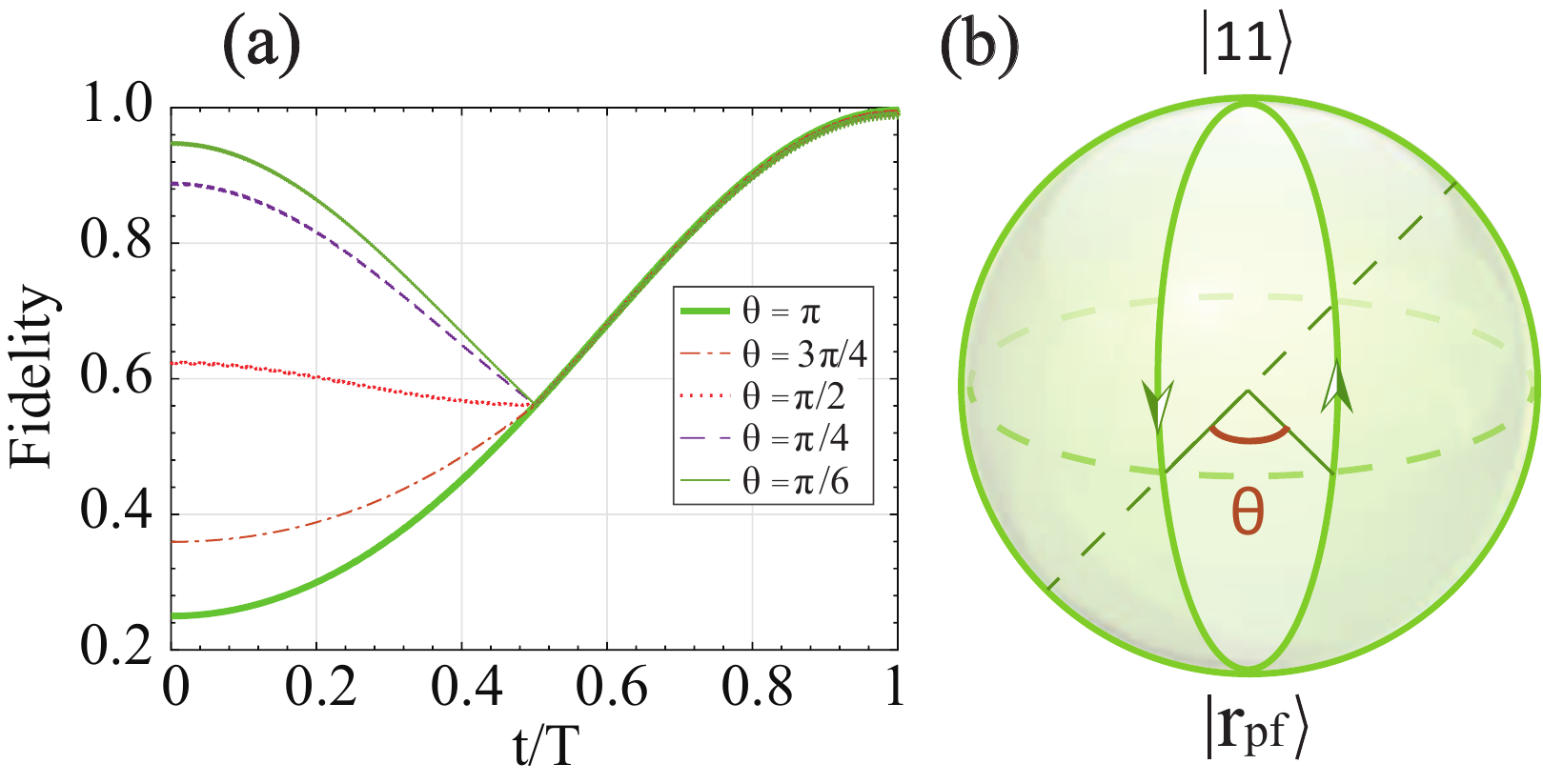}
	\caption{(a)~Evolution of the fidelity of the geometric controlled-arbitrary-phase gate.  The parameters are the same as that in Fig.~\ref{f002}. (b) Bloch sphere representation of the geometric quantum operation. The coupling to the dressed state via RAB gives rise to the desired phase shift to the computational basis at the end of the gate operation.}\label{f3}
\end{figure}

The fidelity of the gate is shown in Fig.~\ref{f3}(a) by numerically solving the master equation with the original Hamiltonian, in which the initial state is $|\psi(0)\rangle=(|00\rangle + |01\rangle + |10\rangle+ |11\rangle)/4$ and the ideal output state is $|\psi(t)\rangle=\hat{U}|\psi(0)\rangle$. The definition of the fidelity is the same as that in Sec.~\ref{s3.2}. With the consideration of dissipation, the gate fidelity is 0.9969, 0.9962, 0.9949, 0.9938 and 0.9936 when $\theta$ equals to $\pi$, $3\pi/4$, $\pi/2$, $\pi/4$ and $\pi/6$, respectively. 
The geometric feature of the phase can be easily verified since $|11\rangle\rightarrow|r_{pf}\rangle\rightarrow e^{i\theta}|11\rangle$ is achieved and $\langle \Psi_j|\hat{H}_e|\Psi_k\rangle=0$~\cite{Sj_qvist_2012,Xu2012,Xue2016,Zhao2017,Liu2019} is satisfied, where $|\Psi_{j}\rangle$~($|\Psi_{k}\rangle$) is any one of the four states in $\{|00\rangle, |01\rangle, |10\rangle, |11\rangle \}$. Thus, $\theta$ is the non-adiabatic geometric phase, which is half of the solid angle enclosed by the evolution path~\cite{PhysRevLett.58.1593}, as shown in Fig.~\ref{f3}(b). 

\subsection{Steady entanglement}
\begin{figure}[htb]\centering
\includegraphics[width=\linewidth]{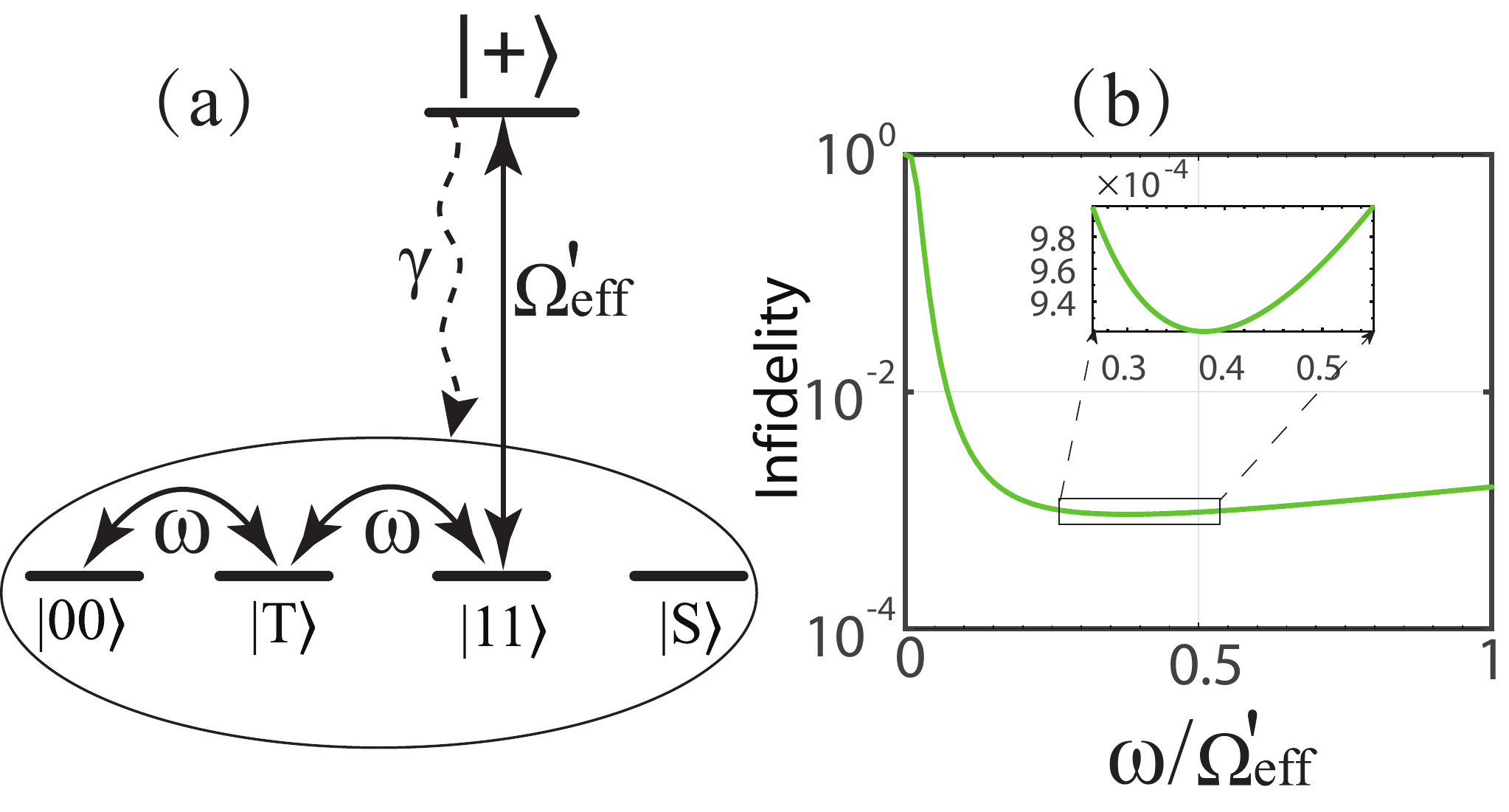}
\caption{(a) Dynamical processes to generate the steady entangled state through combining the unitary and dissipative dynamics. (b) Infidelity of the steady entangled state $(|01\rangle-|10\rangle)/\sqrt{2}$ versus $\omega/\Omega'_{\rm eff}$. The inter-atomic distance is 3~$\mu$m, and the Rabi frequency is s $\Omega=2\pi\times1$~MHz. $\Delta$ is determined through $V_d=\sqrt{2}\Delta$.}\label{f4}
\end{figure}

Steady-state entanglement can be created via dissipation in the strongly interacting Rydberg systems~\cite{PhysRevLett.111.033606, Carr2013}. Following similar ideas, a weak microwave field drives resonantly the transition between two ground states $|0\rangle$ and $|1\rangle$ [Fig.~\ref{f4}(a)],
\begin{eqnarray}\label{e11}
\hat H_{\rm mw}=\frac{\sqrt2\omega}{2}(|00\rangle+|11\rangle)\langle T|+{\rm H.c.},
\end{eqnarray}
where $|T\rangle\equiv(|01\rangle+|10\rangle)/\sqrt2$ is a triplet Bell state. The singlet state $|S\rangle\equiv(|01\rangle-|10\rangle)/\sqrt2$ is decoupled to Hamiltonian~(\ref{e11}) and is the desired steady entangled state. We can learn from Eq.~(\ref{e11}) that the microwave shuffles the states $|00\rangle$, $|T\rangle$, and $|11\rangle$, but keeps $|S\rangle$ invariant. 
Since the stark shifts do not influence the dissipative dynamics,  we thus consider to turn the red-detuned laser off and modify the RAB condition as $V_d=\sqrt{2}\Delta$. The effective Hamiltonian that control unitary dynamics can be written as the form $\hat{H}_{e}'=(\Omega_{\rm eff}'/2)(|11\rangle\langle+| + {\rm H.c.}) +\hat{S}$, where $\Omega_{\rm eff}' = \sqrt{2}\Omega^2/(2\Delta)$ and $\hat{S}$ denotes the stark shift. 

Combining the effective Hamiltonian $\hat{H}_{e}'$ with the microwave Hamiltonian $\hat H_{\rm mw}$ in Eq.~(\ref{e11}), and the dissipative dynamics as depicted in Fig.~\ref{f4}(a), the desired state $|S\rangle$ would be prepared as the steady state of the system. In other words, once $|S\rangle$ is occupied through the dissipative dynamics, the entangled state is created successfully. Otherwise, if the other three states are occupied, the unitary dynamics will excite the two-atom state to $|r_{pf}\rangle$, which would decay to the ground subspace again. In Fig.~\ref{f4}(b), we plot the infidelity $1-F$ of the steady state via numerically solving the master equation~(\ref{master1}) with the full Hamiltonian, and the practical parameters of RRI and atomic spontaneous emission rate. We find that the fidelity of achieving the desired state can be higher than 0.999.

\section{Conclusion}\label{Sec5}

In conclusion, we have proposed three schemes to construct the RAB dynamics with different types of DD interactions that are commonly realized in current Rydberg atom experiments. Based on the dressed state picture, we have derived the effective Hamiltonian that governs the two-atom dynamics. We have verified the validity of the effective Hamiltonian by numerically solving the master equation by taking into account of Rydberg state decay.  In contrast to the vdW-based RAB due to pure energy shifts by the density-density interaction, our study is valid when the inter-atomic distance is relatively small, where the DD interaction dominates. In this regime, we have shown that the DD induced RAB leads to robust dynamics against laser parameter  and interatomic distance fluctuations. 

The DD induced RAB can be applied to realize various quantum information tasks~\cite{Ates2007,Amthor2010,Pohl2009,Qianjun2009,Zuo2010,Lee2012,Li2013,Carr2013,chen2018accelerated,*Li2019,*Yang2019,Lirui2020,PhysRevLett.120.123204,Su2018,*Su2020,*Wu2020,*Zheng2020,Su201702,Gambetta20202,Mazza2020,taylor2019generation,Bai_2020,Beterov2016,Beterov2018,Tretyakov2017,Shi2017,*Shi2019,Klaus2020}, due to the selective, two-body excitation process in the underlying dynamics. As examples, we have shown the proposed RAB can be used in geometric quantum computation, and state entanglement preparation.  Along with the rapid development in optical trapping~\cite{Opticalcontroldipole}, and microwave ~\cite{Afrousheh2004,Bohlouli2007,Sevin_li_2014,Marcuzzi_2015,Gambetta2020,young2020asymmetric} and electric field control~\cite{mmr2001,Walker_2005,Vogt2007,van2008,Ryabtsev2010,Nipper2012,Ravets2014,Ravets2015,Liu2020,Browaeys_2016,Yakshina2016,Gallagher1998,Gorniaczyk2016,Beterov2016,Petrosyan2017,beterov2018resonant,Klaus2020,Paris_Mandoki_2016} of the resonant DD RRI, our schemes and the related applications could be tested and realized in future experiments.

\section*{ACKNOWLEDGEMENTS}
S. L. acknowledges support from Natural Science Foundation of Henan Province (202300410481), National Natural Science Foundation of China (NSFC) under Grant No.11804308, and China Postdoctoral Science Foundation (CPSF) under Grant No. 2018T110735. W. L. acknowledges support from the EPSRC through grant No. EP/R04340X/1 via the QuantERA project “ERyQSenS”, the Royal Society grant No. IEC$\backslash$NSFC$\backslash$181078, and  the UKIERI-UGC Thematic Partnership No. IND/CONT/G/16-17/73.

\appendix
\renewcommand{\thefigure}{A\arabic{figure}}
\section{Derivation of Eq.~(\ref{eq06})}
	We now show the derivation process of the effective Hamiltonian~(\ref{eq06}).

	We start from the rotated Hamiltonian in Eq.~(\ref{e3}), 
	\begin{widetext}
		\begin{eqnarray}\label{eadd3}
		\hat {\mathcal{H}}&=&\big\{\frac{\Omega}{2}\big[\sqrt2\left(e^{i\Delta t}+e^{-i\Delta t}\right)|11\rangle\langle \Psi|+\left(e^{i(\Delta-\sqrt{2}V_{d}) t}+e^{-i(\Delta+\sqrt{2}V_{d}) t}\right)|\Psi\rangle\langle +|+\left(e^{i(\Delta+\sqrt{2}V_{d}) t}+e^{-i(\Delta-\sqrt{2}V_{d}) t}\right)|\Psi\rangle\langle -| \cr\cr&&+ \left(e^{i\Delta t}+e^{-i\Delta t}\right)(|01\rangle\langle 0d|+|10\rangle\langle d0|)\big]+{\rm H.c.}\big\}.
		\end{eqnarray}
\end{widetext}

From Eq.~(\ref{e3}), it can be seen that $|11\rangle$ couple with $|\Psi\rangle$ through two channels with detuning $\Delta$ and $-\Delta$, respectively. Meanwhile, $|\Psi\rangle$ couple with $|\pm\rangle$ through two channels with detuning $\Delta\mp\sqrt{2}V_{d}$ and $-\Delta\mp\sqrt{2}V_{d}$, respectively. To be more clearly, 
in the left panel of Fig.~\ref{a1}, we plot the dynamics of Eq.~(\ref{e3}) with the initial state being $|11\rangle$. It can be readily get that if one want to achieve the coupling between $|11\rangle$ and $|+\rangle$($|-\rangle$), as shown in the right panel in Fig.~\ref{a1}, via second-order perturbation theory, $\sqrt{2}V_{d}-\Delta=\Delta$ should be satisfied. On that basis, Eq.~(\ref{eadd3}) is simplified as
	\begin{widetext}
		\begin{eqnarray}\label{eadd3a}
		\hat {\mathcal{H}}&=&\big\{\frac{\Omega}{2}\big[\sqrt2\left(e^{i\Delta t}+e^{-i\Delta t}\right)|11\rangle\langle \Psi|+\left(e^{-i\Delta t}+e^{-i3\Delta t}\right)|\Psi\rangle\langle +|+\left(e^{i3\Delta t}+e^{i\Delta t}\right)|\Psi\rangle\langle -| \cr\cr&&+ \left(e^{i\Delta t}+e^{-i\Delta t}\right)(|01\rangle\langle 0d|+|10\rangle\langle d0|)\big]+{\rm H.c.}\big\}.
		\end{eqnarray}
\end{widetext}

	\setcounter{figure}{0}  
\begin{figure}
	\includegraphics[width=\linewidth]{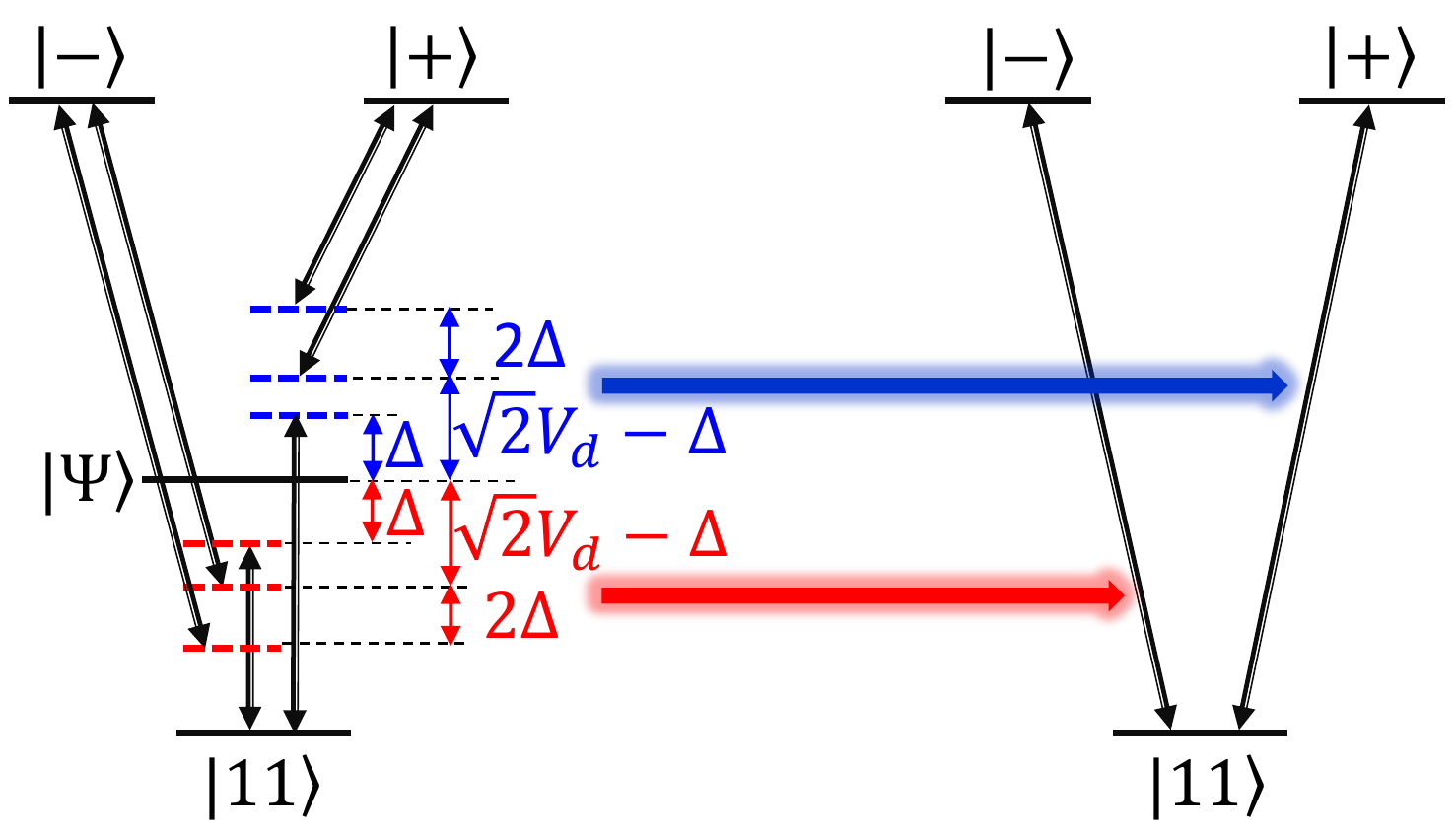}
	\caption{Left panel: Dynamical process of Eq.~(\ref{e3}) when the initial state is $|11\rangle$. Right panel: Dynamical process of Eq.~(\ref{eq06}). The right panel is the effective process of the left one if the antiblockade condition $\sqrt{2}V_{d}-\Delta = \Delta$ is satisfied, which is similar to the "\emph{two-photon process}".}\label{a1}
\end{figure}
We now show the derivation process of Eq.~(\ref{eq06}). Based on Fig.~\ref{a1}, Eqs.~\ref{eadd3} and \ref{eadd3a}, one can see that $|+\rangle$ and $|-\rangle$ can not couple with each other through the intermediate state $|\Psi\rangle$ via the second-order perturbation theory. That is because the coupling between $|+\rangle$ and $|-\rangle$ are oscillating with high frequency $e^{i2\Delta t}$ that should be discarded. The Rabi frequency corresponding to the transition between state $|+\rangle$  and state $|11\rangle$ are calculated as
\begin{eqnarray} 
	\frac{\langle +|\hat {\mathcal{H}}|\Psi\rangle\langle \Psi|\hat{\mathcal{H}}|11\rangle}{\Delta}=\frac{\sqrt{2}\Omega^2}{4\Delta}\cr\cr
	\frac{\langle 11|\hat {\mathcal{H}}|\Psi\rangle\langle \Psi|\hat{\mathcal{H}}|+\rangle}{\Delta}=\frac{\sqrt{2}\Omega^2}{4\Delta}
	\label{Eqs1}
	\end{eqnarray}
Similarly, the Rabi frequency corresponding to the transition frequency between state$|-\rangle$  and state $|11\rangle$ can be calculated as
\begin{eqnarray}
	\frac{\langle -|\hat {\mathcal{H}}|\Psi\rangle\langle \Psi|\hat {\mathcal{H}}|11\rangle}{-\Delta}=\frac{\sqrt{2}\Omega^2}{-4\Delta}
	\cr\cr \frac{\langle 11|\hat {\mathcal{H}}|\Psi\rangle\langle \Psi|\hat{\mathcal{H}}|-\rangle}{-\Delta}=\frac{\sqrt{2}\Omega^2}{-4\Delta}
	\end{eqnarray}
The stark shifts of state $|+\rangle$, $|-\rangle$, $|10\rangle$ and $|01\rangle$ are

\begin{widetext}
       \begin{eqnarray}
		&&\frac{\langle +|(\frac{\Omega}{2}e^{i\Delta t}|+\rangle\langle\Psi|)|\Psi\rangle\langle \Psi|(\frac{\Omega}{2}e^{-i\Delta t}|\Psi\rangle\langle +|)|+\rangle}{\Delta}+ \frac{\langle +|(\frac{\Omega}{2}e^{i3\Delta t}|+\rangle\langle\Psi|)|\Psi\rangle\langle \Psi|(\frac{\Omega}{2}e^{-i3\Delta t}|\Psi\rangle\langle +|)|+\rangle}{3\Delta} =\frac{\Omega^2}{3\Delta},
		\cr\cr&& \frac{\langle -|(\frac{\Omega}{2}e^{-i\Delta t}|-\rangle\langle\Psi|)|\Psi\rangle\langle \Psi|(\frac{\Omega}{2}e^{i\Delta t}|\Psi\rangle\langle +|)|+\rangle}{-\Delta}+ \frac{\langle +|(\frac{\Omega}{2}e^{-i3\Delta t}|-\rangle\langle\Psi|)|\Psi\rangle\langle \Psi|(\frac{\Omega}{2}e^{i3\Delta t}|\Psi\rangle\langle -|)|-\rangle}{-3\Delta} =\frac{\Omega^2}{-3\Delta},\cr\cr&&
		\frac{\langle 10|(\frac{\Omega}{2}e^{i\Delta t}|10\rangle\langle d0|)|d0\rangle\langle d0|(\frac{\Omega}{2}e^{-i\Delta t}|d0\rangle\langle 10|)|10\rangle}{\Delta}+ \frac{\langle 10|(\frac{\Omega}{2}e^{-i\Delta t}|10\rangle\langle d0|)|d0\rangle\langle d0|(\frac{\Omega}{2}e^{i\Delta t}|d0\rangle\langle 10|)|10\rangle}{-\Delta} =0 \cr\cr&&
		{\rm and} \cr\cr&&
		\frac{\langle 01|(\frac{\Omega}{2}e^{i\Delta t}|01\rangle\langle 0d|)|0d\rangle\langle 0d|(\frac{\Omega}{2}e^{-i\Delta t}|0d\rangle\langle 01|)|01\rangle}{\Delta}+ \frac{\langle 01|(\frac{\Omega}{2}e^{-i\Delta t}|01\rangle\langle 0d|)|0d\rangle\langle 0d|(\frac{\Omega}{2}e^{i\Delta t}|0d\rangle\langle 01|)|01\rangle}{-\Delta} =0,
		\end{eqnarray}
		respectively.
\end{widetext}
Besides, the coupling between $|10\rangle$ ($|01\rangle$) and single-excited state are large detuned and should be discarded. Besides, it should be noted that it is needless to calculate the stark shift of $|\Psi\rangle$ since it is also discarded due to the large detuning condition. So the effective Hamiltonian of the system can be written as
\begin{eqnarray}
	{\hat H_e}&=&\frac{\sqrt{2}\Omega^2}{4\Delta}\left(|11\rangle\langle+|-|11\rangle\langle-|+{\rm H.c.}
	\right)\cr\cr&&+\frac{\Omega^2}{3\Delta}(|+\rangle\langle+|-|-\rangle\langle-|),
	\end{eqnarray}
which is exactly Eq.~(\ref{eq06}).

\section{Derivation of Eq.~(\ref{eq09})}\label{appendixb}
Following the similar process, Eqs.~(\ref{eq09}) and (\ref{e10}) can also be achieved with the corresponding given antiblockade conditions. Alternatively, here we use another method~\cite{effective1,effective2,effective3,effective4} based on time-averaging to calculate the effective Hamiltonian~(\ref{eq09}). Firstly, we rotate the whole Hamiltonian with respect to $\hat{U}=e^{i2\Delta(|\tilde{+}\rangle\langle\tilde{+}|-|\tilde{-}\rangle\langle\tilde{-}|)t}$. The total Hamiltonian in the rotated frame is then changed to
\begin{widetext}
\begin{eqnarray}\label{b1}
\hat H_{\rm rotate}&=&\frac{\Omega}{\sqrt2}\left(e^{i\Delta t}+e^{-i\Delta t}\right)|11\rangle\langle \Phi|+\frac{\Omega}{2}\left(e^{-i\Delta t}+e^{-i3\Delta t}\right)|\Phi\rangle\langle \widetilde{+}|+\frac{\Omega}{2}\left(e^{i3\Delta t}+e^{i\Delta t}\right)|\Phi\rangle\langle \widetilde{-}|\cr\cr&&+\frac{\Omega}{2}\left(e^{i\Delta t}+e^{-i\Delta t}\right)(|01\rangle\langle 0d|+|10\rangle\langle d0|)
+{\rm H.c.}+(V_{d}-2\Delta)(|\widetilde{+}\rangle\langle \widetilde{+}|-|\widetilde{-}\rangle\langle \widetilde{-}|).
\end{eqnarray}
\end{widetext}
We now briefly review the effective Hamiltonian formula in Refs.~\cite{effective1,effective2,effective3,effective4}.
For a Hamiltonian in the interaction picture
\begin{equation}\label{b3}
\hat{H}=\sum_{n=1}^{N}\hat{h}^\dagger_ne^{i\omega_nt}+\hat{h}_{n}e^{-i\omega_nt},
\end{equation}
if the large-detuning condition is satisfied, the effective Hamiltonian would be 
\begin{equation}
\hat{H}_{\rm eff}=\sum_{m,n=1}^{N}\frac{1}{\hbar\overline{\omega_{mn}}}\left[\hat{h}^\dagger_m,\hat{h}_n\right]e^{i(\omega_m-\omega_n)t},
\end{equation}
where $\overline{\omega_{mn}}=2\omega_m\omega_n/(\omega_m+\omega_n)$. After using Eq.~(\ref{b3}), the processes to calculate the effective form of Eq.~(\ref{b1}) are listed as follows
\begin{widetext}
	\begin{eqnarray}\label{b4}
	&&\frac{\Omega^2}{2\sqrt{2}\Delta}[|11\rangle\langle\Phi|e^{i\Delta t}, |\Phi\rangle\langle\widetilde{+}|e^{-i\Delta t}]e^{i(\Delta-\Delta)t}=\frac{\Omega^2}{2\sqrt{2}\Delta}|11\rangle\langle\widetilde{+}|
,~~ \frac{\Omega^2}{2\sqrt{2}\Delta}[|\Phi\rangle\langle11|e^{i\Delta t},
	|\widetilde{-}\rangle\langle\Phi|e^{-i\Delta t}]e^{i(\Delta-\Delta)t}=-\frac{\Omega^2}{2\sqrt{2}\Delta}|\widetilde{-}\rangle\langle 11|,
	\cr\cr
	&&\frac{\Omega^2}{2\sqrt{2}\Delta}[|\widetilde{+}\rangle\langle\Phi|e^{i\Delta t}, |\Phi\rangle\langle11|e^{-i\Delta t}]e^{i(\Delta-\Delta)t}=\frac{\Omega^2}{2\sqrt{2}\Delta}|\widetilde{+}\rangle\langle11|
,~~\frac{\Omega^2}{2\sqrt{2}\Delta}[|\Phi\rangle\langle\widetilde{-}|e^{i\Delta t}, |11\rangle\langle\Phi|e^{-i\Delta t}]e^{i(\Delta-\Delta)t}=-\frac{\Omega^2}{2\sqrt{2}\Delta}|11\rangle\langle\widetilde{-}|,
	\cr\cr
	&&~~~~~~~~~~~~~~~~~~~~~~~~~~~~~~~~~~~~~~~~\frac{\Omega^2}{2\Delta}[|11\rangle\langle\Phi|e^{i\Delta t}, |\Phi\rangle\langle11|e^{-i\Delta t}]e^{i(\Delta-\Delta)t}=\frac{\Omega^2}{2\Delta}(|11\rangle\langle11|-|\Phi\rangle\langle\Phi|),
	\cr\cr
	&&~~~~~~~~~~~~~~~~~~~~~~~~~~~~~~~~~~~~~~~~\frac{\Omega^2}{2\Delta}[|\Phi\rangle\langle11|e^{i\Delta t}, |11\rangle\langle\Phi|e^{-i\Delta t}]e^{i(\Delta-\Delta)t}=\frac{\Omega^2}{2\Delta}(|\Phi\rangle\langle\Phi|-|11\rangle\langle11|),
	\cr\cr
	&&~~~~~~~~~~~~~~~~~~~~~~~~~~~~~~~~~~~~~~~~\frac{\Omega^2}{4\Delta}[|\widetilde{+}\rangle\langle\Phi|e^{i\Delta t}, |\Phi\rangle|\langle\widetilde{+}|e^{-i\Delta t}]e^{i(\Delta-\Delta)t}=\frac{\Omega^2}{4\Delta}(|\widetilde{+}\rangle\langle\widetilde{+}|-|\Phi\rangle\langle\Phi|),
	\cr\cr
	&&~~~~~~~~~~~~~~~~~~~~~~~~~~~~~~~~~~~~~~~~\frac{\Omega^2}{12\Delta}[|\widetilde{+}\rangle\langle\Phi|e^{i3\Delta t}, |\Phi\rangle|\widetilde{+}\rangle|e^{-i3\Delta t}]e^{i(3\Delta-3\Delta)t}=\frac{\Omega^2}{12\Delta}(|\widetilde{+}\rangle\langle\widetilde{+}|-|\Phi\rangle\langle\Phi|),
	\cr\cr
	&&~~~~~~~~~~~~~~~~~~~~~~~~~~~~~~~~~~~~~~~~\frac{\Omega^2}{12\Delta}[|\Phi\rangle\langle\widetilde{-}|e^{i3\Delta t}, |\widetilde{-}\rangle\langle\Phi|e^{-i3\Delta t}]e^{i(3\Delta-3\Delta)t}=\frac{\Omega^2}{12\Delta}(|\Phi\rangle\langle\Phi|-|\widetilde{-}\rangle\langle\widetilde{-}|),
	\cr\cr
	&&~~~~~~~~~~~~~~~~~~~~~~~~~~~~~~~~~~~~~~~~\frac{\Omega^2}{4\Delta}[|\Phi\rangle\langle\widetilde{-}|e^{i\Delta t}, |\widetilde{-}\rangle\langle\Phi|e^{-i\Delta t}]e^{i(\Delta-\Delta)t}=\frac{\Omega^2}{4\Delta}(|\Phi\rangle\langle\Phi|-|\widetilde{-}\rangle\langle\widetilde{-}|),
	\cr\cr
	&&~~~~~~~~~~~~~~~~~~~~\frac{\Omega^2}{4\Delta}[(|01\rangle\langle0d|+|10\rangle\langle d0|)e^{i\Delta t}, (|0d\rangle\langle 01|+|d0\rangle\langle 10|)e^{-i\Delta t}]e^{i(\Delta-\Delta)t}=\frac{\Omega^2}{4\Delta}(|01\rangle\langle01|-|0d\rangle\langle0d|),
	\cr\cr
	&&~~~~~~~~~~~~~~~~~~~~\frac{\Omega^2}{4\Delta}[(|0d\rangle\langle01|+|d0\rangle\langle 10|)e^{i\Delta t}, (|01\rangle\langle 0d|+|10\rangle\langle d0|)e^{-i\Delta t}]e^{i(\Delta-\Delta)t}=\frac{\Omega^2}{4\Delta}(|0d\rangle\langle0d|-|01\rangle\langle01|).
	\end{eqnarray}
It should be noted that the high-frequency oscillation terms are discarded and were not shown in Eq.~(\ref{b4}). The sum of the terms in Eq.~(\ref{b4}) induces the effective Hamiltonian as
	\begin{eqnarray}
	{\hat H_e}&=&\frac{\Omega^2}{2\sqrt{2}\Delta}\left(|11\rangle\langle\widetilde{+}|-|11\rangle\langle\widetilde{-}|+{\rm H.c.}
	\right)+\frac{\Omega^2}{3\Delta}(|\widetilde{+}\rangle\langle\widetilde{+}|-|\widetilde{-}\rangle\langle\widetilde{-}|)+(V_{d}-2\Delta)(|\widetilde{+}\rangle\langle \widetilde{+}|-|\widetilde{-}\rangle\langle \widetilde{-}|).
	\end{eqnarray}
\end{widetext}	
In which, the stark shift terms of state $|\widetilde{+}\rangle$ and state $|\widetilde{-}\rangle$ exactly cancels out $(V_{d}-2\Delta)(|\widetilde{+}\rangle\langle \widetilde{+}|-|\widetilde{-}\rangle\langle \widetilde{-}|) $ when the antiblockade condition $V_d=2\Delta-\Omega^2/(3\Delta)$ are satisfied. After using $|\widetilde{\pm}\rangle\equiv(|pd\rangle\pm|dp\rangle)/\sqrt{2}$ the total effective Hamiltonian of the system in the rotated frame becomes
\begin{eqnarray}
\hat H_{\rm e}=\frac{\Omega^2}{2\Delta}|11\rangle\langle dp|+{\rm H.c.},
\end{eqnarray} 
which means Eq.~(\ref{eq09}) is achieved. Similarly, to deviate Eq.~(\ref{e10}), either one of the above two methods are feasible.

\bibliography{apssamp1}

\begin{thebibliography}{122}%
\makeatletter
\providecommand \@ifxundefined [1]{%
 \@ifx{#1\undefined}
}%
\providecommand \@ifnum [1]{%
 \ifnum #1\expandafter \@firstoftwo
 \else \expandafter \@secondoftwo
 \fi
}%
\providecommand \@ifx [1]{%
 \ifx #1\expandafter \@firstoftwo
 \else \expandafter \@secondoftwo
 \fi
}%
\providecommand \natexlab [1]{#1}%
\providecommand \enquote  [1]{``#1''}%
\providecommand \bibnamefont  [1]{#1}%
\providecommand \bibfnamefont [1]{#1}%
\providecommand \citenamefont [1]{#1}%
\providecommand \href@noop [0]{\@secondoftwo}%
\providecommand \href [0]{\begingroup \@sanitize@url \@href}%
\providecommand \@href[1]{\@@startlink{#1}\@@href}%
\providecommand \@@href[1]{\endgroup#1\@@endlink}%
\providecommand \@sanitize@url [0]{\catcode `\\12\catcode `\$12\catcode
  `\&12\catcode `\#12\catcode `\^12\catcode `\_12\catcode `\%12\relax}%
\providecommand \@@startlink[1]{}%
\providecommand \@@endlink[0]{}%
\providecommand \url  [0]{\begingroup\@sanitize@url \@url }%
\providecommand \@url [1]{\endgroup\@href {#1}{\urlprefix }}%
\providecommand \urlprefix  [0]{URL }%
\providecommand \Eprint [0]{\href }%
\providecommand \doibase [0]{http://dx.doi.org/}%
\providecommand \selectlanguage [0]{\@gobble}%
\providecommand \bibinfo  [0]{\@secondoftwo}%
\providecommand \bibfield  [0]{\@secondoftwo}%
\providecommand \translation [1]{[#1]}%
\providecommand \BibitemOpen [0]{}%
\providecommand \bibitemStop [0]{}%
\providecommand \bibitemNoStop [0]{.\EOS\space}%
\providecommand \EOS [0]{\spacefactor3000\relax}%
\providecommand \BibitemShut  [1]{\csname bibitem#1\endcsname}%
\let\auto@bib@innerbib\@empty
\bibitem [{\citenamefont {Gallagher}(2005)}]{Gallagher1994}%
  \BibitemOpen
  \bibfield  {author} {\bibinfo {author} {\bibfnamefont {T.~F.}\ \bibnamefont
  {Gallagher}},\ }\href@noop {} {\emph {\bibinfo {title} {Rydberg atoms}}}\
  (\bibinfo  {publisher} {Cambridge University Press},\ \bibinfo {year}
  {2005})\BibitemShut {NoStop}%
\bibitem [{\citenamefont {Jaksch}\ \emph {et~al.}(2000)\citenamefont {Jaksch},
  \citenamefont {Cirac}, \citenamefont {Zoller}, \citenamefont {Rolston},
  \citenamefont {C\^ot\'e},\ and\ \citenamefont {Lukin}}]{djp2000}%
  \BibitemOpen
  \bibfield  {author} {\bibinfo {author} {\bibfnamefont {D.}~\bibnamefont
  {Jaksch}}, \bibinfo {author} {\bibfnamefont {J.~I.}\ \bibnamefont {Cirac}},
  \bibinfo {author} {\bibfnamefont {P.}~\bibnamefont {Zoller}}, \bibinfo
  {author} {\bibfnamefont {S.~L.}\ \bibnamefont {Rolston}}, \bibinfo {author}
  {\bibfnamefont {R.}~\bibnamefont {C\^ot\'e}}, \ and\ \bibinfo {author}
  {\bibfnamefont {M.~D.}\ \bibnamefont {Lukin}},\ }\href {\doibase
  10.1103/PhysRevLett.85.2208} {\bibfield  {journal} {\bibinfo  {journal}
  {Phys. Rev. Lett.}\ }\textbf {\bibinfo {volume} {85}},\ \bibinfo {pages}
  {2208} (\bibinfo {year} {2000})}\BibitemShut {NoStop}%
\bibitem [{\citenamefont {Lukin}\ \emph {et~al.}(2001)\citenamefont {Lukin},
  \citenamefont {Fleischhauer}, \citenamefont {Cote}, \citenamefont {Duan},
  \citenamefont {Jaksch}, \citenamefont {Cirac},\ and\ \citenamefont
  {Zoller}}]{mmr2001}%
  \BibitemOpen
  \bibfield  {author} {\bibinfo {author} {\bibfnamefont {M.~D.}\ \bibnamefont
  {Lukin}}, \bibinfo {author} {\bibfnamefont {M.}~\bibnamefont {Fleischhauer}},
  \bibinfo {author} {\bibfnamefont {R.}~\bibnamefont {Cote}}, \bibinfo {author}
  {\bibfnamefont {L.~M.}\ \bibnamefont {Duan}}, \bibinfo {author}
  {\bibfnamefont {D.}~\bibnamefont {Jaksch}}, \bibinfo {author} {\bibfnamefont
  {J.~I.}\ \bibnamefont {Cirac}}, \ and\ \bibinfo {author} {\bibfnamefont
  {P.}~\bibnamefont {Zoller}},\ }\href {\doibase 10.1103/PhysRevLett.87.037901}
  {\bibfield  {journal} {\bibinfo  {journal} {Phys. Rev. Lett.}\ }\textbf
  {\bibinfo {volume} {87}},\ \bibinfo {pages} {037901} (\bibinfo {year}
  {2001})}\BibitemShut {NoStop}%
\bibitem [{\citenamefont {Saffman}\ \emph {et~al.}(2010)\citenamefont
  {Saffman}, \citenamefont {Walker},\ and\ \citenamefont
  {M\o{}lmer}}]{mtk2010}%
  \BibitemOpen
  \bibfield  {author} {\bibinfo {author} {\bibfnamefont {M.}~\bibnamefont
  {Saffman}}, \bibinfo {author} {\bibfnamefont {T.~G.}\ \bibnamefont {Walker}},
  \ and\ \bibinfo {author} {\bibfnamefont {K.}~\bibnamefont {M\o{}lmer}},\
  }\href {\doibase 10.1103/RevModPhys.82.2313} {\bibfield  {journal} {\bibinfo
  {journal} {Rev. Mod. Phys.}\ }\textbf {\bibinfo {volume} {82}},\ \bibinfo
  {pages} {2313} (\bibinfo {year} {2010})}\BibitemShut {NoStop}%
\bibitem [{\citenamefont {Comparat}\ and\ \citenamefont
  {Pillet}(2010)}]{Comparat:10}%
  \BibitemOpen
  \bibfield  {author} {\bibinfo {author} {\bibfnamefont {D.}~\bibnamefont
  {Comparat}}\ and\ \bibinfo {author} {\bibfnamefont {P.}~\bibnamefont
  {Pillet}},\ }\href {\doibase 10.1364/JOSAB.27.00A208} {\bibfield  {journal}
  {\bibinfo  {journal} {J. Opt. Soc. Am. B}\ }\textbf {\bibinfo {volume}
  {27}},\ \bibinfo {pages} {A208} (\bibinfo {year} {2010})}\BibitemShut
  {NoStop}%
\bibitem [{\citenamefont {Li}\ and\ \citenamefont
  {Lesanovsky}(2014)}]{li_entangling_2014}%
  \BibitemOpen
  \bibfield  {author} {\bibinfo {author} {\bibfnamefont {W.}~\bibnamefont
  {Li}}\ and\ \bibinfo {author} {\bibfnamefont {I.}~\bibnamefont
  {Lesanovsky}},\ }\href {\doibase 10.1007/s00340-013-5709-6} {\bibfield
  {journal} {\bibinfo  {journal} {Appl. Phys. B}\ }\textbf {\bibinfo {volume}
  {114}},\ \bibinfo {pages} {37} (\bibinfo {year} {2014})}\BibitemShut
  {NoStop}%
\bibitem [{\citenamefont {Saffman}(2016)}]{Saffman_2016}%
  \BibitemOpen
  \bibfield  {author} {\bibinfo {author} {\bibfnamefont {M.}~\bibnamefont
  {Saffman}},\ }\href {\doibase 10.1088/0953-4075/49/20/202001} {\bibfield
  {journal} {\bibinfo  {journal} {J. Phys. B: Atom. Mol. Opt. Phys.}\ }\textbf
  {\bibinfo {volume} {49}},\ \bibinfo {pages} {202001} (\bibinfo {year}
  {2016})}\BibitemShut {NoStop}%
\bibitem [{\citenamefont {Isenhower}\ \emph {et~al.}(2010)\citenamefont
  {Isenhower}, \citenamefont {Urban}, \citenamefont {Zhang}, \citenamefont
  {Gill}, \citenamefont {Henage}, \citenamefont {Johnson}, \citenamefont
  {Walker},\ and\ \citenamefont {Saffman}}]{lex2010}%
  \BibitemOpen
  \bibfield  {author} {\bibinfo {author} {\bibfnamefont {L.}~\bibnamefont
  {Isenhower}}, \bibinfo {author} {\bibfnamefont {E.}~\bibnamefont {Urban}},
  \bibinfo {author} {\bibfnamefont {X.~L.}\ \bibnamefont {Zhang}}, \bibinfo
  {author} {\bibfnamefont {A.~T.}\ \bibnamefont {Gill}}, \bibinfo {author}
  {\bibfnamefont {T.}~\bibnamefont {Henage}}, \bibinfo {author} {\bibfnamefont
  {T.~A.}\ \bibnamefont {Johnson}}, \bibinfo {author} {\bibfnamefont {T.~G.}\
  \bibnamefont {Walker}}, \ and\ \bibinfo {author} {\bibfnamefont
  {M.}~\bibnamefont {Saffman}},\ }\href {\doibase
  10.1103/PhysRevLett.104.010503} {\bibfield  {journal} {\bibinfo  {journal}
  {Phys. Rev. Lett.}\ }\textbf {\bibinfo {volume} {104}},\ \bibinfo {pages}
  {010503} (\bibinfo {year} {2010})}\BibitemShut {NoStop}%
\bibitem [{\citenamefont {Zhang}\ \emph {et~al.}(2010)\citenamefont {Zhang},
  \citenamefont {Isenhower}, \citenamefont {Gill}, \citenamefont {Walker},\
  and\ \citenamefont {Saffman}}]{xla2010}%
  \BibitemOpen
  \bibfield  {author} {\bibinfo {author} {\bibfnamefont {X.~L.}\ \bibnamefont
  {Zhang}}, \bibinfo {author} {\bibfnamefont {L.}~\bibnamefont {Isenhower}},
  \bibinfo {author} {\bibfnamefont {A.~T.}\ \bibnamefont {Gill}}, \bibinfo
  {author} {\bibfnamefont {T.~G.}\ \bibnamefont {Walker}}, \ and\ \bibinfo
  {author} {\bibfnamefont {M.}~\bibnamefont {Saffman}},\ }\href {\doibase
  10.1103/PhysRevA.82.030306} {\bibfield  {journal} {\bibinfo  {journal} {Phys.
  Rev. A}\ }\textbf {\bibinfo {volume} {82}},\ \bibinfo {pages} {030306(R)}
  (\bibinfo {year} {2010})}\BibitemShut {NoStop}%
\bibitem [{\citenamefont {Wilk}\ \emph {et~al.}(2010)\citenamefont {Wilk},
  \citenamefont {Ga\"etan}, \citenamefont {Evellin}, \citenamefont {Wolters},
  \citenamefont {Miroshnychenko}, \citenamefont {Grangier},\ and\ \citenamefont
  {Browaeys}}]{tac2010}%
  \BibitemOpen
  \bibfield  {author} {\bibinfo {author} {\bibfnamefont {T.}~\bibnamefont
  {Wilk}}, \bibinfo {author} {\bibfnamefont {A.}~\bibnamefont {Ga\"etan}},
  \bibinfo {author} {\bibfnamefont {C.}~\bibnamefont {Evellin}}, \bibinfo
  {author} {\bibfnamefont {J.}~\bibnamefont {Wolters}}, \bibinfo {author}
  {\bibfnamefont {Y.}~\bibnamefont {Miroshnychenko}}, \bibinfo {author}
  {\bibfnamefont {P.}~\bibnamefont {Grangier}}, \ and\ \bibinfo {author}
  {\bibfnamefont {A.}~\bibnamefont {Browaeys}},\ }\href {\doibase
  10.1103/PhysRevLett.104.010502} {\bibfield  {journal} {\bibinfo  {journal}
  {Phys. Rev. Lett.}\ }\textbf {\bibinfo {volume} {104}},\ \bibinfo {pages}
  {010502} (\bibinfo {year} {2010})}\BibitemShut {NoStop}%
\bibitem [{\citenamefont {Maller}\ \emph {et~al.}(2015)\citenamefont {Maller},
  \citenamefont {Lichtman}, \citenamefont {Xia}, \citenamefont {Sun},
  \citenamefont {Piotrowicz}, \citenamefont {Carr}, \citenamefont {Isenhower},\
  and\ \citenamefont {Saffman}}]{kmt2015}%
  \BibitemOpen
  \bibfield  {author} {\bibinfo {author} {\bibfnamefont {K.~M.}\ \bibnamefont
  {Maller}}, \bibinfo {author} {\bibfnamefont {M.~T.}\ \bibnamefont
  {Lichtman}}, \bibinfo {author} {\bibfnamefont {T.}~\bibnamefont {Xia}},
  \bibinfo {author} {\bibfnamefont {Y.}~\bibnamefont {Sun}}, \bibinfo {author}
  {\bibfnamefont {M.~J.}\ \bibnamefont {Piotrowicz}}, \bibinfo {author}
  {\bibfnamefont {A.~W.}\ \bibnamefont {Carr}}, \bibinfo {author}
  {\bibfnamefont {L.}~\bibnamefont {Isenhower}}, \ and\ \bibinfo {author}
  {\bibfnamefont {M.}~\bibnamefont {Saffman}},\ }\href {\doibase
  10.1103/PhysRevA.92.022336} {\bibfield  {journal} {\bibinfo  {journal} {Phys.
  Rev. A}\ }\textbf {\bibinfo {volume} {92}},\ \bibinfo {pages} {022336}
  (\bibinfo {year} {2015})}\BibitemShut {NoStop}%
\bibitem [{\citenamefont {Zeng}\ \emph {et~al.}(2017)\citenamefont {Zeng},
  \citenamefont {Xu}, \citenamefont {He}, \citenamefont {Liu}, \citenamefont
  {Liu}, \citenamefont {Wang}, \citenamefont {Papoular}, \citenamefont
  {Shlyapnikov},\ and\ \citenamefont {Zhan}}]{ypx2017}%
  \BibitemOpen
  \bibfield  {author} {\bibinfo {author} {\bibfnamefont {Y.}~\bibnamefont
  {Zeng}}, \bibinfo {author} {\bibfnamefont {P.}~\bibnamefont {Xu}}, \bibinfo
  {author} {\bibfnamefont {X.}~\bibnamefont {He}}, \bibinfo {author}
  {\bibfnamefont {Y.}~\bibnamefont {Liu}}, \bibinfo {author} {\bibfnamefont
  {M.}~\bibnamefont {Liu}}, \bibinfo {author} {\bibfnamefont {J.}~\bibnamefont
  {Wang}}, \bibinfo {author} {\bibfnamefont {D.~J.}\ \bibnamefont {Papoular}},
  \bibinfo {author} {\bibfnamefont {G.~V.}\ \bibnamefont {Shlyapnikov}}, \ and\
  \bibinfo {author} {\bibfnamefont {M.}~\bibnamefont {Zhan}},\ }\href {\doibase
  10.1103/PhysRevLett.119.160502} {\bibfield  {journal} {\bibinfo  {journal}
  {Phys. Rev. Lett.}\ }\textbf {\bibinfo {volume} {119}},\ \bibinfo {pages}
  {160502} (\bibinfo {year} {2017})}\BibitemShut {NoStop}%
\bibitem [{\citenamefont {Picken}\ \emph {et~al.}(2018)\citenamefont {Picken},
  \citenamefont {Legaie}, \citenamefont {McDonnell},\ and\ \citenamefont
  {Pritchard}}]{Picken_2018}%
  \BibitemOpen
  \bibfield  {author} {\bibinfo {author} {\bibfnamefont {C.~J.}\ \bibnamefont
  {Picken}}, \bibinfo {author} {\bibfnamefont {R.}~\bibnamefont {Legaie}},
  \bibinfo {author} {\bibfnamefont {K.}~\bibnamefont {McDonnell}}, \ and\
  \bibinfo {author} {\bibfnamefont {J.~D.}\ \bibnamefont {Pritchard}},\ }\href
  {\doibase 10.1088/2058-9565/aaf019} {\bibfield  {journal} {\bibinfo
  {journal} {Quan. Sci. Tech.}\ }\textbf {\bibinfo {volume} {4}},\ \bibinfo
  {pages} {015011} (\bibinfo {year} {2018})}\BibitemShut {NoStop}%
\bibitem [{\citenamefont {Levine}\ \emph {et~al.}(2018)\citenamefont {Levine},
  \citenamefont {Keesling}, \citenamefont {Omran}, \citenamefont {Bernien},
  \citenamefont {Schwartz}, \citenamefont {Zibrov}, \citenamefont {Endres},
  \citenamefont {Greiner}, \citenamefont {Vuleti\ifmmode~\acute{c}\else
  \'{c}\fi{}},\ and\ \citenamefont {Lukin}}]{saa2018}%
  \BibitemOpen
  \bibfield  {author} {\bibinfo {author} {\bibfnamefont {H.}~\bibnamefont
  {Levine}}, \bibinfo {author} {\bibfnamefont {A.}~\bibnamefont {Keesling}},
  \bibinfo {author} {\bibfnamefont {A.}~\bibnamefont {Omran}}, \bibinfo
  {author} {\bibfnamefont {H.}~\bibnamefont {Bernien}}, \bibinfo {author}
  {\bibfnamefont {S.}~\bibnamefont {Schwartz}}, \bibinfo {author}
  {\bibfnamefont {A.~S.}\ \bibnamefont {Zibrov}}, \bibinfo {author}
  {\bibfnamefont {M.}~\bibnamefont {Endres}}, \bibinfo {author} {\bibfnamefont
  {M.}~\bibnamefont {Greiner}}, \bibinfo {author} {\bibfnamefont
  {V.}~\bibnamefont {Vuleti\ifmmode~\acute{c}\else \'{c}\fi{}}}, \ and\
  \bibinfo {author} {\bibfnamefont {M.~D.}\ \bibnamefont {Lukin}},\ }\href
  {\doibase 10.1103/PhysRevLett.121.123603} {\bibfield  {journal} {\bibinfo
  {journal} {Phys. Rev. Lett.}\ }\textbf {\bibinfo {volume} {121}},\ \bibinfo
  {pages} {123603} (\bibinfo {year} {2018})}\BibitemShut {NoStop}%
\bibitem [{\citenamefont {Levine}\ \emph {et~al.}(2019)\citenamefont {Levine},
  \citenamefont {Keesling}, \citenamefont {Semeghini}, \citenamefont {Omran},
  \citenamefont {Wang}, \citenamefont {Ebadi}, \citenamefont {Bernien},
  \citenamefont {Greiner}, \citenamefont {Vuleti\ifmmode~\acute{c}\else
  \'{c}\fi{}}, \citenamefont {Pichler},\ and\ \citenamefont
  {Lukin}}]{levine2019parallel}%
  \BibitemOpen
  \bibfield  {author} {\bibinfo {author} {\bibfnamefont {H.}~\bibnamefont
  {Levine}}, \bibinfo {author} {\bibfnamefont {A.}~\bibnamefont {Keesling}},
  \bibinfo {author} {\bibfnamefont {G.}~\bibnamefont {Semeghini}}, \bibinfo
  {author} {\bibfnamefont {A.}~\bibnamefont {Omran}}, \bibinfo {author}
  {\bibfnamefont {T.~T.}\ \bibnamefont {Wang}}, \bibinfo {author}
  {\bibfnamefont {S.}~\bibnamefont {Ebadi}}, \bibinfo {author} {\bibfnamefont
  {H.}~\bibnamefont {Bernien}}, \bibinfo {author} {\bibfnamefont
  {M.}~\bibnamefont {Greiner}}, \bibinfo {author} {\bibfnamefont
  {V.}~\bibnamefont {Vuleti\ifmmode~\acute{c}\else \'{c}\fi{}}}, \bibinfo
  {author} {\bibfnamefont {H.}~\bibnamefont {Pichler}}, \ and\ \bibinfo
  {author} {\bibfnamefont {M.~D.}\ \bibnamefont {Lukin}},\ }\href {\doibase
  10.1103/PhysRevLett.123.170503} {\bibfield  {journal} {\bibinfo  {journal}
  {Phys. Rev. Lett.}\ }\textbf {\bibinfo {volume} {123}},\ \bibinfo {pages}
  {170503} (\bibinfo {year} {2019})}\BibitemShut {NoStop}%
\bibitem [{\citenamefont {Graham}\ \emph {et~al.}(2019)\citenamefont {Graham},
  \citenamefont {Kwon}, \citenamefont {Grinkemeyer}, \citenamefont {Marra},
  \citenamefont {Jiang}, \citenamefont {Lichtman}, \citenamefont {Sun},
  \citenamefont {Ebert},\ and\ \citenamefont {Saffman}}]{graham2019rydberg}%
  \BibitemOpen
  \bibfield  {author} {\bibinfo {author} {\bibfnamefont {T.~M.}\ \bibnamefont
  {Graham}}, \bibinfo {author} {\bibfnamefont {M.}~\bibnamefont {Kwon}},
  \bibinfo {author} {\bibfnamefont {B.}~\bibnamefont {Grinkemeyer}}, \bibinfo
  {author} {\bibfnamefont {Z.}~\bibnamefont {Marra}}, \bibinfo {author}
  {\bibfnamefont {X.}~\bibnamefont {Jiang}}, \bibinfo {author} {\bibfnamefont
  {M.~T.}\ \bibnamefont {Lichtman}}, \bibinfo {author} {\bibfnamefont
  {Y.}~\bibnamefont {Sun}}, \bibinfo {author} {\bibfnamefont {M.}~\bibnamefont
  {Ebert}}, \ and\ \bibinfo {author} {\bibfnamefont {M.}~\bibnamefont
  {Saffman}},\ }\href {\doibase 10.1103/PhysRevLett.123.230501} {\bibfield
  {journal} {\bibinfo  {journal} {Phys. Rev. Lett.}\ }\textbf {\bibinfo
  {volume} {123}},\ \bibinfo {pages} {230501} (\bibinfo {year}
  {2019})}\BibitemShut {NoStop}%
\bibitem [{\citenamefont {Omran}\ \emph {et~al.}(2019)\citenamefont {Omran},
  \citenamefont {Levine}, \citenamefont {Keesling}, \citenamefont {Semeghini},
  \citenamefont {Wang}, \citenamefont {Ebadi}, \citenamefont {Bernien},
  \citenamefont {Zibrov}, \citenamefont {Pichler}, \citenamefont {Choi},
  \citenamefont {Cui}, \citenamefont {Rossignolo}, \citenamefont {Rembold},
  \citenamefont {Montangero}, \citenamefont {Calarco}, \citenamefont {Endres},
  \citenamefont {Greiner}, \citenamefont {Vuleti{\'c}},\ and\ \citenamefont
  {Lukin}}]{Omran570}%
  \BibitemOpen
  \bibfield  {author} {\bibinfo {author} {\bibfnamefont {A.}~\bibnamefont
  {Omran}}, \bibinfo {author} {\bibfnamefont {H.}~\bibnamefont {Levine}},
  \bibinfo {author} {\bibfnamefont {A.}~\bibnamefont {Keesling}}, \bibinfo
  {author} {\bibfnamefont {G.}~\bibnamefont {Semeghini}}, \bibinfo {author}
  {\bibfnamefont {T.~T.}\ \bibnamefont {Wang}}, \bibinfo {author}
  {\bibfnamefont {S.}~\bibnamefont {Ebadi}}, \bibinfo {author} {\bibfnamefont
  {H.}~\bibnamefont {Bernien}}, \bibinfo {author} {\bibfnamefont {A.~S.}\
  \bibnamefont {Zibrov}}, \bibinfo {author} {\bibfnamefont {H.}~\bibnamefont
  {Pichler}}, \bibinfo {author} {\bibfnamefont {S.}~\bibnamefont {Choi}},
  \bibinfo {author} {\bibfnamefont {J.}~\bibnamefont {Cui}}, \bibinfo {author}
  {\bibfnamefont {M.}~\bibnamefont {Rossignolo}}, \bibinfo {author}
  {\bibfnamefont {P.}~\bibnamefont {Rembold}}, \bibinfo {author} {\bibfnamefont
  {S.}~\bibnamefont {Montangero}}, \bibinfo {author} {\bibfnamefont
  {T.}~\bibnamefont {Calarco}}, \bibinfo {author} {\bibfnamefont
  {M.}~\bibnamefont {Endres}}, \bibinfo {author} {\bibfnamefont
  {M.}~\bibnamefont {Greiner}}, \bibinfo {author} {\bibfnamefont
  {V.}~\bibnamefont {Vuleti{\'c}}}, \ and\ \bibinfo {author} {\bibfnamefont
  {M.~D.}\ \bibnamefont {Lukin}},\ }\href {\doibase 10.1126/science.aax9743}
  {\bibfield  {journal} {\bibinfo  {journal} {Science}\ }\textbf {\bibinfo
  {volume} {365}},\ \bibinfo {pages} {570} (\bibinfo {year}
  {2019})}\BibitemShut {NoStop}%
\bibitem [{\citenamefont {Ates}\ \emph {et~al.}(2007)\citenamefont {Ates},
  \citenamefont {Pohl}, \citenamefont {Pattard},\ and\ \citenamefont
  {Rost}}]{Ates2007}%
  \BibitemOpen
  \bibfield  {author} {\bibinfo {author} {\bibfnamefont {C.}~\bibnamefont
  {Ates}}, \bibinfo {author} {\bibfnamefont {T.}~\bibnamefont {Pohl}}, \bibinfo
  {author} {\bibfnamefont {T.}~\bibnamefont {Pattard}}, \ and\ \bibinfo
  {author} {\bibfnamefont {J.~M.}\ \bibnamefont {Rost}},\ }\href {\doibase
  10.1103/PhysRevLett.98.023002} {\bibfield  {journal} {\bibinfo  {journal}
  {Phys. Rev. Lett.}\ }\textbf {\bibinfo {volume} {98}},\ \bibinfo {pages}
  {023002} (\bibinfo {year} {2007})}\BibitemShut {NoStop}%
\bibitem [{\citenamefont {Amthor}\ \emph {et~al.}(2010)\citenamefont {Amthor},
  \citenamefont {Giese}, \citenamefont {Hofmann},\ and\ \citenamefont
  {Weidem\"uller}}]{Amthor2010}%
  \BibitemOpen
  \bibfield  {author} {\bibinfo {author} {\bibfnamefont {T.}~\bibnamefont
  {Amthor}}, \bibinfo {author} {\bibfnamefont {C.}~\bibnamefont {Giese}},
  \bibinfo {author} {\bibfnamefont {C.~S.}\ \bibnamefont {Hofmann}}, \ and\
  \bibinfo {author} {\bibfnamefont {M.}~\bibnamefont {Weidem\"uller}},\ }\href
  {\doibase 10.1103/PhysRevLett.104.013001} {\bibfield  {journal} {\bibinfo
  {journal} {Phys. Rev. Lett.}\ }\textbf {\bibinfo {volume} {104}},\ \bibinfo
  {pages} {013001} (\bibinfo {year} {2010})}\BibitemShut {NoStop}%
\bibitem [{\citenamefont {Zuo}\ and\ \citenamefont {Nakagawa}(2010)}]{Zuo2010}%
  \BibitemOpen
  \bibfield  {author} {\bibinfo {author} {\bibfnamefont {Z.}~\bibnamefont
  {Zuo}}\ and\ \bibinfo {author} {\bibfnamefont {K.}~\bibnamefont {Nakagawa}},\
  }\href {\doibase 10.1103/PhysRevA.82.062328} {\bibfield  {journal} {\bibinfo
  {journal} {Phys. Rev. A}\ }\textbf {\bibinfo {volume} {82}},\ \bibinfo
  {pages} {062328} (\bibinfo {year} {2010})}\BibitemShut {NoStop}%
\bibitem [{\citenamefont {Lee}\ \emph {et~al.}(2012)\citenamefont {Lee},
  \citenamefont {H\"affner},\ and\ \citenamefont {Cross}}]{Lee2012}%
  \BibitemOpen
  \bibfield  {author} {\bibinfo {author} {\bibfnamefont {T.~E.}\ \bibnamefont
  {Lee}}, \bibinfo {author} {\bibfnamefont {H.}~\bibnamefont {H\"affner}}, \
  and\ \bibinfo {author} {\bibfnamefont {M.~C.}\ \bibnamefont {Cross}},\ }\href
  {\doibase 10.1103/PhysRevLett.108.023602} {\bibfield  {journal} {\bibinfo
  {journal} {Phys. Rev. Lett.}\ }\textbf {\bibinfo {volume} {108}},\ \bibinfo
  {pages} {023602} (\bibinfo {year} {2012})}\BibitemShut {NoStop}%
\bibitem [{\citenamefont {Ates}\ \emph {et~al.}(2012)\citenamefont {Ates},
  \citenamefont {Olmos}, \citenamefont {Li},\ and\ \citenamefont
  {Lesanovsky}}]{PhysRevLett.109.233003}%
  \BibitemOpen
  \bibfield  {author} {\bibinfo {author} {\bibfnamefont {C.}~\bibnamefont
  {Ates}}, \bibinfo {author} {\bibfnamefont {B.}~\bibnamefont {Olmos}},
  \bibinfo {author} {\bibfnamefont {W.}~\bibnamefont {Li}}, \ and\ \bibinfo
  {author} {\bibfnamefont {I.}~\bibnamefont {Lesanovsky}},\ }\href {\doibase
  10.1103/PhysRevLett.109.233003} {\bibfield  {journal} {\bibinfo  {journal}
  {Phys. Rev. Lett.}\ }\textbf {\bibinfo {volume} {109}},\ \bibinfo {pages}
  {233003} (\bibinfo {year} {2012})}\BibitemShut {NoStop}%
\bibitem [{\citenamefont {Li}\ \emph {et~al.}(2013)\citenamefont {Li},
  \citenamefont {Ates},\ and\ \citenamefont {Lesanovsky}}]{Li2013}%
  \BibitemOpen
  \bibfield  {author} {\bibinfo {author} {\bibfnamefont {W.}~\bibnamefont
  {Li}}, \bibinfo {author} {\bibfnamefont {C.}~\bibnamefont {Ates}}, \ and\
  \bibinfo {author} {\bibfnamefont {I.}~\bibnamefont {Lesanovsky}},\ }\href
  {\doibase 10.1103/PhysRevLett.110.213005} {\bibfield  {journal} {\bibinfo
  {journal} {Phys. Rev. Lett.}\ }\textbf {\bibinfo {volume} {110}},\ \bibinfo
  {pages} {213005} (\bibinfo {year} {2013})}\BibitemShut {NoStop}%
\bibitem [{\citenamefont {Carr}\ and\ \citenamefont
  {Saffman}(2013)}]{Carr2013}%
  \BibitemOpen
  \bibfield  {author} {\bibinfo {author} {\bibfnamefont {A.~W.}\ \bibnamefont
  {Carr}}\ and\ \bibinfo {author} {\bibfnamefont {M.}~\bibnamefont {Saffman}},\
  }\href {\doibase 10.1103/PhysRevLett.111.033607} {\bibfield  {journal}
  {\bibinfo  {journal} {Phys. Rev. Lett.}\ }\textbf {\bibinfo {volume} {111}},\
  \bibinfo {pages} {033607} (\bibinfo {year} {2013})}\BibitemShut {NoStop}%
\bibitem [{\citenamefont {Chen}\ \emph {et~al.}(2018)\citenamefont {Chen},
  \citenamefont {Shi}, \citenamefont {Song}, \citenamefont {Xia},\ and\
  \citenamefont {Zheng}}]{chen2018accelerated}%
  \BibitemOpen
  \bibfield  {author} {\bibinfo {author} {\bibfnamefont {Y.-H.}\ \bibnamefont
  {Chen}}, \bibinfo {author} {\bibfnamefont {Z.-C.}\ \bibnamefont {Shi}},
  \bibinfo {author} {\bibfnamefont {J.}~\bibnamefont {Song}}, \bibinfo {author}
  {\bibfnamefont {Y.}~\bibnamefont {Xia}}, \ and\ \bibinfo {author}
  {\bibfnamefont {S.-B.}\ \bibnamefont {Zheng}},\ }\href {\doibase
  10.1103/PhysRevA.97.032328} {\bibfield  {journal} {\bibinfo  {journal} {Phys.
  Rev. A}\ }\textbf {\bibinfo {volume} {97}},\ \bibinfo {pages} {032328}
  (\bibinfo {year} {2018})}\BibitemShut {NoStop}%
\bibitem [{\citenamefont {Li}\ and\ \citenamefont {Shao}(2019)}]{Li2019}%
  \BibitemOpen
  \bibfield  {author} {\bibinfo {author} {\bibfnamefont {D.~X.}\ \bibnamefont
  {Li}}\ and\ \bibinfo {author} {\bibfnamefont {X.~Q.}\ \bibnamefont {Shao}},\
  }\href {\doibase 10.1103/PhysRevA.99.032348} {\bibfield  {journal} {\bibinfo
  {journal} {Phys. Rev. A}\ }\textbf {\bibinfo {volume} {99}},\ \bibinfo
  {pages} {032348} (\bibinfo {year} {2019})}\BibitemShut {NoStop}%
\bibitem [{\citenamefont {Yang}\ \emph {et~al.}(2019)\citenamefont {Yang},
  \citenamefont {Li},\ and\ \citenamefont {Shao}}]{Yang2019}%
  \BibitemOpen
  \bibfield  {author} {\bibinfo {author} {\bibfnamefont {C.}~\bibnamefont
  {Yang}}, \bibinfo {author} {\bibfnamefont {D.}~\bibnamefont {Li}}, \ and\
  \bibinfo {author} {\bibfnamefont {X.}~\bibnamefont {Shao}},\ }\href {\doibase
  10.1007/s11433-019-1431-0} {\bibfield  {journal} {\bibinfo  {journal} {Sci.
  China Phys., Mech.}\ }\textbf {\bibinfo {volume} {62}},\ \bibinfo {pages}
  {110312} (\bibinfo {year} {2019})}\BibitemShut {NoStop}%
\bibitem [{\citenamefont {Li}\ \emph {et~al.}(2020{\natexlab{a}})\citenamefont
  {Li}, \citenamefont {Yu}, \citenamefont {Su},\ and\ \citenamefont
  {Qian}}]{Lirui2020}%
  \BibitemOpen
  \bibfield  {author} {\bibinfo {author} {\bibfnamefont {R.}~\bibnamefont
  {Li}}, \bibinfo {author} {\bibfnamefont {D.}~\bibnamefont {Yu}}, \bibinfo
  {author} {\bibfnamefont {S.-L.}\ \bibnamefont {Su}}, \ and\ \bibinfo {author}
  {\bibfnamefont {J.}~\bibnamefont {Qian}},\ }\href {\doibase
  10.1103/PhysRevA.101.042328} {\bibfield  {journal} {\bibinfo  {journal}
  {Phys. Rev. A}\ }\textbf {\bibinfo {volume} {101}},\ \bibinfo {pages}
  {042328} (\bibinfo {year} {2020}{\natexlab{a}})}\BibitemShut {NoStop}%
\bibitem [{\citenamefont {Basak}\ \emph {et~al.}(2018)\citenamefont {Basak},
  \citenamefont {Chougale},\ and\ \citenamefont
  {Nath}}]{PhysRevLett.120.123204}%
  \BibitemOpen
  \bibfield  {author} {\bibinfo {author} {\bibfnamefont {S.}~\bibnamefont
  {Basak}}, \bibinfo {author} {\bibfnamefont {Y.}~\bibnamefont {Chougale}}, \
  and\ \bibinfo {author} {\bibfnamefont {R.}~\bibnamefont {Nath}},\ }\href
  {\doibase 10.1103/PhysRevLett.120.123204} {\bibfield  {journal} {\bibinfo
  {journal} {Phys. Rev. Lett.}\ }\textbf {\bibinfo {volume} {120}},\ \bibinfo
  {pages} {123204} (\bibinfo {year} {2018})}\BibitemShut {NoStop}%
\bibitem [{\citenamefont {Su}\ \emph {et~al.}(2018)\citenamefont {Su},
  \citenamefont {Shen}, \citenamefont {Liang},\ and\ \citenamefont
  {Zhang}}]{Su2018}%
  \BibitemOpen
  \bibfield  {author} {\bibinfo {author} {\bibfnamefont {S.~L.}\ \bibnamefont
  {Su}}, \bibinfo {author} {\bibfnamefont {H.~Z.}\ \bibnamefont {Shen}},
  \bibinfo {author} {\bibfnamefont {E.}~\bibnamefont {Liang}}, \ and\ \bibinfo
  {author} {\bibfnamefont {S.}~\bibnamefont {Zhang}},\ }\href {\doibase
  10.1103/PhysRevA.98.032306} {\bibfield  {journal} {\bibinfo  {journal} {Phys.
  Rev. A}\ }\textbf {\bibinfo {volume} {98}},\ \bibinfo {pages} {032306}
  (\bibinfo {year} {2018})}\BibitemShut {NoStop}%
\bibitem [{\citenamefont {Su}\ \emph {et~al.}(2020{\natexlab{a}})\citenamefont
  {Su}, \citenamefont {Guo}, \citenamefont {Tian}, \citenamefont {Zhu},
  \citenamefont {Yan}, \citenamefont {Liang},\ and\ \citenamefont
  {Feng}}]{Su2020}%
  \BibitemOpen
  \bibfield  {author} {\bibinfo {author} {\bibfnamefont {S.-L.}\ \bibnamefont
  {Su}}, \bibinfo {author} {\bibfnamefont {F.-Q.}\ \bibnamefont {Guo}},
  \bibinfo {author} {\bibfnamefont {L.}~\bibnamefont {Tian}}, \bibinfo {author}
  {\bibfnamefont {X.-Y.}\ \bibnamefont {Zhu}}, \bibinfo {author} {\bibfnamefont
  {L.-L.}\ \bibnamefont {Yan}}, \bibinfo {author} {\bibfnamefont {E.-J.}\
  \bibnamefont {Liang}}, \ and\ \bibinfo {author} {\bibfnamefont
  {M.}~\bibnamefont {Feng}},\ }\href {\doibase 10.1103/PhysRevA.101.012347}
  {\bibfield  {journal} {\bibinfo  {journal} {Phys. Rev. A}\ }\textbf {\bibinfo
  {volume} {101}},\ \bibinfo {pages} {012347} (\bibinfo {year}
  {2020}{\natexlab{a}})}\BibitemShut {NoStop}%
\bibitem [{\citenamefont {Wu}\ \emph {et~al.}(2020)\citenamefont {Wu},
  \citenamefont {Su}, \citenamefont {Wang}, \citenamefont {Song}, \citenamefont
  {Xia},\ and\ \citenamefont {Jiang}}]{Wu2020}%
  \BibitemOpen
  \bibfield  {author} {\bibinfo {author} {\bibfnamefont {J.-L.}\ \bibnamefont
  {Wu}}, \bibinfo {author} {\bibfnamefont {S.-L.}\ \bibnamefont {Su}}, \bibinfo
  {author} {\bibfnamefont {Y.}~\bibnamefont {Wang}}, \bibinfo {author}
  {\bibfnamefont {J.}~\bibnamefont {Song}}, \bibinfo {author} {\bibfnamefont
  {Y.}~\bibnamefont {Xia}}, \ and\ \bibinfo {author} {\bibfnamefont {Y.-Y.}\
  \bibnamefont {Jiang}},\ }\href {\doibase 10.1364/OL.386765} {\bibfield
  {journal} {\bibinfo  {journal} {Opt. Lett.}\ }\textbf {\bibinfo {volume}
  {45}},\ \bibinfo {pages} {1200} (\bibinfo {year} {2020})}\BibitemShut
  {NoStop}%
\bibitem [{\citenamefont {Zheng}\ \emph {et~al.}(2020)\citenamefont {Zheng},
  \citenamefont {Kang}, \citenamefont {Su}, \citenamefont {Song},\ and\
  \citenamefont {Xia}}]{Zheng2020}%
  \BibitemOpen
  \bibfield  {author} {\bibinfo {author} {\bibfnamefont {R.-H.}\ \bibnamefont
  {Zheng}}, \bibinfo {author} {\bibfnamefont {Y.-H.}\ \bibnamefont {Kang}},
  \bibinfo {author} {\bibfnamefont {S.-L.}\ \bibnamefont {Su}}, \bibinfo
  {author} {\bibfnamefont {J.}~\bibnamefont {Song}}, \ and\ \bibinfo {author}
  {\bibfnamefont {Y.}~\bibnamefont {Xia}},\ }\href {\doibase
  10.1103/PhysRevA.102.012609} {\bibfield  {journal} {\bibinfo  {journal}
  {Phys. Rev. A}\ }\textbf {\bibinfo {volume} {102}},\ \bibinfo {pages}
  {012609} (\bibinfo {year} {2020})}\BibitemShut {NoStop}%
\bibitem [{\citenamefont {Su}\ \emph {et~al.}(2017)\citenamefont {Su},
  \citenamefont {Gao}, \citenamefont {Liang},\ and\ \citenamefont
  {Zhang}}]{Su201702}%
  \BibitemOpen
  \bibfield  {author} {\bibinfo {author} {\bibfnamefont {S.-L.}\ \bibnamefont
  {Su}}, \bibinfo {author} {\bibfnamefont {Y.}~\bibnamefont {Gao}}, \bibinfo
  {author} {\bibfnamefont {E.}~\bibnamefont {Liang}}, \ and\ \bibinfo {author}
  {\bibfnamefont {S.}~\bibnamefont {Zhang}},\ }\href {\doibase
  10.1103/PhysRevA.95.022319} {\bibfield  {journal} {\bibinfo  {journal} {Phys.
  Rev. A}\ }\textbf {\bibinfo {volume} {95}},\ \bibinfo {pages} {022319}
  (\bibinfo {year} {2017})}\BibitemShut {NoStop}%
\bibitem [{\citenamefont {Gambetta}\ \emph
  {et~al.}(2020{\natexlab{a}})\citenamefont {Gambetta}, \citenamefont {Zhang},
  \citenamefont {Hennrich}, \citenamefont {Lesanovsky},\ and\ \citenamefont
  {Li}}]{Gambetta20202}%
  \BibitemOpen
  \bibfield  {author} {\bibinfo {author} {\bibfnamefont {F.~M.}\ \bibnamefont
  {Gambetta}}, \bibinfo {author} {\bibfnamefont {C.}~\bibnamefont {Zhang}},
  \bibinfo {author} {\bibfnamefont {M.}~\bibnamefont {Hennrich}}, \bibinfo
  {author} {\bibfnamefont {I.}~\bibnamefont {Lesanovsky}}, \ and\ \bibinfo
  {author} {\bibfnamefont {W.}~\bibnamefont {Li}},\ }\href {\doibase
  10.1103/PhysRevLett.125.133602} {\bibfield  {journal} {\bibinfo  {journal}
  {Phys. Rev. Lett.}\ }\textbf {\bibinfo {volume} {125}},\ \bibinfo {pages}
  {133602} (\bibinfo {year} {2020}{\natexlab{a}})}\BibitemShut {NoStop}%
\bibitem [{\citenamefont {Mazza}\ \emph {et~al.}(2020)\citenamefont {Mazza},
  \citenamefont {Schmidt},\ and\ \citenamefont {Lesanovsky}}]{Mazza2020}%
  \BibitemOpen
  \bibfield  {author} {\bibinfo {author} {\bibfnamefont {P.~P.}\ \bibnamefont
  {Mazza}}, \bibinfo {author} {\bibfnamefont {R.}~\bibnamefont {Schmidt}}, \
  and\ \bibinfo {author} {\bibfnamefont {I.}~\bibnamefont {Lesanovsky}},\
  }\href {\doibase 10.1103/PhysRevLett.125.033602} {\bibfield  {journal}
  {\bibinfo  {journal} {Phys. Rev. Lett.}\ }\textbf {\bibinfo {volume} {125}},\
  \bibinfo {pages} {033602} (\bibinfo {year} {2020})}\BibitemShut {NoStop}%
\bibitem [{\citenamefont {Taylor}\ \emph {et~al.}(2019)\citenamefont {Taylor},
  \citenamefont {Sinclair}, \citenamefont {Bonsma-Fisher}, \citenamefont
  {England}, \citenamefont {Spanner},\ and\ \citenamefont
  {Heshami}}]{taylor2019generation}%
  \BibitemOpen
  \bibfield  {author} {\bibinfo {author} {\bibfnamefont {J.}~\bibnamefont
  {Taylor}}, \bibinfo {author} {\bibfnamefont {J.}~\bibnamefont {Sinclair}},
  \bibinfo {author} {\bibfnamefont {K.}~\bibnamefont {Bonsma-Fisher}}, \bibinfo
  {author} {\bibfnamefont {D.}~\bibnamefont {England}}, \bibinfo {author}
  {\bibfnamefont {M.}~\bibnamefont {Spanner}}, \ and\ \bibinfo {author}
  {\bibfnamefont {K.}~\bibnamefont {Heshami}},\ }\href@noop {} {} (\bibinfo
  {year} {2019}),\ \Eprint {http://arxiv.org/abs/1912.05675} {arXiv:1912.05675
  [physics.atom-ph]} \BibitemShut {NoStop}%
\bibitem [{\citenamefont {Bai}\ \emph {et~al.}(2020)\citenamefont {Bai},
  \citenamefont {Tian}, \citenamefont {Han}, \citenamefont {Jiao},
  \citenamefont {Wu}, \citenamefont {Zhao},\ and\ \citenamefont
  {Jia}}]{Bai_2020}%
  \BibitemOpen
  \bibfield  {author} {\bibinfo {author} {\bibfnamefont {S.}~\bibnamefont
  {Bai}}, \bibinfo {author} {\bibfnamefont {X.}~\bibnamefont {Tian}}, \bibinfo
  {author} {\bibfnamefont {X.}~\bibnamefont {Han}}, \bibinfo {author}
  {\bibfnamefont {Y.}~\bibnamefont {Jiao}}, \bibinfo {author} {\bibfnamefont
  {J.}~\bibnamefont {Wu}}, \bibinfo {author} {\bibfnamefont {J.}~\bibnamefont
  {Zhao}}, \ and\ \bibinfo {author} {\bibfnamefont {S.}~\bibnamefont {Jia}},\
  }\href {\doibase 10.1088/1367-2630/ab6575} {\bibfield  {journal} {\bibinfo
  {journal} {New J. Phys.}\ }\textbf {\bibinfo {volume} {22}},\ \bibinfo
  {pages} {013004} (\bibinfo {year} {2020})}\BibitemShut {NoStop}%
\bibitem [{\citenamefont {Cidrim}\ \emph {et~al.}(2020)\citenamefont {Cidrim},
  \citenamefont {do~Espirito~Santo}, \citenamefont {Schachenmayer},
  \citenamefont {Kaiser},\ and\ \citenamefont {Bachelard}}]{Cidrim2020}%
  \BibitemOpen
  \bibfield  {author} {\bibinfo {author} {\bibfnamefont {A.}~\bibnamefont
  {Cidrim}}, \bibinfo {author} {\bibfnamefont {T.~S.}\ \bibnamefont
  {do~Espirito~Santo}}, \bibinfo {author} {\bibfnamefont {J.}~\bibnamefont
  {Schachenmayer}}, \bibinfo {author} {\bibfnamefont {R.}~\bibnamefont
  {Kaiser}}, \ and\ \bibinfo {author} {\bibfnamefont {R.}~\bibnamefont
  {Bachelard}},\ }\href {\doibase 10.1103/PhysRevLett.125.073601} {\bibfield
  {journal} {\bibinfo  {journal} {Phys. Rev. Lett.}\ }\textbf {\bibinfo
  {volume} {125}},\ \bibinfo {pages} {073601} (\bibinfo {year}
  {2020})}\BibitemShut {NoStop}%
\bibitem [{\citenamefont {Williamson}\ \emph {et~al.}(2020)\citenamefont
  {Williamson}, \citenamefont {Borgh},\ and\ \citenamefont
  {Ruostekoski}}]{Williamson2020}%
  \BibitemOpen
  \bibfield  {author} {\bibinfo {author} {\bibfnamefont {L.~A.}\ \bibnamefont
  {Williamson}}, \bibinfo {author} {\bibfnamefont {M.~O.}\ \bibnamefont
  {Borgh}}, \ and\ \bibinfo {author} {\bibfnamefont {J.}~\bibnamefont
  {Ruostekoski}},\ }\href {\doibase 10.1103/PhysRevLett.125.073602} {\bibfield
  {journal} {\bibinfo  {journal} {Phys. Rev. Lett.}\ }\textbf {\bibinfo
  {volume} {125}},\ \bibinfo {pages} {073602} (\bibinfo {year}
  {2020})}\BibitemShut {NoStop}%
\bibitem [{\citenamefont {Orioli}\ \emph {et~al.}(2018)\citenamefont {Orioli},
  \citenamefont {Signoles}, \citenamefont {Wildhagen}, \citenamefont
  {G{\"u}nter}, \citenamefont {Berges}, \citenamefont {Whitlock},\ and\
  \citenamefont {Weidem{\"u}ller}}]{orioli_relaxation_2018}%
  \BibitemOpen
  \bibfield  {author} {\bibinfo {author} {\bibfnamefont {A.~P.}\ \bibnamefont
  {Orioli}}, \bibinfo {author} {\bibfnamefont {A.}~\bibnamefont {Signoles}},
  \bibinfo {author} {\bibfnamefont {H.}~\bibnamefont {Wildhagen}}, \bibinfo
  {author} {\bibfnamefont {G.}~\bibnamefont {G{\"u}nter}}, \bibinfo {author}
  {\bibfnamefont {J.}~\bibnamefont {Berges}}, \bibinfo {author} {\bibfnamefont
  {S.}~\bibnamefont {Whitlock}}, \ and\ \bibinfo {author} {\bibfnamefont
  {M.}~\bibnamefont {Weidem{\"u}ller}},\ }\href {\doibase
  10.1103/PhysRevLett.120.063601} {\bibfield  {journal} {\bibinfo  {journal}
  {Phys. Rev. Lett.}\ }\textbf {\bibinfo {volume} {120}},\ \bibinfo {pages}
  {063601} (\bibinfo {year} {2018})}\BibitemShut {NoStop}%
\bibitem [{\citenamefont {Beterov}\ \emph {et~al.}(2016)\citenamefont
  {Beterov}, \citenamefont {Saffman}, \citenamefont {Yakshina}, \citenamefont
  {Tretyakov}, \citenamefont {Entin}, \citenamefont {Bergamini}, \citenamefont
  {Kuznetsova},\ and\ \citenamefont {Ryabtsev}}]{Beterov2016}%
  \BibitemOpen
  \bibfield  {author} {\bibinfo {author} {\bibfnamefont {I.~I.}\ \bibnamefont
  {Beterov}}, \bibinfo {author} {\bibfnamefont {M.}~\bibnamefont {Saffman}},
  \bibinfo {author} {\bibfnamefont {E.~A.}\ \bibnamefont {Yakshina}}, \bibinfo
  {author} {\bibfnamefont {D.~B.}\ \bibnamefont {Tretyakov}}, \bibinfo {author}
  {\bibfnamefont {V.~M.}\ \bibnamefont {Entin}}, \bibinfo {author}
  {\bibfnamefont {S.}~\bibnamefont {Bergamini}}, \bibinfo {author}
  {\bibfnamefont {E.~A.}\ \bibnamefont {Kuznetsova}}, \ and\ \bibinfo {author}
  {\bibfnamefont {I.~I.}\ \bibnamefont {Ryabtsev}},\ }\href {\doibase
  10.1103/PhysRevA.94.062307} {\bibfield  {journal} {\bibinfo  {journal} {Phys.
  Rev. A}\ }\textbf {\bibinfo {volume} {94}},\ \bibinfo {pages} {062307}
  (\bibinfo {year} {2016})}\BibitemShut {NoStop}%
\bibitem [{\citenamefont {Beterov}\ \emph
  {et~al.}(2018{\natexlab{a}})\citenamefont {Beterov}, \citenamefont
  {Ashkarin}, \citenamefont {Yakshina}, \citenamefont {Tretyakov},
  \citenamefont {Entin}, \citenamefont {Ryabtsev}, \citenamefont {Cheinet},
  \citenamefont {Pillet},\ and\ \citenamefont {Saffman}}]{Beterov2018}%
  \BibitemOpen
  \bibfield  {author} {\bibinfo {author} {\bibfnamefont {I.~I.}\ \bibnamefont
  {Beterov}}, \bibinfo {author} {\bibfnamefont {I.~N.}\ \bibnamefont
  {Ashkarin}}, \bibinfo {author} {\bibfnamefont {E.~A.}\ \bibnamefont
  {Yakshina}}, \bibinfo {author} {\bibfnamefont {D.~B.}\ \bibnamefont
  {Tretyakov}}, \bibinfo {author} {\bibfnamefont {V.~M.}\ \bibnamefont
  {Entin}}, \bibinfo {author} {\bibfnamefont {I.~I.}\ \bibnamefont {Ryabtsev}},
  \bibinfo {author} {\bibfnamefont {P.}~\bibnamefont {Cheinet}}, \bibinfo
  {author} {\bibfnamefont {P.}~\bibnamefont {Pillet}}, \ and\ \bibinfo {author}
  {\bibfnamefont {M.}~\bibnamefont {Saffman}},\ }\href {\doibase
  10.1103/PhysRevA.98.042704} {\bibfield  {journal} {\bibinfo  {journal} {Phys.
  Rev. A}\ }\textbf {\bibinfo {volume} {98}},\ \bibinfo {pages} {042704}
  (\bibinfo {year} {2018}{\natexlab{a}})}\BibitemShut {NoStop}%
\bibitem [{\citenamefont {Tretyakov}\ \emph {et~al.}(2017)\citenamefont
  {Tretyakov}, \citenamefont {Beterov}, \citenamefont {Yakshina}, \citenamefont
  {Entin}, \citenamefont {Ryabtsev}, \citenamefont {Cheinet},\ and\
  \citenamefont {Pillet}}]{Tretyakov2017}%
  \BibitemOpen
  \bibfield  {author} {\bibinfo {author} {\bibfnamefont {D.~B.}\ \bibnamefont
  {Tretyakov}}, \bibinfo {author} {\bibfnamefont {I.~I.}\ \bibnamefont
  {Beterov}}, \bibinfo {author} {\bibfnamefont {E.~A.}\ \bibnamefont
  {Yakshina}}, \bibinfo {author} {\bibfnamefont {V.~M.}\ \bibnamefont {Entin}},
  \bibinfo {author} {\bibfnamefont {I.~I.}\ \bibnamefont {Ryabtsev}}, \bibinfo
  {author} {\bibfnamefont {P.}~\bibnamefont {Cheinet}}, \ and\ \bibinfo
  {author} {\bibfnamefont {P.}~\bibnamefont {Pillet}},\ }\href {\doibase
  10.1103/PhysRevLett.119.173402} {\bibfield  {journal} {\bibinfo  {journal}
  {Phys. Rev. Lett.}\ }\textbf {\bibinfo {volume} {119}},\ \bibinfo {pages}
  {173402} (\bibinfo {year} {2017})}\BibitemShut {NoStop}%
\bibitem [{\citenamefont {Shi}(2017)}]{Shi2017}%
  \BibitemOpen
  \bibfield  {author} {\bibinfo {author} {\bibfnamefont {X.-F.}\ \bibnamefont
  {Shi}},\ }\href {\doibase 10.1103/PhysRevApplied.7.064017} {\bibfield
  {journal} {\bibinfo  {journal} {Phys. Rev. Applied}\ }\textbf {\bibinfo
  {volume} {7}},\ \bibinfo {pages} {064017} (\bibinfo {year}
  {2017})}\BibitemShut {NoStop}%
\bibitem [{\citenamefont {Khazali}\ and\ \citenamefont
  {M\o{}lmer}(2020)}]{Klaus2020}%
  \BibitemOpen
  \bibfield  {author} {\bibinfo {author} {\bibfnamefont {M.}~\bibnamefont
  {Khazali}}\ and\ \bibinfo {author} {\bibfnamefont {K.}~\bibnamefont
  {M\o{}lmer}},\ }\href {\doibase 10.1103/PhysRevX.10.021054} {\bibfield
  {journal} {\bibinfo  {journal} {Phys. Rev. X}\ }\textbf {\bibinfo {volume}
  {10}},\ \bibinfo {pages} {021054} (\bibinfo {year} {2020})}\BibitemShut
  {NoStop}%
\bibitem [{\citenamefont {Gambetta}\ \emph {et~al.}(2021)\citenamefont
  {Gambetta}, \citenamefont {Zhang}, \citenamefont {Hennrich}, \citenamefont
  {Lesanovsky},\ and\ \citenamefont {Li}}]{PhysRevLett.126.233404}%
  \BibitemOpen
  \bibfield  {author} {\bibinfo {author} {\bibfnamefont {F.~M.}\ \bibnamefont
  {Gambetta}}, \bibinfo {author} {\bibfnamefont {C.}~\bibnamefont {Zhang}},
  \bibinfo {author} {\bibfnamefont {M.}~\bibnamefont {Hennrich}}, \bibinfo
  {author} {\bibfnamefont {I.}~\bibnamefont {Lesanovsky}}, \ and\ \bibinfo
  {author} {\bibfnamefont {W.}~\bibnamefont {Li}},\ }\href {\doibase
  10.1103/PhysRevLett.126.233404} {\bibfield  {journal} {\bibinfo  {journal}
  {Phys. Rev. Lett.}\ }\textbf {\bibinfo {volume} {126}},\ \bibinfo {pages}
  {233404} (\bibinfo {year} {2021})}\BibitemShut {NoStop}%
\bibitem [{\citenamefont {Pohl}\ and\ \citenamefont {Berman}(2009)}]{Pohl2009}%
  \BibitemOpen
  \bibfield  {author} {\bibinfo {author} {\bibfnamefont {T.}~\bibnamefont
  {Pohl}}\ and\ \bibinfo {author} {\bibfnamefont {P.~R.}\ \bibnamefont
  {Berman}},\ }\href {\doibase 10.1103/PhysRevLett.102.013004} {\bibfield
  {journal} {\bibinfo  {journal} {Phys. Rev. Lett.}\ }\textbf {\bibinfo
  {volume} {102}},\ \bibinfo {pages} {013004} (\bibinfo {year}
  {2009})}\BibitemShut {NoStop}%
\bibitem [{\citenamefont {Nipper}\ \emph {et~al.}(2012)\citenamefont {Nipper},
  \citenamefont {Balewski}, \citenamefont {Krupp}, \citenamefont {Butscher},
  \citenamefont {L\"ow},\ and\ \citenamefont {Pfau}}]{Nipper2012}%
  \BibitemOpen
  \bibfield  {author} {\bibinfo {author} {\bibfnamefont {J.}~\bibnamefont
  {Nipper}}, \bibinfo {author} {\bibfnamefont {J.~B.}\ \bibnamefont
  {Balewski}}, \bibinfo {author} {\bibfnamefont {A.~T.}\ \bibnamefont {Krupp}},
  \bibinfo {author} {\bibfnamefont {B.}~\bibnamefont {Butscher}}, \bibinfo
  {author} {\bibfnamefont {R.}~\bibnamefont {L\"ow}}, \ and\ \bibinfo {author}
  {\bibfnamefont {T.}~\bibnamefont {Pfau}},\ }\href {\doibase
  10.1103/PhysRevLett.108.113001} {\bibfield  {journal} {\bibinfo  {journal}
  {Phys. Rev. Lett.}\ }\textbf {\bibinfo {volume} {108}},\ \bibinfo {pages}
  {113001} (\bibinfo {year} {2012})}\BibitemShut {NoStop}%
\bibitem [{\citenamefont {Ravets}\ \emph {et~al.}(2014)\citenamefont {Ravets},
  \citenamefont {Labuhn}, \citenamefont {Barredo}, \citenamefont {B{\'e}guin},
  \citenamefont {Lahaye},\ and\ \citenamefont {Browaeys}}]{Ravets2014}%
  \BibitemOpen
  \bibfield  {author} {\bibinfo {author} {\bibfnamefont {S.}~\bibnamefont
  {Ravets}}, \bibinfo {author} {\bibfnamefont {H.}~\bibnamefont {Labuhn}},
  \bibinfo {author} {\bibfnamefont {D.}~\bibnamefont {Barredo}}, \bibinfo
  {author} {\bibfnamefont {L.}~\bibnamefont {B{\'e}guin}}, \bibinfo {author}
  {\bibfnamefont {T.}~\bibnamefont {Lahaye}}, \ and\ \bibinfo {author}
  {\bibfnamefont {A.}~\bibnamefont {Browaeys}},\ }\href {\doibase
  10.1038/nphys3119} {\bibfield  {journal} {\bibinfo  {journal} {Nat. Phys.}\
  }\textbf {\bibinfo {volume} {10}},\ \bibinfo {pages} {914} (\bibinfo {year}
  {2014})}\BibitemShut {NoStop}%
\bibitem [{\citenamefont {Ravets}\ \emph {et~al.}(2015)\citenamefont {Ravets},
  \citenamefont {Labuhn}, \citenamefont {Barredo}, \citenamefont {Lahaye},\
  and\ \citenamefont {Browaeys}}]{Ravets2015}%
  \BibitemOpen
  \bibfield  {author} {\bibinfo {author} {\bibfnamefont {S.}~\bibnamefont
  {Ravets}}, \bibinfo {author} {\bibfnamefont {H.}~\bibnamefont {Labuhn}},
  \bibinfo {author} {\bibfnamefont {D.}~\bibnamefont {Barredo}}, \bibinfo
  {author} {\bibfnamefont {T.}~\bibnamefont {Lahaye}}, \ and\ \bibinfo {author}
  {\bibfnamefont {A.}~\bibnamefont {Browaeys}},\ }\href {\doibase
  10.1103/PhysRevA.92.020701} {\bibfield  {journal} {\bibinfo  {journal} {Phys.
  Rev. A}\ }\textbf {\bibinfo {volume} {92}},\ \bibinfo {pages} {020701(R)}
  (\bibinfo {year} {2015})}\BibitemShut {NoStop}%
\bibitem [{\citenamefont {Bohlouli-Zanjani}\ \emph {et~al.}(2007)\citenamefont
  {Bohlouli-Zanjani}, \citenamefont {Petrus},\ and\ \citenamefont
  {Martin}}]{Bohlouli2007}%
  \BibitemOpen
  \bibfield  {author} {\bibinfo {author} {\bibfnamefont {P.}~\bibnamefont
  {Bohlouli-Zanjani}}, \bibinfo {author} {\bibfnamefont {J.~A.}\ \bibnamefont
  {Petrus}}, \ and\ \bibinfo {author} {\bibfnamefont {J.~D.~D.}\ \bibnamefont
  {Martin}},\ }\href {\doibase 10.1103/PhysRevLett.98.203005} {\bibfield
  {journal} {\bibinfo  {journal} {Phys. Rev. Lett.}\ }\textbf {\bibinfo
  {volume} {98}},\ \bibinfo {pages} {203005} (\bibinfo {year}
  {2007})}\BibitemShut {NoStop}%
\bibitem [{\citenamefont {Yakshina}\ \emph {et~al.}(2016)\citenamefont
  {Yakshina}, \citenamefont {Tretyakov}, \citenamefont {Beterov}, \citenamefont
  {Entin}, \citenamefont {Andreeva}, \citenamefont {Cinins}, \citenamefont
  {Markovski}, \citenamefont {Iftikhar}, \citenamefont {Ekers},\ and\
  \citenamefont {Ryabtsev}}]{Yakshina2016}%
  \BibitemOpen
  \bibfield  {author} {\bibinfo {author} {\bibfnamefont {E.~A.}\ \bibnamefont
  {Yakshina}}, \bibinfo {author} {\bibfnamefont {D.~B.}\ \bibnamefont
  {Tretyakov}}, \bibinfo {author} {\bibfnamefont {I.~I.}\ \bibnamefont
  {Beterov}}, \bibinfo {author} {\bibfnamefont {V.~M.}\ \bibnamefont {Entin}},
  \bibinfo {author} {\bibfnamefont {C.}~\bibnamefont {Andreeva}}, \bibinfo
  {author} {\bibfnamefont {A.}~\bibnamefont {Cinins}}, \bibinfo {author}
  {\bibfnamefont {A.}~\bibnamefont {Markovski}}, \bibinfo {author}
  {\bibfnamefont {Z.}~\bibnamefont {Iftikhar}}, \bibinfo {author}
  {\bibfnamefont {A.}~\bibnamefont {Ekers}}, \ and\ \bibinfo {author}
  {\bibfnamefont {I.~I.}\ \bibnamefont {Ryabtsev}},\ }\href {\doibase
  10.1103/PhysRevA.94.043417} {\bibfield  {journal} {\bibinfo  {journal} {Phys.
  Rev. A}\ }\textbf {\bibinfo {volume} {94}},\ \bibinfo {pages} {043417}
  (\bibinfo {year} {2016})}\BibitemShut {NoStop}%
\bibitem [{\citenamefont {Liu}\ \emph {et~al.}(2020)\citenamefont {Liu},
  \citenamefont {Inman}, \citenamefont {Carroll},\ and\ \citenamefont
  {Noel}}]{Liu2020}%
  \BibitemOpen
  \bibfield  {author} {\bibinfo {author} {\bibfnamefont {Z.~C.}\ \bibnamefont
  {Liu}}, \bibinfo {author} {\bibfnamefont {N.~P.}\ \bibnamefont {Inman}},
  \bibinfo {author} {\bibfnamefont {T.~J.}\ \bibnamefont {Carroll}}, \ and\
  \bibinfo {author} {\bibfnamefont {M.~W.}\ \bibnamefont {Noel}},\ }\href
  {\doibase 10.1103/PhysRevLett.124.133402} {\bibfield  {journal} {\bibinfo
  {journal} {Phys. Rev. Lett.}\ }\textbf {\bibinfo {volume} {124}},\ \bibinfo
  {pages} {133402} (\bibinfo {year} {2020})}\BibitemShut {NoStop}%
\bibitem [{\citenamefont {Browaeys}\ \emph {et~al.}(2016)\citenamefont
  {Browaeys}, \citenamefont {Barredo},\ and\ \citenamefont
  {Lahaye}}]{Browaeys_2016}%
  \BibitemOpen
  \bibfield  {author} {\bibinfo {author} {\bibfnamefont {A.}~\bibnamefont
  {Browaeys}}, \bibinfo {author} {\bibfnamefont {D.}~\bibnamefont {Barredo}}, \
  and\ \bibinfo {author} {\bibfnamefont {T.}~\bibnamefont {Lahaye}},\ }\href
  {\doibase 10.1088/0953-4075/49/15/152001} {\bibfield  {journal} {\bibinfo
  {journal} {J. Phys. B: Atom., Mol. Opt. Phys.}\ }\textbf {\bibinfo {volume}
  {49}},\ \bibinfo {pages} {152001} (\bibinfo {year} {2016})}\BibitemShut
  {NoStop}%
\bibitem [{\citenamefont {Ates}\ \emph {et~al.}(2008)\citenamefont {Ates},
  \citenamefont {Eisfeld},\ and\ \citenamefont {Rost}}]{Ates_2008}%
  \BibitemOpen
  \bibfield  {author} {\bibinfo {author} {\bibfnamefont {C.}~\bibnamefont
  {Ates}}, \bibinfo {author} {\bibfnamefont {A.}~\bibnamefont {Eisfeld}}, \
  and\ \bibinfo {author} {\bibfnamefont {J.~M.}\ \bibnamefont {Rost}},\ }\href
  {\doibase 10.1088/1367-2630/10/4/045030} {\bibfield  {journal} {\bibinfo
  {journal} {New J. Phys.}\ }\textbf {\bibinfo {volume} {10}},\ \bibinfo
  {pages} {045030} (\bibinfo {year} {2008})}\BibitemShut {NoStop}%
\bibitem [{\citenamefont {Barredo}\ \emph {et~al.}(2015)\citenamefont
  {Barredo}, \citenamefont {Labuhn}, \citenamefont {Ravets}, \citenamefont
  {Lahaye}, \citenamefont {Browaeys},\ and\ \citenamefont
  {Adams}}]{Barredo2015}%
  \BibitemOpen
  \bibfield  {author} {\bibinfo {author} {\bibfnamefont {D.}~\bibnamefont
  {Barredo}}, \bibinfo {author} {\bibfnamefont {H.}~\bibnamefont {Labuhn}},
  \bibinfo {author} {\bibfnamefont {S.}~\bibnamefont {Ravets}}, \bibinfo
  {author} {\bibfnamefont {T.}~\bibnamefont {Lahaye}}, \bibinfo {author}
  {\bibfnamefont {A.}~\bibnamefont {Browaeys}}, \ and\ \bibinfo {author}
  {\bibfnamefont {C.~S.}\ \bibnamefont {Adams}},\ }\href {\doibase
  10.1103/PhysRevLett.114.113002} {\bibfield  {journal} {\bibinfo  {journal}
  {Phys. Rev. Lett.}\ }\textbf {\bibinfo {volume} {114}},\ \bibinfo {pages}
  {113002} (\bibinfo {year} {2015})}\BibitemShut {NoStop}%
\bibitem [{\citenamefont {Young}\ \emph {et~al.}(2020)\citenamefont {Young},
  \citenamefont {Bienias}, \citenamefont {Belyansky}, \citenamefont {Kaufman},\
  and\ \citenamefont {Gorshkov}}]{young2020asymmetric}%
  \BibitemOpen
  \bibfield  {author} {\bibinfo {author} {\bibfnamefont {J.~T.}\ \bibnamefont
  {Young}}, \bibinfo {author} {\bibfnamefont {P.}~\bibnamefont {Bienias}},
  \bibinfo {author} {\bibfnamefont {R.}~\bibnamefont {Belyansky}}, \bibinfo
  {author} {\bibfnamefont {A.~M.}\ \bibnamefont {Kaufman}}, \ and\ \bibinfo
  {author} {\bibfnamefont {A.~V.}\ \bibnamefont {Gorshkov}},\ }\href@noop {}
  {\enquote {\bibinfo {title} {Asymmetric blockade and multi-qubit gates via
  dipole-dipole interactions},}\ } (\bibinfo {year} {2020}),\ \Eprint
  {http://arxiv.org/abs/2006.02486} {arXiv:2006.02486 [quant-ph]} \BibitemShut
  {NoStop}%
\bibitem [{\citenamefont {Anderson}\ \emph {et~al.}(1998)\citenamefont
  {Anderson}, \citenamefont {Veale},\ and\ \citenamefont
  {Gallagher}}]{Gallagher1998}%
  \BibitemOpen
  \bibfield  {author} {\bibinfo {author} {\bibfnamefont {W.~R.}\ \bibnamefont
  {Anderson}}, \bibinfo {author} {\bibfnamefont {J.~R.}\ \bibnamefont {Veale}},
  \ and\ \bibinfo {author} {\bibfnamefont {T.~F.}\ \bibnamefont {Gallagher}},\
  }\href {\doibase 10.1103/PhysRevLett.80.249} {\bibfield  {journal} {\bibinfo
  {journal} {Phys. Rev. Lett.}\ }\textbf {\bibinfo {volume} {80}},\ \bibinfo
  {pages} {249} (\bibinfo {year} {1998})}\BibitemShut {NoStop}%
\bibitem [{\citenamefont {Gorniaczyk}\ \emph {et~al.}(2016)\citenamefont
  {Gorniaczyk}, \citenamefont {Tresp}, \citenamefont {Bienias}, \citenamefont
  {Paris-Mandoki}, \citenamefont {Li}, \citenamefont {Mirgorodskiy},
  \citenamefont {B{\"u}chler}, \citenamefont {Lesanovsky},\ and\ \citenamefont
  {Hofferberth}}]{Gorniaczyk2016}%
  \BibitemOpen
  \bibfield  {author} {\bibinfo {author} {\bibfnamefont {H.}~\bibnamefont
  {Gorniaczyk}}, \bibinfo {author} {\bibfnamefont {C.}~\bibnamefont {Tresp}},
  \bibinfo {author} {\bibfnamefont {P.}~\bibnamefont {Bienias}}, \bibinfo
  {author} {\bibfnamefont {A.}~\bibnamefont {Paris-Mandoki}}, \bibinfo {author}
  {\bibfnamefont {W.}~\bibnamefont {Li}}, \bibinfo {author} {\bibfnamefont
  {I.}~\bibnamefont {Mirgorodskiy}}, \bibinfo {author} {\bibfnamefont {H.~P.}\
  \bibnamefont {B{\"u}chler}}, \bibinfo {author} {\bibfnamefont
  {I.}~\bibnamefont {Lesanovsky}}, \ and\ \bibinfo {author} {\bibfnamefont
  {S.}~\bibnamefont {Hofferberth}},\ }\href {\doibase 10.1038/ncomms12480}
  {\bibfield  {journal} {\bibinfo  {journal} {Nat. Commu.}\ }\textbf {\bibinfo
  {volume} {7}},\ \bibinfo {pages} {12480} (\bibinfo {year}
  {2016})}\BibitemShut {NoStop}%
\bibitem [{\citenamefont {Petrosyan}\ \emph {et~al.}(2017)\citenamefont
  {Petrosyan}, \citenamefont {Motzoi}, \citenamefont {Saffman},\ and\
  \citenamefont {M\o{}lmer}}]{Petrosyan2017}%
  \BibitemOpen
  \bibfield  {author} {\bibinfo {author} {\bibfnamefont {D.}~\bibnamefont
  {Petrosyan}}, \bibinfo {author} {\bibfnamefont {F.}~\bibnamefont {Motzoi}},
  \bibinfo {author} {\bibfnamefont {M.}~\bibnamefont {Saffman}}, \ and\
  \bibinfo {author} {\bibfnamefont {K.}~\bibnamefont {M\o{}lmer}},\ }\href
  {\doibase 10.1103/PhysRevA.96.042306} {\bibfield  {journal} {\bibinfo
  {journal} {Phys. Rev. A}\ }\textbf {\bibinfo {volume} {96}},\ \bibinfo
  {pages} {042306} (\bibinfo {year} {2017})}\BibitemShut {NoStop}%
\bibitem [{\citenamefont {Beterov}\ \emph
  {et~al.}(2018{\natexlab{b}})\citenamefont {Beterov}, \citenamefont
  {Khamzina}, \citenamefont {Tret’yakov}, \citenamefont {Entin},
  \citenamefont {Yakshina},\ and\ \citenamefont
  {Ryabtsev}}]{beterov2018resonant}%
  \BibitemOpen
  \bibfield  {author} {\bibinfo {author} {\bibfnamefont {I.~I.}\ \bibnamefont
  {Beterov}}, \bibinfo {author} {\bibfnamefont {G.~N.}\ \bibnamefont
  {Khamzina}}, \bibinfo {author} {\bibfnamefont {D.~B.}\ \bibnamefont
  {Tret’yakov}}, \bibinfo {author} {\bibfnamefont {V.~M.}\ \bibnamefont
  {Entin}}, \bibinfo {author} {\bibfnamefont {E.~A.}\ \bibnamefont {Yakshina}},
  \ and\ \bibinfo {author} {\bibfnamefont {I.~I.}\ \bibnamefont {Ryabtsev}},\
  }\href {\doibase https://doi.org/10.1070/QEL16663} {\bibfield  {journal}
  {\bibinfo  {journal} {Quant. Electronics}\ }\textbf {\bibinfo {volume}
  {48}},\ \bibinfo {pages} {453} (\bibinfo {year}
  {2018}{\natexlab{b}})}\BibitemShut {NoStop}%
\bibitem [{\citenamefont {Singer}\ \emph {et~al.}(2005)\citenamefont {Singer},
  \citenamefont {Stanojevic}, \citenamefont {Weidemüller},\ and\ \citenamefont
  {C{\^{o}}t{\'{e}}}}]{Singer_2005}%
  \BibitemOpen
  \bibfield  {author} {\bibinfo {author} {\bibfnamefont {K.}~\bibnamefont
  {Singer}}, \bibinfo {author} {\bibfnamefont {J.}~\bibnamefont {Stanojevic}},
  \bibinfo {author} {\bibfnamefont {M.}~\bibnamefont {Weidemüller}}, \ and\
  \bibinfo {author} {\bibfnamefont {R.}~\bibnamefont {C{\^{o}}t{\'{e}}}},\
  }\href {\doibase 10.1088/0953-4075/38/2/021} {\bibfield  {journal} {\bibinfo
  {journal} {Journal of Physics B: Atomic, Molecular and Optical Physics}\
  }\textbf {\bibinfo {volume} {38}},\ \bibinfo {pages} {S295} (\bibinfo {year}
  {2005})}\BibitemShut {NoStop}%
\bibitem [{\citenamefont {Olmos}\ \emph {et~al.}(2011)\citenamefont {Olmos},
  \citenamefont {Li}, \citenamefont {Hofferberth},\ and\ \citenamefont
  {Lesanovsky}}]{PhysRevA.84.041607}%
  \BibitemOpen
  \bibfield  {author} {\bibinfo {author} {\bibfnamefont {B.}~\bibnamefont
  {Olmos}}, \bibinfo {author} {\bibfnamefont {W.}~\bibnamefont {Li}}, \bibinfo
  {author} {\bibfnamefont {S.}~\bibnamefont {Hofferberth}}, \ and\ \bibinfo
  {author} {\bibfnamefont {I.}~\bibnamefont {Lesanovsky}},\ }\href {\doibase
  10.1103/PhysRevA.84.041607} {\bibfield  {journal} {\bibinfo  {journal} {Phys.
  Rev. A}\ }\textbf {\bibinfo {volume} {84}},\ \bibinfo {pages} {041607}
  (\bibinfo {year} {2011})}\BibitemShut {NoStop}%
\bibitem [{\citenamefont {Šibalić}\ \emph {et~al.}(2017)\citenamefont
  {Šibalić}, \citenamefont {Pritchard}, \citenamefont {Adams},\ and\
  \citenamefont {Weatherill}}]{SIBALIC2017319}%
  \BibitemOpen
  \bibfield  {author} {\bibinfo {author} {\bibfnamefont {N.}~\bibnamefont
  {Šibalić}}, \bibinfo {author} {\bibfnamefont {J.}~\bibnamefont
  {Pritchard}}, \bibinfo {author} {\bibfnamefont {C.}~\bibnamefont {Adams}}, \
  and\ \bibinfo {author} {\bibfnamefont {K.}~\bibnamefont {Weatherill}},\
  }\href {\doibase https://doi.org/10.1016/j.cpc.2017.06.015} {\bibfield
  {journal} {\bibinfo  {journal} {Computer Physics Communications}\ }\textbf
  {\bibinfo {volume} {220}},\ \bibinfo {pages} {319} (\bibinfo {year}
  {2017})}\BibitemShut {NoStop}%
\bibitem [{\citenamefont {Pupillo}\ \emph {et~al.}(2010)\citenamefont
  {Pupillo}, \citenamefont {Micheli}, \citenamefont {Boninsegni}, \citenamefont
  {Lesanovsky},\ and\ \citenamefont {Zoller}}]{PhysRevLett.104.223002}%
  \BibitemOpen
  \bibfield  {author} {\bibinfo {author} {\bibfnamefont {G.}~\bibnamefont
  {Pupillo}}, \bibinfo {author} {\bibfnamefont {A.}~\bibnamefont {Micheli}},
  \bibinfo {author} {\bibfnamefont {M.}~\bibnamefont {Boninsegni}}, \bibinfo
  {author} {\bibfnamefont {I.}~\bibnamefont {Lesanovsky}}, \ and\ \bibinfo
  {author} {\bibfnamefont {P.}~\bibnamefont {Zoller}},\ }\href {\doibase
  10.1103/PhysRevLett.104.223002} {\bibfield  {journal} {\bibinfo  {journal}
  {Phys. Rev. Lett.}\ }\textbf {\bibinfo {volume} {104}},\ \bibinfo {pages}
  {223002} (\bibinfo {year} {2010})}\BibitemShut {NoStop}%
\bibitem [{\citenamefont {Jau}\ \emph {et~al.}(2016)\citenamefont {Jau},
  \citenamefont {Hankin}, \citenamefont {Keating}, \citenamefont {Deutsch},\
  and\ \citenamefont {Biedermann}}]{Jau2016}%
  \BibitemOpen
  \bibfield  {author} {\bibinfo {author} {\bibfnamefont {Y.-Y.}\ \bibnamefont
  {Jau}}, \bibinfo {author} {\bibfnamefont {A.~M.}\ \bibnamefont {Hankin}},
  \bibinfo {author} {\bibfnamefont {T.}~\bibnamefont {Keating}}, \bibinfo
  {author} {\bibfnamefont {I.~H.}\ \bibnamefont {Deutsch}}, \ and\ \bibinfo
  {author} {\bibfnamefont {G.~W.}\ \bibnamefont {Biedermann}},\ }\href
  {\doibase 10.1038/nphys3487} {\bibfield  {journal} {\bibinfo  {journal}
  {Nature Physics}\ }\textbf {\bibinfo {volume} {12}},\ \bibinfo {pages} {71}
  (\bibinfo {year} {2016})}\BibitemShut {NoStop}%
\bibitem [{\citenamefont {Zeiher}\ \emph {et~al.}(2016)\citenamefont {Zeiher},
  \citenamefont {van Bijnen}, \citenamefont {Schau{\ss}}, \citenamefont {Hild},
  \citenamefont {Choi}, \citenamefont {Pohl}, \citenamefont {Bloch},\ and\
  \citenamefont {Gross}}]{Zeiher2016}%
  \BibitemOpen
  \bibfield  {author} {\bibinfo {author} {\bibfnamefont {J.}~\bibnamefont
  {Zeiher}}, \bibinfo {author} {\bibfnamefont {R.}~\bibnamefont {van Bijnen}},
  \bibinfo {author} {\bibfnamefont {P.}~\bibnamefont {Schau{\ss}}}, \bibinfo
  {author} {\bibfnamefont {S.}~\bibnamefont {Hild}}, \bibinfo {author}
  {\bibfnamefont {J.-y.}\ \bibnamefont {Choi}}, \bibinfo {author}
  {\bibfnamefont {T.}~\bibnamefont {Pohl}}, \bibinfo {author} {\bibfnamefont
  {I.}~\bibnamefont {Bloch}}, \ and\ \bibinfo {author} {\bibfnamefont
  {C.}~\bibnamefont {Gross}},\ }\href {\doibase 10.1038/nphys3835} {\bibfield
  {journal} {\bibinfo  {journal} {Nature Physics}\ }\textbf {\bibinfo {volume}
  {12}},\ \bibinfo {pages} {1095} (\bibinfo {year} {2016})}\BibitemShut
  {NoStop}%
\bibitem [{\citenamefont {Honer}\ \emph {et~al.}(2010)\citenamefont {Honer},
  \citenamefont {Weimer}, \citenamefont {Pfau},\ and\ \citenamefont
  {B\"uchler}}]{PhysRevLett.105.160404}%
  \BibitemOpen
  \bibfield  {author} {\bibinfo {author} {\bibfnamefont {J.}~\bibnamefont
  {Honer}}, \bibinfo {author} {\bibfnamefont {H.}~\bibnamefont {Weimer}},
  \bibinfo {author} {\bibfnamefont {T.}~\bibnamefont {Pfau}}, \ and\ \bibinfo
  {author} {\bibfnamefont {H.~P.}\ \bibnamefont {B\"uchler}},\ }\href {\doibase
  10.1103/PhysRevLett.105.160404} {\bibfield  {journal} {\bibinfo  {journal}
  {Phys. Rev. Lett.}\ }\textbf {\bibinfo {volume} {105}},\ \bibinfo {pages}
  {160404} (\bibinfo {year} {2010})}\BibitemShut {NoStop}%
\bibitem [{\citenamefont {Glaetzle}\ \emph {et~al.}(2015)\citenamefont
  {Glaetzle}, \citenamefont {Dalmonte}, \citenamefont {Nath}, \citenamefont
  {Gross}, \citenamefont {Bloch},\ and\ \citenamefont
  {Zoller}}]{PhysRevLett.114.173002}%
  \BibitemOpen
  \bibfield  {author} {\bibinfo {author} {\bibfnamefont {A.~W.}\ \bibnamefont
  {Glaetzle}}, \bibinfo {author} {\bibfnamefont {M.}~\bibnamefont {Dalmonte}},
  \bibinfo {author} {\bibfnamefont {R.}~\bibnamefont {Nath}}, \bibinfo {author}
  {\bibfnamefont {C.}~\bibnamefont {Gross}}, \bibinfo {author} {\bibfnamefont
  {I.}~\bibnamefont {Bloch}}, \ and\ \bibinfo {author} {\bibfnamefont
  {P.}~\bibnamefont {Zoller}},\ }\href {\doibase
  10.1103/PhysRevLett.114.173002} {\bibfield  {journal} {\bibinfo  {journal}
  {Phys. Rev. Lett.}\ }\textbf {\bibinfo {volume} {114}},\ \bibinfo {pages}
  {173002} (\bibinfo {year} {2015})}\BibitemShut {NoStop}%
\bibitem [{\citenamefont {Johnson}\ and\ \citenamefont
  {Rolston}(2010)}]{PhysRevA.82.033412}%
  \BibitemOpen
  \bibfield  {author} {\bibinfo {author} {\bibfnamefont {J.~E.}\ \bibnamefont
  {Johnson}}\ and\ \bibinfo {author} {\bibfnamefont {S.~L.}\ \bibnamefont
  {Rolston}},\ }\href {\doibase 10.1103/PhysRevA.82.033412} {\bibfield
  {journal} {\bibinfo  {journal} {Phys. Rev. A}\ }\textbf {\bibinfo {volume}
  {82}},\ \bibinfo {pages} {033412} (\bibinfo {year} {2010})}\BibitemShut
  {NoStop}%
\bibitem [{\citenamefont {van Bijnen}\ and\ \citenamefont
  {Pohl}(2015)}]{PhysRevLett.114.243002}%
  \BibitemOpen
  \bibfield  {author} {\bibinfo {author} {\bibfnamefont {R.~M.~W.}\
  \bibnamefont {van Bijnen}}\ and\ \bibinfo {author} {\bibfnamefont
  {T.}~\bibnamefont {Pohl}},\ }\href {\doibase 10.1103/PhysRevLett.114.243002}
  {\bibfield  {journal} {\bibinfo  {journal} {Phys. Rev. Lett.}\ }\textbf
  {\bibinfo {volume} {114}},\ \bibinfo {pages} {243002} (\bibinfo {year}
  {2015})}\BibitemShut {NoStop}%
\bibitem [{\citenamefont {Balewski}\ \emph {et~al.}(2014)\citenamefont
  {Balewski}, \citenamefont {Krupp}, \citenamefont {Gaj}, \citenamefont
  {Hofferberth}, \citenamefont {Löw},\ and\ \citenamefont
  {Pfau}}]{Balewski_2014}%
  \BibitemOpen
  \bibfield  {author} {\bibinfo {author} {\bibfnamefont {J.~B.}\ \bibnamefont
  {Balewski}}, \bibinfo {author} {\bibfnamefont {A.~T.}\ \bibnamefont {Krupp}},
  \bibinfo {author} {\bibfnamefont {A.}~\bibnamefont {Gaj}}, \bibinfo {author}
  {\bibfnamefont {S.}~\bibnamefont {Hofferberth}}, \bibinfo {author}
  {\bibfnamefont {R.}~\bibnamefont {Löw}}, \ and\ \bibinfo {author}
  {\bibfnamefont {T.}~\bibnamefont {Pfau}},\ }\href {\doibase
  10.1088/1367-2630/16/6/063012} {\bibfield  {journal} {\bibinfo  {journal}
  {New Journal of Physics}\ }\textbf {\bibinfo {volume} {16}},\ \bibinfo
  {pages} {063012} (\bibinfo {year} {2014})}\BibitemShut {NoStop}%
\bibitem [{\citenamefont {Henkel}\ \emph {et~al.}(2012)\citenamefont {Henkel},
  \citenamefont {Cinti}, \citenamefont {Jain}, \citenamefont {Pupillo},\ and\
  \citenamefont {Pohl}}]{PhysRevLett.108.265301}%
  \BibitemOpen
  \bibfield  {author} {\bibinfo {author} {\bibfnamefont {N.}~\bibnamefont
  {Henkel}}, \bibinfo {author} {\bibfnamefont {F.}~\bibnamefont {Cinti}},
  \bibinfo {author} {\bibfnamefont {P.}~\bibnamefont {Jain}}, \bibinfo {author}
  {\bibfnamefont {G.}~\bibnamefont {Pupillo}}, \ and\ \bibinfo {author}
  {\bibfnamefont {T.}~\bibnamefont {Pohl}},\ }\href {\doibase
  10.1103/PhysRevLett.108.265301} {\bibfield  {journal} {\bibinfo  {journal}
  {Phys. Rev. Lett.}\ }\textbf {\bibinfo {volume} {108}},\ \bibinfo {pages}
  {265301} (\bibinfo {year} {2012})}\BibitemShut {NoStop}%
\bibitem [{\citenamefont {Schempp}\ \emph {et~al.}(2015)\citenamefont
  {Schempp}, \citenamefont {G\"unter}, \citenamefont {W\"uster}, \citenamefont
  {Weidem\"uller},\ and\ \citenamefont {Whitlock}}]{PhysRevLett.115.093002}%
  \BibitemOpen
  \bibfield  {author} {\bibinfo {author} {\bibfnamefont {H.}~\bibnamefont
  {Schempp}}, \bibinfo {author} {\bibfnamefont {G.}~\bibnamefont {G\"unter}},
  \bibinfo {author} {\bibfnamefont {S.}~\bibnamefont {W\"uster}}, \bibinfo
  {author} {\bibfnamefont {M.}~\bibnamefont {Weidem\"uller}}, \ and\ \bibinfo
  {author} {\bibfnamefont {S.}~\bibnamefont {Whitlock}},\ }\href {\doibase
  10.1103/PhysRevLett.115.093002} {\bibfield  {journal} {\bibinfo  {journal}
  {Phys. Rev. Lett.}\ }\textbf {\bibinfo {volume} {115}},\ \bibinfo {pages}
  {093002} (\bibinfo {year} {2015})}\BibitemShut {NoStop}%
\bibitem [{\citenamefont {Tanasittikosol}\ \emph {et~al.}(2011)\citenamefont
  {Tanasittikosol}, \citenamefont {Pritchard}, \citenamefont {Maxwell},
  \citenamefont {Gauguet}, \citenamefont {Weatherill}, \citenamefont
  {Potvliege},\ and\ \citenamefont {Adams}}]{Tanasittikosol_2011}%
  \BibitemOpen
  \bibfield  {author} {\bibinfo {author} {\bibfnamefont {M.}~\bibnamefont
  {Tanasittikosol}}, \bibinfo {author} {\bibfnamefont {J.~D.}\ \bibnamefont
  {Pritchard}}, \bibinfo {author} {\bibfnamefont {D.}~\bibnamefont {Maxwell}},
  \bibinfo {author} {\bibfnamefont {A.}~\bibnamefont {Gauguet}}, \bibinfo
  {author} {\bibfnamefont {K.~J.}\ \bibnamefont {Weatherill}}, \bibinfo
  {author} {\bibfnamefont {R.~M.}\ \bibnamefont {Potvliege}}, \ and\ \bibinfo
  {author} {\bibfnamefont {C.~S.}\ \bibnamefont {Adams}},\ }\href {\doibase
  10.1088/0953-4075/44/18/184020} {\bibfield  {journal} {\bibinfo  {journal}
  {Journal of Physics B: Atomic, Molecular and Optical Physics}\ }\textbf
  {\bibinfo {volume} {44}},\ \bibinfo {pages} {184020} (\bibinfo {year}
  {2011})}\BibitemShut {NoStop}%
\bibitem [{\citenamefont {Macr\`{\i}}\ and\ \citenamefont
  {Pohl}(2014)}]{PhysRevA.89.011402}%
  \BibitemOpen
  \bibfield  {author} {\bibinfo {author} {\bibfnamefont {T.}~\bibnamefont
  {Macr\`{\i}}}\ and\ \bibinfo {author} {\bibfnamefont {T.}~\bibnamefont
  {Pohl}},\ }\href {\doibase 10.1103/PhysRevA.89.011402} {\bibfield  {journal}
  {\bibinfo  {journal} {Phys. Rev. A}\ }\textbf {\bibinfo {volume} {89}},\
  \bibinfo {pages} {011402} (\bibinfo {year} {2014})}\BibitemShut {NoStop}%
\bibitem [{\citenamefont {Keating}\ \emph {et~al.}(2013)\citenamefont
  {Keating}, \citenamefont {Goyal}, \citenamefont {Jau}, \citenamefont
  {Biedermann}, \citenamefont {Landahl},\ and\ \citenamefont
  {Deutsch}}]{PhysRevA.87.052314}%
  \BibitemOpen
  \bibfield  {author} {\bibinfo {author} {\bibfnamefont {T.}~\bibnamefont
  {Keating}}, \bibinfo {author} {\bibfnamefont {K.}~\bibnamefont {Goyal}},
  \bibinfo {author} {\bibfnamefont {Y.-Y.}\ \bibnamefont {Jau}}, \bibinfo
  {author} {\bibfnamefont {G.~W.}\ \bibnamefont {Biedermann}}, \bibinfo
  {author} {\bibfnamefont {A.~J.}\ \bibnamefont {Landahl}}, \ and\ \bibinfo
  {author} {\bibfnamefont {I.~H.}\ \bibnamefont {Deutsch}},\ }\href {\doibase
  10.1103/PhysRevA.87.052314} {\bibfield  {journal} {\bibinfo  {journal} {Phys.
  Rev. A}\ }\textbf {\bibinfo {volume} {87}},\ \bibinfo {pages} {052314}
  (\bibinfo {year} {2013})}\BibitemShut {NoStop}%
\bibitem [{\citenamefont {Wüster}\ \emph {et~al.}(2011)\citenamefont
  {Wüster}, \citenamefont {Ates}, \citenamefont {Eisfeld},\ and\ \citenamefont
  {Rost}}]{W_ster_2011}%
  \BibitemOpen
  \bibfield  {author} {\bibinfo {author} {\bibfnamefont {S.}~\bibnamefont
  {Wüster}}, \bibinfo {author} {\bibfnamefont {C.}~\bibnamefont {Ates}},
  \bibinfo {author} {\bibfnamefont {A.}~\bibnamefont {Eisfeld}}, \ and\
  \bibinfo {author} {\bibfnamefont {J.~M.}\ \bibnamefont {Rost}},\ }\href
  {\doibase 10.1088/1367-2630/13/7/073044} {\bibfield  {journal} {\bibinfo
  {journal} {New Journal of Physics}\ }\textbf {\bibinfo {volume} {13}},\
  \bibinfo {pages} {073044} (\bibinfo {year} {2011})}\BibitemShut {NoStop}%
\bibitem [{\citenamefont {Keating}\ \emph {et~al.}(2015)\citenamefont
  {Keating}, \citenamefont {Cook}, \citenamefont {Hankin}, \citenamefont {Jau},
  \citenamefont {Biedermann},\ and\ \citenamefont
  {Deutsch}}]{PhysRevA.91.012337}%
  \BibitemOpen
  \bibfield  {author} {\bibinfo {author} {\bibfnamefont {T.}~\bibnamefont
  {Keating}}, \bibinfo {author} {\bibfnamefont {R.~L.}\ \bibnamefont {Cook}},
  \bibinfo {author} {\bibfnamefont {A.~M.}\ \bibnamefont {Hankin}}, \bibinfo
  {author} {\bibfnamefont {Y.-Y.}\ \bibnamefont {Jau}}, \bibinfo {author}
  {\bibfnamefont {G.~W.}\ \bibnamefont {Biedermann}}, \ and\ \bibinfo {author}
  {\bibfnamefont {I.~H.}\ \bibnamefont {Deutsch}},\ }\href {\doibase
  10.1103/PhysRevA.91.012337} {\bibfield  {journal} {\bibinfo  {journal} {Phys.
  Rev. A}\ }\textbf {\bibinfo {volume} {91}},\ \bibinfo {pages} {012337}
  (\bibinfo {year} {2015})}\BibitemShut {NoStop}%
\bibitem [{\citenamefont {Mattioli}\ \emph {et~al.}(2013)\citenamefont
  {Mattioli}, \citenamefont {Dalmonte}, \citenamefont {Lechner},\ and\
  \citenamefont {Pupillo}}]{PhysRevLett.111.165302}%
  \BibitemOpen
  \bibfield  {author} {\bibinfo {author} {\bibfnamefont {M.}~\bibnamefont
  {Mattioli}}, \bibinfo {author} {\bibfnamefont {M.}~\bibnamefont {Dalmonte}},
  \bibinfo {author} {\bibfnamefont {W.}~\bibnamefont {Lechner}}, \ and\
  \bibinfo {author} {\bibfnamefont {G.}~\bibnamefont {Pupillo}},\ }\href
  {\doibase 10.1103/PhysRevLett.111.165302} {\bibfield  {journal} {\bibinfo
  {journal} {Phys. Rev. Lett.}\ }\textbf {\bibinfo {volume} {111}},\ \bibinfo
  {pages} {165302} (\bibinfo {year} {2013})}\BibitemShut {NoStop}%
\bibitem [{\citenamefont {Li}\ \emph {et~al.}(2012)\citenamefont {Li},
  \citenamefont {Hamadeh},\ and\ \citenamefont {Lesanovsky}}]{li_probing_2012}%
  \BibitemOpen
  \bibfield  {author} {\bibinfo {author} {\bibfnamefont {W.}~\bibnamefont
  {Li}}, \bibinfo {author} {\bibfnamefont {L.}~\bibnamefont {Hamadeh}}, \ and\
  \bibinfo {author} {\bibfnamefont {I.}~\bibnamefont {Lesanovsky}},\ }\href
  {\doibase 10.1103/PhysRevA.85.053615} {\bibfield  {journal} {\bibinfo
  {journal} {Phys. Rev. A}\ }\textbf {\bibinfo {volume} {85}},\ \bibinfo
  {pages} {053615} (\bibinfo {year} {2012})}\BibitemShut {NoStop}%
\bibitem [{\citenamefont {Gaul}\ \emph {et~al.}(2016)\citenamefont {Gaul},
  \citenamefont {DeSalvo}, \citenamefont {Aman}, \citenamefont {Dunning},
  \citenamefont {Killian},\ and\ \citenamefont
  {Pohl}}]{PhysRevLett.116.243001}%
  \BibitemOpen
  \bibfield  {author} {\bibinfo {author} {\bibfnamefont {C.}~\bibnamefont
  {Gaul}}, \bibinfo {author} {\bibfnamefont {B.~J.}\ \bibnamefont {DeSalvo}},
  \bibinfo {author} {\bibfnamefont {J.~A.}\ \bibnamefont {Aman}}, \bibinfo
  {author} {\bibfnamefont {F.~B.}\ \bibnamefont {Dunning}}, \bibinfo {author}
  {\bibfnamefont {T.~C.}\ \bibnamefont {Killian}}, \ and\ \bibinfo {author}
  {\bibfnamefont {T.}~\bibnamefont {Pohl}},\ }\href {\doibase
  10.1103/PhysRevLett.116.243001} {\bibfield  {journal} {\bibinfo  {journal}
  {Phys. Rev. Lett.}\ }\textbf {\bibinfo {volume} {116}},\ \bibinfo {pages}
  {243001} (\bibinfo {year} {2016})}\BibitemShut {NoStop}%
\bibitem [{\citenamefont {Petrosyan}\ and\ \citenamefont
  {M\o{}lmer}(2014)}]{PhysRevLett.113.123003}%
  \BibitemOpen
  \bibfield  {author} {\bibinfo {author} {\bibfnamefont {D.}~\bibnamefont
  {Petrosyan}}\ and\ \bibinfo {author} {\bibfnamefont {K.}~\bibnamefont
  {M\o{}lmer}},\ }\href {\doibase 10.1103/PhysRevLett.113.123003} {\bibfield
  {journal} {\bibinfo  {journal} {Phys. Rev. Lett.}\ }\textbf {\bibinfo
  {volume} {113}},\ \bibinfo {pages} {123003} (\bibinfo {year}
  {2014})}\BibitemShut {NoStop}%
\bibitem [{\citenamefont {Mukherjee}\ \emph {et~al.}(2015)\citenamefont
  {Mukherjee}, \citenamefont {Ates}, \citenamefont {Li},\ and\ \citenamefont
  {W{\"u}ster}}]{mukherjee_phase-imprinting_2015}%
  \BibitemOpen
  \bibfield  {author} {\bibinfo {author} {\bibfnamefont {R.}~\bibnamefont
  {Mukherjee}}, \bibinfo {author} {\bibfnamefont {C.}~\bibnamefont {Ates}},
  \bibinfo {author} {\bibfnamefont {W.}~\bibnamefont {Li}}, \ and\ \bibinfo
  {author} {\bibfnamefont {S.}~\bibnamefont {W{\"u}ster}},\ }\href {\doibase
  10.1103/PhysRevLett.115.040401} {\bibfield  {journal} {\bibinfo  {journal}
  {Phys. Rev. Lett.}\ }\textbf {\bibinfo {volume} {115}},\ \bibinfo {pages}
  {040401} (\bibinfo {year} {2015})}\BibitemShut {NoStop}%
\bibitem [{\citenamefont {Buchmann}\ \emph {et~al.}(2017)\citenamefont
  {Buchmann}, \citenamefont {M\o{}lmer},\ and\ \citenamefont
  {Petrosyan}}]{PhysRevA.95.013403}%
  \BibitemOpen
  \bibfield  {author} {\bibinfo {author} {\bibfnamefont {L.~F.}\ \bibnamefont
  {Buchmann}}, \bibinfo {author} {\bibfnamefont {K.}~\bibnamefont {M\o{}lmer}},
  \ and\ \bibinfo {author} {\bibfnamefont {D.}~\bibnamefont {Petrosyan}},\
  }\href {\doibase 10.1103/PhysRevA.95.013403} {\bibfield  {journal} {\bibinfo
  {journal} {Phys. Rev. A}\ }\textbf {\bibinfo {volume} {95}},\ \bibinfo
  {pages} {013403} (\bibinfo {year} {2017})}\BibitemShut {NoStop}%
\bibitem [{\citenamefont {Li}\ \emph {et~al.}(2018)\citenamefont {Li},
  \citenamefont {Gei{\ss}ler}, \citenamefont {Hofstetter},\ and\ \citenamefont
  {Li}}]{li_supersolidity_2018}%
  \BibitemOpen
  \bibfield  {author} {\bibinfo {author} {\bibfnamefont {Y.}~\bibnamefont
  {Li}}, \bibinfo {author} {\bibfnamefont {A.}~\bibnamefont {Gei{\ss}ler}},
  \bibinfo {author} {\bibfnamefont {W.}~\bibnamefont {Hofstetter}}, \ and\
  \bibinfo {author} {\bibfnamefont {W.}~\bibnamefont {Li}},\ }\href {\doibase
  10.1103/PhysRevA.97.023619} {\bibfield  {journal} {\bibinfo  {journal} {Phys.
  Rev. A}\ }\textbf {\bibinfo {volume} {97}},\ \bibinfo {pages} {023619}
  (\bibinfo {year} {2018})}\BibitemShut {NoStop}%
\bibitem [{\citenamefont {Lee}\ \emph {et~al.}(2017)\citenamefont {Lee},
  \citenamefont {Martin}, \citenamefont {Jau}, \citenamefont {Keating},
  \citenamefont {Deutsch},\ and\ \citenamefont
  {Biedermann}}]{PhysRevA.95.041801}%
  \BibitemOpen
  \bibfield  {author} {\bibinfo {author} {\bibfnamefont {J.}~\bibnamefont
  {Lee}}, \bibinfo {author} {\bibfnamefont {M.~J.}\ \bibnamefont {Martin}},
  \bibinfo {author} {\bibfnamefont {Y.-Y.}\ \bibnamefont {Jau}}, \bibinfo
  {author} {\bibfnamefont {T.}~\bibnamefont {Keating}}, \bibinfo {author}
  {\bibfnamefont {I.~H.}\ \bibnamefont {Deutsch}}, \ and\ \bibinfo {author}
  {\bibfnamefont {G.~W.}\ \bibnamefont {Biedermann}},\ }\href {\doibase
  10.1103/PhysRevA.95.041801} {\bibfield  {journal} {\bibinfo  {journal} {Phys.
  Rev. A}\ }\textbf {\bibinfo {volume} {95}},\ \bibinfo {pages} {041801}
  (\bibinfo {year} {2017})}\BibitemShut {NoStop}%
\bibitem [{\citenamefont {Zhou}\ \emph {et~al.}(2020)\citenamefont {Zhou},
  \citenamefont {Li}, \citenamefont {Nath},\ and\ \citenamefont
  {Li}}]{zhou_quench_2020}%
  \BibitemOpen
  \bibfield  {author} {\bibinfo {author} {\bibfnamefont {Y.}~\bibnamefont
  {Zhou}}, \bibinfo {author} {\bibfnamefont {Y.}~\bibnamefont {Li}}, \bibinfo
  {author} {\bibfnamefont {R.}~\bibnamefont {Nath}}, \ and\ \bibinfo {author}
  {\bibfnamefont {W.}~\bibnamefont {Li}},\ }\href {\doibase
  10.1103/PhysRevA.101.013427} {\bibfield  {journal} {\bibinfo  {journal}
  {Phys. Rev. A}\ }\textbf {\bibinfo {volume} {101}},\ \bibinfo {pages}
  {013427} (\bibinfo {year} {2020})}\BibitemShut {NoStop}%
\bibitem [{\citenamefont {McCormack}\ \emph {et~al.}(2020)\citenamefont
  {McCormack}, \citenamefont {Nath},\ and\ \citenamefont
  {Li}}]{mccormack_dynamical_2020}%
  \BibitemOpen
  \bibfield  {author} {\bibinfo {author} {\bibfnamefont {G.}~\bibnamefont
  {McCormack}}, \bibinfo {author} {\bibfnamefont {R.}~\bibnamefont {Nath}}, \
  and\ \bibinfo {author} {\bibfnamefont {W.}~\bibnamefont {Li}},\ }\href
  {\doibase 10.1103/PhysRevA.102.023319} {\bibfield  {journal} {\bibinfo
  {journal} {Phys. Rev. A}\ }\textbf {\bibinfo {volume} {102}},\ \bibinfo
  {pages} {023319} (\bibinfo {year} {2020})}\BibitemShut {NoStop}%
\bibitem [{\citenamefont {Li}\ \emph {et~al.}(2020{\natexlab{b}})\citenamefont
  {Li}, \citenamefont {Cai}, \citenamefont {Wang}, \citenamefont {Li},
  \citenamefont {Yuan},\ and\ \citenamefont {Li}}]{li_many-body_2020}%
  \BibitemOpen
  \bibfield  {author} {\bibinfo {author} {\bibfnamefont {Y.}~\bibnamefont
  {Li}}, \bibinfo {author} {\bibfnamefont {H.}~\bibnamefont {Cai}}, \bibinfo
  {author} {\bibfnamefont {D.-w.}\ \bibnamefont {Wang}}, \bibinfo {author}
  {\bibfnamefont {L.}~\bibnamefont {Li}}, \bibinfo {author} {\bibfnamefont
  {J.}~\bibnamefont {Yuan}}, \ and\ \bibinfo {author} {\bibfnamefont
  {W.}~\bibnamefont {Li}},\ }\href {\doibase 10.1103/PhysRevLett.124.140401}
  {\bibfield  {journal} {\bibinfo  {journal} {Phys. Rev. Lett.}\ }\textbf
  {\bibinfo {volume} {124}},\ \bibinfo {pages} {140401} (\bibinfo {year}
  {2020}{\natexlab{b}})}\BibitemShut {NoStop}%
\bibitem [{\citenamefont {Su}\ \emph {et~al.}(2020{\natexlab{b}})\citenamefont
  {Su}, \citenamefont {Guo}, \citenamefont {Wu}, \citenamefont {Jin},
  \citenamefont {Shao},\ and\ \citenamefont {Zhang}}]{Su_2020epl}%
  \BibitemOpen
  \bibfield  {author} {\bibinfo {author} {\bibfnamefont {S.-L.}\ \bibnamefont
  {Su}}, \bibinfo {author} {\bibfnamefont {F.-Q.}\ \bibnamefont {Guo}},
  \bibinfo {author} {\bibfnamefont {J.-L.}\ \bibnamefont {Wu}}, \bibinfo
  {author} {\bibfnamefont {Z.}~\bibnamefont {Jin}}, \bibinfo {author}
  {\bibfnamefont {X.~Q.}\ \bibnamefont {Shao}}, \ and\ \bibinfo {author}
  {\bibfnamefont {S.}~\bibnamefont {Zhang}},\ }\href {\doibase
  10.1209/0295-5075/131/53001} {\bibfield  {journal} {\bibinfo  {journal}
  {{EPL} (Europhysics Letters)}\ }\textbf {\bibinfo {volume} {131}},\ \bibinfo
  {pages} {53001} (\bibinfo {year} {2020}{\natexlab{b}})}\BibitemShut {NoStop}%
\bibitem [{\citenamefont {Su}\ \emph {et~al.}(2016)\citenamefont {Su},
  \citenamefont {Liang}, \citenamefont {Zhang}, \citenamefont {Wen},
  \citenamefont {Sun}, \citenamefont {Jin},\ and\ \citenamefont
  {Zhu}}]{Su2016}%
  \BibitemOpen
  \bibfield  {author} {\bibinfo {author} {\bibfnamefont {S.-L.}\ \bibnamefont
  {Su}}, \bibinfo {author} {\bibfnamefont {E.}~\bibnamefont {Liang}}, \bibinfo
  {author} {\bibfnamefont {S.}~\bibnamefont {Zhang}}, \bibinfo {author}
  {\bibfnamefont {J.-J.}\ \bibnamefont {Wen}}, \bibinfo {author} {\bibfnamefont
  {L.-L.}\ \bibnamefont {Sun}}, \bibinfo {author} {\bibfnamefont
  {Z.}~\bibnamefont {Jin}}, \ and\ \bibinfo {author} {\bibfnamefont {A.-D.}\
  \bibnamefont {Zhu}},\ }\href {\doibase 10.1103/PhysRevA.93.012306} {\bibfield
   {journal} {\bibinfo  {journal} {Phys. Rev. A}\ }\textbf {\bibinfo {volume}
  {93}},\ \bibinfo {pages} {012306} (\bibinfo {year} {2016})}\BibitemShut
  {NoStop}%
\bibitem [{\citenamefont {James}(2000)}]{effective1}%
  \BibitemOpen
  \bibfield  {author} {\bibinfo {author} {\bibfnamefont {D.}~\bibnamefont
  {James}},\ }\href@noop {} {\bibfield  {journal} {\bibinfo  {journal}
  {Fortschritte der Physik}\ }\textbf {\bibinfo {volume} {48}},\ \bibinfo
  {pages} {823} (\bibinfo {year} {2000})}\BibitemShut {NoStop}%
\bibitem [{\citenamefont {S\o{}rensen}\ and\ \citenamefont
  {M\o{}lmer}(2002)}]{effective2}%
  \BibitemOpen
  \bibfield  {author} {\bibinfo {author} {\bibfnamefont {A.~S.}\ \bibnamefont
  {S\o{}rensen}}\ and\ \bibinfo {author} {\bibfnamefont {K.}~\bibnamefont
  {M\o{}lmer}},\ }\href {\doibase 10.1103/PhysRevA.66.022314} {\bibfield
  {journal} {\bibinfo  {journal} {Phys. Rev. A}\ }\textbf {\bibinfo {volume}
  {66}},\ \bibinfo {pages} {022314} (\bibinfo {year} {2002})}\BibitemShut
  {NoStop}%
\bibitem [{\citenamefont {James}\ and\ \citenamefont
  {Jerke}(2007)}]{effective3}%
  \BibitemOpen
  \bibfield  {author} {\bibinfo {author} {\bibfnamefont {D.~F.}\ \bibnamefont
  {James}}\ and\ \bibinfo {author} {\bibfnamefont {J.}~\bibnamefont {Jerke}},\
  }\href {\doibase 10.1139/p07-060} {\bibfield  {journal} {\bibinfo  {journal}
  {Canadian Journal of Physics}\ }\textbf {\bibinfo {volume} {85}},\ \bibinfo
  {pages} {625} (\bibinfo {year} {2007})}\BibitemShut {NoStop}%
\bibitem [{\citenamefont {Gamel}\ and\ \citenamefont
  {James}(2010)}]{effective4}%
  \BibitemOpen
  \bibfield  {author} {\bibinfo {author} {\bibfnamefont {O.}~\bibnamefont
  {Gamel}}\ and\ \bibinfo {author} {\bibfnamefont {D.~F.~V.}\ \bibnamefont
  {James}},\ }\href {\doibase 10.1103/PhysRevA.82.052106} {\bibfield  {journal}
  {\bibinfo  {journal} {Phys. Rev. A}\ }\textbf {\bibinfo {volume} {82}},\
  \bibinfo {pages} {052106} (\bibinfo {year} {2010})}\BibitemShut {NoStop}%
\bibitem [{\citenamefont {Theodosiou}(1984)}]{lifetime1}%
  \BibitemOpen
  \bibfield  {author} {\bibinfo {author} {\bibfnamefont {C.~E.}\ \bibnamefont
  {Theodosiou}},\ }\href {\doibase 10.1103/PhysRevA.30.2881} {\bibfield
  {journal} {\bibinfo  {journal} {Phys. Rev. A}\ }\textbf {\bibinfo {volume}
  {30}},\ \bibinfo {pages} {2881} (\bibinfo {year} {1984})}\BibitemShut
  {NoStop}%
\bibitem [{\citenamefont {Beterov}\ \emph {et~al.}(2009)\citenamefont
  {Beterov}, \citenamefont {Ryabtsev}, \citenamefont {Tretyakov},\ and\
  \citenamefont {Entin}}]{lifetime2}%
  \BibitemOpen
  \bibfield  {author} {\bibinfo {author} {\bibfnamefont {I.~I.}\ \bibnamefont
  {Beterov}}, \bibinfo {author} {\bibfnamefont {I.~I.}\ \bibnamefont
  {Ryabtsev}}, \bibinfo {author} {\bibfnamefont {D.~B.}\ \bibnamefont
  {Tretyakov}}, \ and\ \bibinfo {author} {\bibfnamefont {V.~M.}\ \bibnamefont
  {Entin}},\ }\href {\doibase 10.1103/PhysRevA.79.052504} {\bibfield  {journal}
  {\bibinfo  {journal} {Phys. Rev. A}\ }\textbf {\bibinfo {volume} {79}},\
  \bibinfo {pages} {052504} (\bibinfo {year} {2009})}\BibitemShut {NoStop}%
\bibitem [{\citenamefont {van Ditzhuijzen}\ \emph {et~al.}(2008)\citenamefont
  {van Ditzhuijzen}, \citenamefont {Koenderink}, \citenamefont {Hern\'andez},
  \citenamefont {Robicheaux}, \citenamefont {Noordam},\ and\ \citenamefont
  {van~den Heuvell}}]{van2008}%
  \BibitemOpen
  \bibfield  {author} {\bibinfo {author} {\bibfnamefont {C.~S.~E.}\
  \bibnamefont {van Ditzhuijzen}}, \bibinfo {author} {\bibfnamefont {A.~F.}\
  \bibnamefont {Koenderink}}, \bibinfo {author} {\bibfnamefont {J.~V.}\
  \bibnamefont {Hern\'andez}}, \bibinfo {author} {\bibfnamefont
  {F.}~\bibnamefont {Robicheaux}}, \bibinfo {author} {\bibfnamefont {L.~D.}\
  \bibnamefont {Noordam}}, \ and\ \bibinfo {author} {\bibfnamefont {H.~B.
  v.~L.}\ \bibnamefont {van~den Heuvell}},\ }\href {\doibase
  10.1103/PhysRevLett.100.243201} {\bibfield  {journal} {\bibinfo  {journal}
  {Phys. Rev. Lett.}\ }\textbf {\bibinfo {volume} {100}},\ \bibinfo {pages}
  {243201} (\bibinfo {year} {2008})}\BibitemShut {NoStop}%
\bibitem [{\citenamefont {Paris-Mandoki}\ \emph {et~al.}(2016)\citenamefont
  {Paris-Mandoki}, \citenamefont {Gorniaczyk}, \citenamefont {Tresp},
  \citenamefont {Mirgorodskiy},\ and\ \citenamefont
  {Hofferberth}}]{Paris_Mandoki_2016}%
  \BibitemOpen
  \bibfield  {author} {\bibinfo {author} {\bibfnamefont {A.}~\bibnamefont
  {Paris-Mandoki}}, \bibinfo {author} {\bibfnamefont {H.}~\bibnamefont
  {Gorniaczyk}}, \bibinfo {author} {\bibfnamefont {C.}~\bibnamefont {Tresp}},
  \bibinfo {author} {\bibfnamefont {I.}~\bibnamefont {Mirgorodskiy}}, \ and\
  \bibinfo {author} {\bibfnamefont {S.}~\bibnamefont {Hofferberth}},\ }\href
  {\doibase 10.1088/0953-4075/49/16/164001} {\bibfield  {journal} {\bibinfo
  {journal} {J. Phys. B: Atom., Mol. Opt. Phys.}\ }\textbf {\bibinfo {volume}
  {49}},\ \bibinfo {pages} {164001} (\bibinfo {year} {2016})}\BibitemShut
  {NoStop}%
\bibitem [{\citenamefont {Walker}\ and\ \citenamefont
  {Saffman}(2008)}]{Walker2008}%
  \BibitemOpen
  \bibfield  {author} {\bibinfo {author} {\bibfnamefont {T.~G.}\ \bibnamefont
  {Walker}}\ and\ \bibinfo {author} {\bibfnamefont {M.}~\bibnamefont
  {Saffman}},\ }\href {\doibase 10.1103/PhysRevA.77.032723} {\bibfield
  {journal} {\bibinfo  {journal} {Phys. Rev. A}\ }\textbf {\bibinfo {volume}
  {77}},\ \bibinfo {pages} {032723} (\bibinfo {year} {2008})}\BibitemShut
  {NoStop}%
\bibitem [{\citenamefont {Gambetta}\ \emph
  {et~al.}(2020{\natexlab{b}})\citenamefont {Gambetta}, \citenamefont {Li},
  \citenamefont {Schmidt-Kaler},\ and\ \citenamefont
  {Lesanovsky}}]{gambetta_engineering_2020}%
  \BibitemOpen
  \bibfield  {author} {\bibinfo {author} {\bibfnamefont {F.~M.}\ \bibnamefont
  {Gambetta}}, \bibinfo {author} {\bibfnamefont {W.}~\bibnamefont {Li}},
  \bibinfo {author} {\bibfnamefont {F.}~\bibnamefont {Schmidt-Kaler}}, \ and\
  \bibinfo {author} {\bibfnamefont {I.}~\bibnamefont {Lesanovsky}},\ }\href
  {\doibase 10.1103/PhysRevLett.124.043402} {\bibfield  {journal} {\bibinfo
  {journal} {Phys. Rev. Lett.}\ }\textbf {\bibinfo {volume} {124}},\ \bibinfo
  {pages} {043402} (\bibinfo {year} {2020}{\natexlab{b}})}\BibitemShut
  {NoStop}%
\bibitem [{\citenamefont {Gambetta}\ \emph
  {et~al.}(2020{\natexlab{c}})\citenamefont {Gambetta}, \citenamefont {Zhang},
  \citenamefont {Hennrich}, \citenamefont {Lesanovsky},\ and\ \citenamefont
  {Li}}]{gambetta_long-range_2020}%
  \BibitemOpen
  \bibfield  {author} {\bibinfo {author} {\bibfnamefont {F.~M.}\ \bibnamefont
  {Gambetta}}, \bibinfo {author} {\bibfnamefont {C.}~\bibnamefont {Zhang}},
  \bibinfo {author} {\bibfnamefont {M.}~\bibnamefont {Hennrich}}, \bibinfo
  {author} {\bibfnamefont {I.}~\bibnamefont {Lesanovsky}}, \ and\ \bibinfo
  {author} {\bibfnamefont {W.}~\bibnamefont {Li}},\ }\href {\doibase
  10.1103/PhysRevLett.125.133602} {\bibfield  {journal} {\bibinfo  {journal}
  {Phys. Rev. Lett.}\ }\textbf {\bibinfo {volume} {125}},\ \bibinfo {pages}
  {133602} (\bibinfo {year} {2020}{\natexlab{c}})}\BibitemShut {NoStop}%
\bibitem [{\citenamefont {Zhang}\ \emph {et~al.}(2020)\citenamefont {Zhang},
  \citenamefont {Pokorny}, \citenamefont {Li}, \citenamefont {Higgins},
  \citenamefont {P{\"o}schl}, \citenamefont {Lesanovsky},\ and\ \citenamefont
  {Hennrich}}]{zhang_submicrosecond_2020}%
  \BibitemOpen
  \bibfield  {author} {\bibinfo {author} {\bibfnamefont {C.}~\bibnamefont
  {Zhang}}, \bibinfo {author} {\bibfnamefont {F.}~\bibnamefont {Pokorny}},
  \bibinfo {author} {\bibfnamefont {W.}~\bibnamefont {Li}}, \bibinfo {author}
  {\bibfnamefont {G.}~\bibnamefont {Higgins}}, \bibinfo {author} {\bibfnamefont
  {A.}~\bibnamefont {P{\"o}schl}}, \bibinfo {author} {\bibfnamefont
  {I.}~\bibnamefont {Lesanovsky}}, \ and\ \bibinfo {author} {\bibfnamefont
  {M.}~\bibnamefont {Hennrich}},\ }\href {\doibase 10.1038/s41586-020-2152-9}
  {\bibfield  {journal} {\bibinfo  {journal} {Nature}\ }\textbf {\bibinfo
  {volume} {580}},\ \bibinfo {pages} {345} (\bibinfo {year}
  {2020})}\BibitemShut {NoStop}%
\bibitem [{\citenamefont {Qian}\ \emph {et~al.}(2009)\citenamefont {Qian},
  \citenamefont {Qian}, \citenamefont {Ke}, \citenamefont {Feng}, \citenamefont
  {Oh},\ and\ \citenamefont {Wang}}]{Qianjun2009}%
  \BibitemOpen
  \bibfield  {author} {\bibinfo {author} {\bibfnamefont {J.}~\bibnamefont
  {Qian}}, \bibinfo {author} {\bibfnamefont {Y.}~\bibnamefont {Qian}}, \bibinfo
  {author} {\bibfnamefont {M.}~\bibnamefont {Ke}}, \bibinfo {author}
  {\bibfnamefont {X.-L.}\ \bibnamefont {Feng}}, \bibinfo {author}
  {\bibfnamefont {C.~H.}\ \bibnamefont {Oh}}, \ and\ \bibinfo {author}
  {\bibfnamefont {Y.}~\bibnamefont {Wang}},\ }\href {\doibase
  10.1103/PhysRevA.80.053413} {\bibfield  {journal} {\bibinfo  {journal} {Phys.
  Rev. A}\ }\textbf {\bibinfo {volume} {80}},\ \bibinfo {pages} {053413}
  (\bibinfo {year} {2009})}\BibitemShut {NoStop}%
\bibitem [{\citenamefont {Shi}(2019)}]{Shi2019}%
  \BibitemOpen
  \bibfield  {author} {\bibinfo {author} {\bibfnamefont {X.-F.}\ \bibnamefont
  {Shi}},\ }\href {\doibase 10.1103/PhysRevApplied.11.044035} {\bibfield
  {journal} {\bibinfo  {journal} {Phys. Rev. Applied}\ }\textbf {\bibinfo
  {volume} {11}},\ \bibinfo {pages} {044035} (\bibinfo {year}
  {2019})}\BibitemShut {NoStop}%
\bibitem [{\citenamefont {Sj\"oqvist}\ \emph {et~al.}(2012)\citenamefont
  {Sj\"oqvist}, \citenamefont {Tong}, \citenamefont {Andersson}, \citenamefont
  {Hessmo}, \citenamefont {Johansson},\ and\ \citenamefont
  {Singh}}]{Sj_qvist_2012}%
  \BibitemOpen
  \bibfield  {author} {\bibinfo {author} {\bibfnamefont {E.}~\bibnamefont
  {Sj\"oqvist}}, \bibinfo {author} {\bibfnamefont {D.~M.}\ \bibnamefont
  {Tong}}, \bibinfo {author} {\bibfnamefont {L.~M.}\ \bibnamefont {Andersson}},
  \bibinfo {author} {\bibfnamefont {B.}~\bibnamefont {Hessmo}}, \bibinfo
  {author} {\bibfnamefont {M.}~\bibnamefont {Johansson}}, \ and\ \bibinfo
  {author} {\bibfnamefont {K.}~\bibnamefont {Singh}},\ }\href {\doibase
  10.1088/1367-2630/14/10/103035} {\bibfield  {journal} {\bibinfo  {journal}
  {New J. Phys.}\ }\textbf {\bibinfo {volume} {14}},\ \bibinfo {pages} {103035}
  (\bibinfo {year} {2012})}\BibitemShut {NoStop}%
\bibitem [{\citenamefont {Xu}\ \emph {et~al.}(2012)\citenamefont {Xu},
  \citenamefont {Zhang}, \citenamefont {Tong}, \citenamefont {Sj\"oqvist},\
  and\ \citenamefont {Kwek}}]{Xu2012}%
  \BibitemOpen
  \bibfield  {author} {\bibinfo {author} {\bibfnamefont {G.~F.}\ \bibnamefont
  {Xu}}, \bibinfo {author} {\bibfnamefont {J.}~\bibnamefont {Zhang}}, \bibinfo
  {author} {\bibfnamefont {D.~M.}\ \bibnamefont {Tong}}, \bibinfo {author}
  {\bibfnamefont {E.}~\bibnamefont {Sj\"oqvist}}, \ and\ \bibinfo {author}
  {\bibfnamefont {L.~C.}\ \bibnamefont {Kwek}},\ }\href {\doibase
  10.1103/PhysRevLett.109.170501} {\bibfield  {journal} {\bibinfo  {journal}
  {Phys. Rev. Lett.}\ }\textbf {\bibinfo {volume} {109}},\ \bibinfo {pages}
  {170501} (\bibinfo {year} {2012})}\BibitemShut {NoStop}%
\bibitem [{\citenamefont {Xue}\ \emph {et~al.}(2016)\citenamefont {Xue},
  \citenamefont {Zhou}, \citenamefont {Chu},\ and\ \citenamefont
  {Hu}}]{Xue2016}%
  \BibitemOpen
  \bibfield  {author} {\bibinfo {author} {\bibfnamefont {Z.-Y.}\ \bibnamefont
  {Xue}}, \bibinfo {author} {\bibfnamefont {J.}~\bibnamefont {Zhou}}, \bibinfo
  {author} {\bibfnamefont {Y.-M.}\ \bibnamefont {Chu}}, \ and\ \bibinfo
  {author} {\bibfnamefont {Y.}~\bibnamefont {Hu}},\ }\href {\doibase
  10.1103/PhysRevA.94.022331} {\bibfield  {journal} {\bibinfo  {journal} {Phys.
  Rev. A}\ }\textbf {\bibinfo {volume} {94}},\ \bibinfo {pages} {022331}
  (\bibinfo {year} {2016})}\BibitemShut {NoStop}%
\bibitem [{\citenamefont {Zhao}\ \emph {et~al.}(2017)\citenamefont {Zhao},
  \citenamefont {Cui}, \citenamefont {Xu}, \citenamefont {Sj\"oqvist},\ and\
  \citenamefont {Tong}}]{Zhao2017}%
  \BibitemOpen
  \bibfield  {author} {\bibinfo {author} {\bibfnamefont {P.~Z.}\ \bibnamefont
  {Zhao}}, \bibinfo {author} {\bibfnamefont {X.-D.}\ \bibnamefont {Cui}},
  \bibinfo {author} {\bibfnamefont {G.~F.}\ \bibnamefont {Xu}}, \bibinfo
  {author} {\bibfnamefont {E.}~\bibnamefont {Sj\"oqvist}}, \ and\ \bibinfo
  {author} {\bibfnamefont {D.~M.}\ \bibnamefont {Tong}},\ }\href {\doibase
  10.1103/PhysRevA.96.052316} {\bibfield  {journal} {\bibinfo  {journal} {Phys.
  Rev. A}\ }\textbf {\bibinfo {volume} {96}},\ \bibinfo {pages} {052316}
  (\bibinfo {year} {2017})}\BibitemShut {NoStop}%
\bibitem [{\citenamefont {Liu}\ \emph {et~al.}(2019)\citenamefont {Liu},
  \citenamefont {Song}, \citenamefont {Xue}, \citenamefont {Wang},\ and\
  \citenamefont {Yung}}]{Liu2019}%
  \BibitemOpen
  \bibfield  {author} {\bibinfo {author} {\bibfnamefont {B.-J.}\ \bibnamefont
  {Liu}}, \bibinfo {author} {\bibfnamefont {X.-K.}\ \bibnamefont {Song}},
  \bibinfo {author} {\bibfnamefont {Z.-Y.}\ \bibnamefont {Xue}}, \bibinfo
  {author} {\bibfnamefont {X.}~\bibnamefont {Wang}}, \ and\ \bibinfo {author}
  {\bibfnamefont {M.-H.}\ \bibnamefont {Yung}},\ }\href {\doibase
  10.1103/PhysRevLett.123.100501} {\bibfield  {journal} {\bibinfo  {journal}
  {Phys. Rev. Lett.}\ }\textbf {\bibinfo {volume} {123}},\ \bibinfo {pages}
  {100501} (\bibinfo {year} {2019})}\BibitemShut {NoStop}%
\bibitem [{\citenamefont {Aharonov}\ and\ \citenamefont
  {Anandan}(1987)}]{PhysRevLett.58.1593}%
  \BibitemOpen
  \bibfield  {author} {\bibinfo {author} {\bibfnamefont {Y.}~\bibnamefont
  {Aharonov}}\ and\ \bibinfo {author} {\bibfnamefont {J.}~\bibnamefont
  {Anandan}},\ }\href {\doibase 10.1103/PhysRevLett.58.1593} {\bibfield
  {journal} {\bibinfo  {journal} {Phys. Rev. Lett.}\ }\textbf {\bibinfo
  {volume} {58}},\ \bibinfo {pages} {1593} (\bibinfo {year}
  {1987})}\BibitemShut {NoStop}%
\bibitem [{\citenamefont {Rao}\ and\ \citenamefont
  {M\o{}lmer}(2013)}]{PhysRevLett.111.033606}%
  \BibitemOpen
  \bibfield  {author} {\bibinfo {author} {\bibfnamefont {D.~D.~B.}\
  \bibnamefont {Rao}}\ and\ \bibinfo {author} {\bibfnamefont {K.}~\bibnamefont
  {M\o{}lmer}},\ }\href {\doibase 10.1103/PhysRevLett.111.033606} {\bibfield
  {journal} {\bibinfo  {journal} {Phys. Rev. Lett.}\ }\textbf {\bibinfo
  {volume} {111}},\ \bibinfo {pages} {033606} (\bibinfo {year}
  {2013})}\BibitemShut {NoStop}%
\bibitem [{\citenamefont {de~L\'es\'eleuc}\ \emph {et~al.}(2017)\citenamefont
  {de~L\'es\'eleuc}, \citenamefont {Barredo}, \citenamefont {Lienhard},
  \citenamefont {Browaeys},\ and\ \citenamefont
  {Lahaye}}]{Opticalcontroldipole}%
  \BibitemOpen
  \bibfield  {author} {\bibinfo {author} {\bibfnamefont {S.}~\bibnamefont
  {de~L\'es\'eleuc}}, \bibinfo {author} {\bibfnamefont {D.}~\bibnamefont
  {Barredo}}, \bibinfo {author} {\bibfnamefont {V.}~\bibnamefont {Lienhard}},
  \bibinfo {author} {\bibfnamefont {A.}~\bibnamefont {Browaeys}}, \ and\
  \bibinfo {author} {\bibfnamefont {T.}~\bibnamefont {Lahaye}},\ }\href
  {\doibase 10.1103/PhysRevLett.119.053202} {\bibfield  {journal} {\bibinfo
  {journal} {Phys. Rev. Lett.}\ }\textbf {\bibinfo {volume} {119}},\ \bibinfo
  {pages} {053202} (\bibinfo {year} {2017})}\BibitemShut {NoStop}%
\bibitem [{\citenamefont {Afrousheh}\ \emph {et~al.}(2004)\citenamefont
  {Afrousheh}, \citenamefont {Bohlouli-Zanjani}, \citenamefont {Vagale},
  \citenamefont {Mugford}, \citenamefont {Fedorov},\ and\ \citenamefont
  {Martin}}]{Afrousheh2004}%
  \BibitemOpen
  \bibfield  {author} {\bibinfo {author} {\bibfnamefont {K.}~\bibnamefont
  {Afrousheh}}, \bibinfo {author} {\bibfnamefont {P.}~\bibnamefont
  {Bohlouli-Zanjani}}, \bibinfo {author} {\bibfnamefont {D.}~\bibnamefont
  {Vagale}}, \bibinfo {author} {\bibfnamefont {A.}~\bibnamefont {Mugford}},
  \bibinfo {author} {\bibfnamefont {M.}~\bibnamefont {Fedorov}}, \ and\
  \bibinfo {author} {\bibfnamefont {J.~D.~D.}\ \bibnamefont {Martin}},\ }\href
  {\doibase 10.1103/PhysRevLett.93.233001} {\bibfield  {journal} {\bibinfo
  {journal} {Phys. Rev. Lett.}\ }\textbf {\bibinfo {volume} {93}},\ \bibinfo
  {pages} {233001} (\bibinfo {year} {2004})}\BibitemShut {NoStop}%
\bibitem [{\citenamefont {Sevin{\c{c}}li}\ and\ \citenamefont
  {Pohl}(2014)}]{Sevin_li_2014}%
  \BibitemOpen
  \bibfield  {author} {\bibinfo {author} {\bibfnamefont {S.}~\bibnamefont
  {Sevin{\c{c}}li}}\ and\ \bibinfo {author} {\bibfnamefont {T.}~\bibnamefont
  {Pohl}},\ }\href {\doibase 10.1088/1367-2630/16/12/123036} {\bibfield
  {journal} {\bibinfo  {journal} {New J. Phys.}\ }\textbf {\bibinfo {volume}
  {16}},\ \bibinfo {pages} {123036} (\bibinfo {year} {2014})}\BibitemShut
  {NoStop}%
\bibitem [{\citenamefont {Marcuzzi}\ \emph {et~al.}(2015)\citenamefont
  {Marcuzzi}, \citenamefont {Levi}, \citenamefont {Li}, \citenamefont
  {Garrahan}, \citenamefont {Olmos},\ and\ \citenamefont
  {Lesanovsky}}]{Marcuzzi_2015}%
  \BibitemOpen
  \bibfield  {author} {\bibinfo {author} {\bibfnamefont {M.}~\bibnamefont
  {Marcuzzi}}, \bibinfo {author} {\bibfnamefont {E.}~\bibnamefont {Levi}},
  \bibinfo {author} {\bibfnamefont {W.}~\bibnamefont {Li}}, \bibinfo {author}
  {\bibfnamefont {J.~P.}\ \bibnamefont {Garrahan}}, \bibinfo {author}
  {\bibfnamefont {B.}~\bibnamefont {Olmos}}, \ and\ \bibinfo {author}
  {\bibfnamefont {I.}~\bibnamefont {Lesanovsky}},\ }\href {\doibase
  10.1088/1367-2630/17/7/072003} {\bibfield  {journal} {\bibinfo  {journal}
  {New J. Phys.}\ }\textbf {\bibinfo {volume} {17}},\ \bibinfo {pages} {072003}
  (\bibinfo {year} {2015})}\BibitemShut {NoStop}%
\bibitem [{\citenamefont {Gambetta}\ \emph
  {et~al.}(2020{\natexlab{d}})\citenamefont {Gambetta}, \citenamefont {Li},
  \citenamefont {Schmidt-Kaler},\ and\ \citenamefont
  {Lesanovsky}}]{Gambetta2020}%
  \BibitemOpen
  \bibfield  {author} {\bibinfo {author} {\bibfnamefont {F.~M.}\ \bibnamefont
  {Gambetta}}, \bibinfo {author} {\bibfnamefont {W.}~\bibnamefont {Li}},
  \bibinfo {author} {\bibfnamefont {F.}~\bibnamefont {Schmidt-Kaler}}, \ and\
  \bibinfo {author} {\bibfnamefont {I.}~\bibnamefont {Lesanovsky}},\ }\href
  {\doibase 10.1103/PhysRevLett.124.043402} {\bibfield  {journal} {\bibinfo
  {journal} {Phys. Rev. Lett.}\ }\textbf {\bibinfo {volume} {124}},\ \bibinfo
  {pages} {043402} (\bibinfo {year} {2020}{\natexlab{d}})}\BibitemShut
  {NoStop}%
\bibitem [{\citenamefont {Walker}\ and\ \citenamefont
  {Saffman}(2005)}]{Walker_2005}%
  \BibitemOpen
  \bibfield  {author} {\bibinfo {author} {\bibfnamefont {T.~G.}\ \bibnamefont
  {Walker}}\ and\ \bibinfo {author} {\bibfnamefont {M.}~\bibnamefont
  {Saffman}},\ }\href {\doibase 10.1088/0953-4075/38/2/022} {\bibfield
  {journal} {\bibinfo  {journal} {J. Phys. B}\ }\textbf {\bibinfo {volume}
  {38}},\ \bibinfo {pages} {S309} (\bibinfo {year} {2005})}\BibitemShut
  {NoStop}%
\bibitem [{\citenamefont {Vogt}\ \emph {et~al.}(2007)\citenamefont {Vogt},
  \citenamefont {Viteau}, \citenamefont {Chotia}, \citenamefont {Zhao},
  \citenamefont {Comparat},\ and\ \citenamefont {Pillet}}]{Vogt2007}%
  \BibitemOpen
  \bibfield  {author} {\bibinfo {author} {\bibfnamefont {T.}~\bibnamefont
  {Vogt}}, \bibinfo {author} {\bibfnamefont {M.}~\bibnamefont {Viteau}},
  \bibinfo {author} {\bibfnamefont {A.}~\bibnamefont {Chotia}}, \bibinfo
  {author} {\bibfnamefont {J.}~\bibnamefont {Zhao}}, \bibinfo {author}
  {\bibfnamefont {D.}~\bibnamefont {Comparat}}, \ and\ \bibinfo {author}
  {\bibfnamefont {P.}~\bibnamefont {Pillet}},\ }\href {\doibase
  10.1103/PhysRevLett.99.073002} {\bibfield  {journal} {\bibinfo  {journal}
  {Phys. Rev. Lett.}\ }\textbf {\bibinfo {volume} {99}},\ \bibinfo {pages}
  {073002} (\bibinfo {year} {2007})}\BibitemShut {NoStop}%
\bibitem [{\citenamefont {Ryabtsev}\ \emph {et~al.}(2010)\citenamefont
  {Ryabtsev}, \citenamefont {Tretyakov}, \citenamefont {Beterov},\ and\
  \citenamefont {Entin}}]{Ryabtsev2010}%
  \BibitemOpen
  \bibfield  {author} {\bibinfo {author} {\bibfnamefont {I.~I.}\ \bibnamefont
  {Ryabtsev}}, \bibinfo {author} {\bibfnamefont {D.~B.}\ \bibnamefont
  {Tretyakov}}, \bibinfo {author} {\bibfnamefont {I.~I.}\ \bibnamefont
  {Beterov}}, \ and\ \bibinfo {author} {\bibfnamefont {V.~M.}\ \bibnamefont
  {Entin}},\ }\href {\doibase 10.1103/PhysRevLett.104.073003} {\bibfield
  {journal} {\bibinfo  {journal} {Phys. Rev. Lett.}\ }\textbf {\bibinfo
  {volume} {104}},\ \bibinfo {pages} {073003} (\bibinfo {year}
  {2010})}\BibitemShut {NoStop}%
\end{thebibliography}%

\end{document}